\documentclass[12pt,article,hyper]{JHEP}
\usepackage{epsfig}
\usepackage{amssymb}
\usepackage{graphics}
\usepackage{feynmf}
\usepackage{amsmath}      
\let\a=\alpha   \let\b=\beta   \let\g=\gamma   \let\d=\delta
\let\e=\epsilon \let\z=\zeta   \let\h=\eta     \let\q=\theta
\let\i=\iota      \let\l=\lambda  \let\m=\mu
\let\n=\nu      \let\x=\xi     \let\p=\pi      \let\r=\rho
\let\s=\sigma  \let\t=\tau      \let\f=\phi
\let\c=\chi     \let\y=\psi    
\let\G=\Gamma  \let\D=\Delta   \let\L=\Lambda
           
\let\F=\Phi    \let\Y=\Psi       \let\W=\Omega
   \let\vf=\varphi
\def\sla{\raise.15ex\hbox{$/$}\kern-.57em}

\def\ie{{\it i.e.\ }}
\newcommand{\Realint}{\mathbb R}
\newcommand{\Zint}{\mathbb{Z}}

\newcommand{\be}{\begin{equation}}
\newcommand{\ee}{\end{equation}}
\newcommand{\bea}{\begin{eqnarray}}
\newcommand{\eea}{\end{eqnarray}}
\newcommand{\ba}{\begin{array}}
\newcommand{\ea}{\end{array}}
\def\nn{\nonumber}

\newcommand{\drawsquare}[2]{\hbox{%
\rule{#2pt}{#1pt}\hskip-#2pt
\rule{#1pt}{#2pt}\hskip-#1pt
\rule[#1pt]{#1pt}{#2pt}}\rule[#1pt]{#2pt}{#2pt}\hskip-#2pt
\rule{#2pt}{#1pt}}
\newcommand{\fund}{\raisebox{-.5pt}{\drawsquare{6.5}{0.4}}}
\newcommand{\Yasymm}{\raisebox{-3.5pt}{\drawsquare{6.5}{0.4}}\hskip-6.9pt%
        \raisebox{3pt}{\drawsquare{6.5}{0.4}}}
\newcommand{\antifund}{\overline{\fund}}
\newcommand{\bYasymm}{\overline{\Yasymm}}
\newcommand{\YasymmS}{\raisebox{-1.9pt}{\drawsquare{4}{0.4}}\hskip-4.4pt%
        \raisebox{2.1pt}{\drawsquare{4}{0.4}}}
\newcommand{\bYasymmS}{\overline{\YasymmS}}
\title{Thesis: Orientifolds, Anomalies and the Standard Model}
\author{Anastasopoulos Pascal\footnote{panasta@physics.uoc.gr}\\
Department of Physics, University of Crete\\
GR-710 03 Heraklion, GREECE\\
\smallskip
and\\
\smallskip
Laboratoire de Physique Th{\'e}orique Ecole
Polytechnique\\
91128, Palaiseau, FRANCE}
\preprint{\hepth{0503055}}
\abstract{In this thesis, we study aspects of D-brane realizations
of the Standard Model. Specifically, we study orientifold models
with rotation and translation elements that break supersymmetry,
provide the general consistency conditions and derive the massless
spectrum for these type of orientifolds. These models contain in
general anomalous $U(1)$ gauge fields. The Green-Schwarz mechanism
cancels the anomaly and provides a mass term for the anomalous
gauge fields. We calculate the bare mass for supersymmetric and
non-supersymmetric vacua and we show that higher dimensional
anomalies can affect the masses of the anomalous $U(1)$s.
Phenomenological aspects are also discussed. We evaluate the
contribution of the extra $U(1)$ fields to the anomalous moments
and it is shown that this imposes constraints on the magnitude of
the string scale.}
\begin{document}
%
%

\section{Introduction}

The Standard Model (SM) of physics has been confirmed to a great
accuracy in many experiments. Despite the fact that the Higgs
particle remains experimentally elusive, few doubt that there will
be major surprises in this direction.

In a related direction, however there is concrete experimental
evidence that neutrinos have (tiny) masses and mixings and the SM
should be extended to accommodate this. Many ideas exist on how
this can be achieved and we are awaiting experimental evidence to
delineate any particular direction.
On the other hand there are some theoretical issues that make
physicists believe that the SM is not the final story. Some of
these are:
\begin{itemize}

\item (Quantum) gravity is not incorporated. It is not a
renormalizable theory.

\item The SM suffers from the $hierarchy$ $problem$. It is
believed that SM particles are coming from a Ground Unified Theory
that spontaneously broke to $SU(3)\times SU(2) \times U(1)$. The
breaking scale of this unified theory turns out to be $M_U\sim
10^{16}$GeV, which is very far from the electroweak breaking
scale. In order to evaluate the potential for the Higgs, we have
to fine-tune the parameters that receive contributions from all
orders in perturbation theory.

\end{itemize}

Several ideas have been put forward to deal with the large
hierarchy of scales. One is the so called theory of $technicolor$
which considers all scalar fields in the SM as bound states of
fermions by a new set of interactions \cite{Farhi:1980xs}.

Another idea is a new symmetry, the so-called $supersymmetry$ that
relates fermions and bosons. Supersymmetry (if it exists) is
obviously broken at low energy. However, it solves the hierarchy
problem, since above the supersymmetry breaking scale there are no
radiative corrections to the masses of the fields.


String theory, after an initial short life as a theory of hadrons,
regained popularity because it was found to be the only consistent
framework known that provided a workable unified theory of quantum
gravity \cite{Green:sp,Polchinski:rq,Kiritsis:1997hj,Johnson:gi}.

Initially, it was the heterotic string theory \cite{het} that
first provided a picture of grand unification, supersymmetry and
quantum gravity.
There were several models which at low energies came close to the
SM \cite{hetmod}.
It predicts that the unification scale is close to the
four-dimensional Planck scale, $M_P\sim 10^{19}$GeV, giving an
answer to the hierarchy problem.


Recently, other ten-dimensional supersymmetric string theories
(type-IIA/B closed and Type I closed and open strings) have come
into focus. Moreover, It has been shown that all superstring
theories are vacuum states of a larger eleven-dimensional theory
so-called $M$-$theory$ \cite{ht,w}, with non-perturbative
dualities relating the strong coupling behavior of one theory to
the weak coupling behavior of another.
The resurgence of interest in these theories is also partly due to
the discovery of solitonic objects (D-branes and NS5-branes) that
are contained in these theories. In particular, D-branes have
provided a new geometrical interpretation of gauge theories
\cite{pol, Bachas:1998rg}.

In this search for new string vacua, a new possibility also
emerged, namely that the string scale could be much lower than the
four-dimensional Planck scale and in particular it could be as low
as a few TeV \cite{low}, opening new avenues for experimental
confirmation of all such theories that had until now been
considered more as mathematical structures than as physical
models. Such ground states are the so called $orientifolds$ that
are generalized compactifications of type I string theory
(compactifications of superstring theories are expected since
their critical dimension is $D=10$). Crucially, these models
contain D-branes whose (localized) fluctuations should describe
the SM particles while gravity is naturally included in the closed
sector.

In such orientifold models the non-abelian couplings of the
D-brane gauge fields and the relation between the four-dimensional
Planck scale and the string scale is given by:
\bea {1\over g^2_{YM}}={V_{||}\over g_s}
~~~~~,~~~~~
{M^2_P\over M^2_s}={V_6\over g_s V_{||}}~,\eea
where $V_6$ and $V_{||}$ are the volumes of the 6-dimensional
compact manifold and the longitudinal sub-manifold of the D-branes
respectively. Therefore, if $V_6/V_{||}\gg 1$ and if the theory is
kept perturbative $g_s<1$, the string scale can be anywhere
between the Planck scale and a few TeV.

Supersymmetry breaking in these models is achieved by various
geometric mechanisms, such as:
\begin{itemize}
\item Intersecting branes \cite{bachas,bdl}.
\item Non-freely acting supersymmetry-breaking orbifolds that
generically induce breaking in the open sector.
\item Freely acting supersymmetry-breaking orbifolds, such as the
Scherk-Schwarz mechanism that we will explore in detail later on.
\cite{ss,sss}
\end{itemize}

In these models there is no hierarchy problem, since above the
string scale there is no field theoretic running of couplings.
However, a low string scale requires some of the internal
dimensions to be larger than the string scale. Therefore, the
``old" problem changes form and maybe rephrased as the ``new"
hierarchy problem: why the minimum of the potential of the moduli
is at $R\gg 1$? This question still remains an open problem in
string theory.

In this thesis we study some aspects of D-brane realizations of
the SM. We start with an introductory chapter to string theory and
superstring theory and also discuss some issues of
compactifications and orbifold constructions.
The following chapter explains the foundations of unoriented open
and closed string theories, the orientifolds. The presence of
extended dynamical objects (D-branes) is necessary to make the
theory consistent \cite{Pradisi:1988xd, Gimon:1996rq,
Berkooz:1996dw, Ibanez:1998qp, Angelantonj:2002ct}.

During my thesis, I worked on this field of research in
collaboration with A.B. Hammou and N. Irges and we provided
general consistency conditions for supersymmetric and
non-supersymmetric orientifolds (Scherk-Schwarz deformation
breaking \cite{sss, Antoniadis:1998ki, Cotrone:1999xs,
Scrucca:2001ni, Ib1, Blum1, Anastasopoulos:2003ha}) and we also
gave the general structure of the massless spectrum of these
models.
%


As we mentioned above, in the orientifold models gauge
interactions are described by open strings whose ends are confined
on the D-branes, while gravity is mediated by closed strings in
the bulk \cite{Aldazabal:2000dg, Blumenhagen:2000ea,
Cvetic:2002qa, Bailin:2000kd, Kokorelis:2002ip,
Antoniadis:2000en}. Ordinary matter is preferably generated by the
fluctuations of the open strings and is thus also localized on the
appropriate D-branes.
Consistency conditions and Wilson lines can provide a D-brane
configuration that will localize the Standard Model gauge group
and massless spectrum on a stack of 3 plus 2 plus 1 at least
D-branes. The rest of the D-branes being further away will not
affect the local properties of the model.


These D-brane configurations naturally provide some extra $U(1)$
gauge fields. Such $U(1)$ fields have generically four-dimensional
anomalies which are cancelled via the Green-Schwarz mechanism
\cite{Green:sg, Sagnotti:1992qw, Ibanez:1998qp, Scrucca:2004jn,
Klein:1999im}. A scalar axionic field (zero-form, or its dual
two-form) is responsible for the cancellation of the anomalies of
each anomalous boson. This mechanism gives a mass to the anomalous
$U(1)$ fields and breaks the associated gauge symmetry.

If the string scale is around a few TeV, observation of such
anomalous $U(1)$ gauge bosons becomes a realistic possibility
\cite{Kiritsis:2002aj, Ghilencea:2002da}.

As was shown in \cite{Antoniadis:2002cs}, it is possible to
compute the bare masses of the anomalous $U(1)$s by evaluating the
ultraviolet tadpole of the one-loop open string diagram with the
insertion of two gauge bosons on different boundaries. In this
limit, the diagrams of the annulus with both gauge bosons in the
same boundary and the M\"obius strip do not contribute when vacua
have cancelled tadpoles.

It turns out that $U(1)$ gauge fields that are free of
four-dimensional anomalies can still be massive
\cite{Ibanez:2001nd, Scrucca:2002is, Antoniadis:2002cs}.
Herein we show that this is due to the presence of mass-generating
six-dimensional anomalies. If there are decompactification limits
in the theory, then six-dimensional anomalies can affect
four-dimensional masses. This work was result of my research:
In six dimensions, two types of field are necessary to cancel the
anomalies, a scalar axion and a two-form. There is also a
four-form field but it is dual to the scalar. Via the
Green-Schwarz mechanism, the pseudoscalar axions give mass to the
anomalous $U(1)$ fields. However, the two-forms are not involved
in mass generation.
It is shown that four-dimensional non-anomalous $U(1)$s can have
masses if their decompactification limits suffer from
six-dimensional anomalies. We calculate the masses of the
anomalous $U(1)$s of various six-dimensional orientifolds and we
compare our results with decompactification limits of the
four-dimensional orientifolds $Z'_6$ and $Z_6$
\cite{Anastasopoulos:2003aj}.


Chapter 5 is result of my research where we are interested
in the masses of the anomalous $U(1)$s in non-supersymmetric
models since such models are of the type that will eventually
represent the low energy physics of the Standard Model. In
particular, intersecting-brane realizations of the Standard Model
are generically non-supersymmetric.
We calculate the mass formulae using the ``background field
method" \cite{Bachas:bh} and find that they are the same as the
supersymmetric ones when we have cancellation of all tadpoles
\cite{Anastasopoulos:2004ga}. In cases where NSNS tadpoles do not
vanish, there are extra contributions proportional to the
non-vanishing tadpole terms.
The mass formulae derived earlier in this section are valid even
if we add Wilson lines that move the branes away from the fixed
points. The Wilson lines generically break the gauge group and
they will affect the masses of the anomalous $U(1)$s through the
traces of the model dependent $\g$ matrices.
The formulae, are applied to a $Z_2$ non-supersymmetric
orientifold model, with RR and NSNS tadpoles to be cancelled,
where supersymmetry is broken by a Scherk-Schwarz deformation
\cite{Anastasopoulos:2003ha}.


The Green-Schwarz mechanism is not the only source for the masses
of anomalous $U(1)$s. In Standard Model realizations, the Higgs is
necessarily charged under one of the anomalous $U(1)$s. As it was
described in \cite{Kiritsis:2003mc}, the Higgs contribution to the
mass of these $U(1)s$ is $g_A \sqrt{M^2+e_H^2 \langle H
\rangle^2}$, where $g_A$ is the gauge coupling of the anomalous
$U(1)$ and $e_H$ is the $U(1)$ charge of the Higgs.
The Higgs contribution to the $U(1)$ mass can be obtained from the
effective field theory unlike the ultraviolet mass which can only
be calculated in string theory.

In the last chapter, based on a D-brane realization of the
Standard Model \cite{Antoniadis:2000en}, we make some
phenomenological predictions and we evaluate the contribution of
the massive anomalous $U(1)$s to the anomalous magnetic moment
(AMM) of muon $\a_{muon}=(g-2)/2$.
These contributions are currently in the range allowed by
experiment. Finally, we use the precise measurement of
$\a_{muon}=(g-2)/2$ from the Brookhaven AGS experiment \cite{BNL}
to provide precise constrains for the masses of the anomalous
$U(1)$s in the TeV range.
This work has been done during my thesis, in collaboration with E.
Kiritsis \cite{Kiritsis:2002aj}.

\newpage

\section{String Theory}

\subsection{Bosonic String}

String theory is a quantum theory where the fundamental object is
a 1-dimensional element: $the$ $string$. The lagrangian that
describes such an object in flat space is the so-called
"Nambu-Goto" action:
\bea 
S_{NG}=-T \int dA \eea
This action is the direct generalization of the relativistic point
particle lagrangian where the mass of the particle has been
replaced by the tension of the string $T$ and the world-line $ds$
by the the world-volume $dA$.

Suppose that $\x_i$ with $i=1,2$ are coordinates of the
world-sheet and $G_{\m\n}$ is the metric of a $D$-dimensional
spacetime where the string propagates. 
%
%
%
%
%
If spacetime is flat $G_{\m\n}= \h_{\m\n}$, the Nambu-Goto action
takes the form:
\bea S_{NG}=-T \int \sqrt{-\det G_{ij}}d^2\x =-T \int
\sqrt{(\dot{X} X')^2- \dot{X}^2 X'^2}d^2\x \label{NGaction} \eea
where $G_{ij}=G_{\m\n} \partial_i X^\m \partial_j X^\n$ the
induced metric.

The square root in the Nambu-Goto action (\ref{NGaction}) makes
the treatment of the quantum theory quite complicated. To overcome
this difficulty, Polyakov introduced an intrinsic fluctuating
metric $g_{\a\b}$ on the worldsheet. For flat spacetime, his
action takes the form:
\bea S_P=-{1\over 4\p \a'}\int d^2\x \sqrt{-\det g}g^{\a\b}
\partial_\a X^\m \partial_\b X_\m + {\langle\vf\rangle \over
4\p}\int d^2\x \sqrt{-\det g}R \label{Paction}\eea
where $\vf$ the dilaton field. This action describes 2-dimensional
gravity coupled to $D$ worldsheet scalars. The last term in
(\ref{Paction}) is a topological invariant, the Euler character of
the $2D$ surface.

The stress tensor of the scalars is defined as the variation of
the matter-action with respect to the metric:
\bea T_{\a\b}=-{2\over T} {1\over \sqrt{-\det g}}{\d S_P \over \d
g^{\a\b}}= \partial_\a X^\m \partial_\b X_\m -{1\over 2} g_{\a\b}
g^{\g\d} \partial_\g X^\m \partial_\d X_\m \eea
The $2D$ Einstein equations give the classical solution for the
metric $g_{\a\b}$:
\bea T_{\a\b}=0 \quad\Rightarrow \quad g_{\a\b}=\partial_\a X^\m
\partial_\b X_\m \label{NG=P}\eea
Notice that the zero in the right part of (\ref{NG=P}) is due to
the fact that $2D$ gravity is Ricci flat.
Substituting back the classical solution to the Polyakov action we
find the Nambu-Goto action, where $T=(2\p \a')^{-1}$. Therefore,
the two actions are equivalent at least classically.

From now on we will take the Polyakov action as the starting point
of our study. The symmetries of this action are:
\begin{itemize}
\item Poincar\'e invariance
%
%
%
%
\item Local two dimensional reparametrization invariance
%
%
%
%
%
%
%
\item Conformal invariance
%
%
\end{itemize}

Using the above symmetries we can give to the Polyakov action a
convenient form. This is the so-called $conformal~gauge$ where the
worldsheet metric becomes flat:
\bea g_{\a\b}=\h_{\a\b}~. \eea
It is convenient to work in Euclidean signature by performing a
Wick rotation $\t\to -i\t$. We also make a conformal
transformation that maps a cylinder to a complex plane:
\bea z=e^{\t-i\s}~,~~~\bar{z}=e^{\t+i\s}~.
\eea
In the $z$ plane, equal times contours are concentric circles. The
$\t\to-\infty$ gets mapped to $z=0$.
%
%
%
%
%
The Polyakov action takes the form:
\bea S_P \sim \int d^2z ~\partial X^\m \bar{\partial} X_\m ~.\eea
The classical equations of motion (EOM) can be evaluated by
varying the action with respect to the fields. In the conformal
gauge, the EOM for the bosons are:
\bea \partial\bar{\partial} X^\m=0 ~.\label{EOM}\eea
Even if we have fixed the gauge, we have to impose the equations
which where found by the variation of the metric $g_{\a\b}$
(\ref{NG=P}):
\bea T_{\a\b}=0 ~.
\label{Virasoro}\eea
The later are known as the $Virasoro$ $constraints$.

\subsubsection{Solving the string equations of motion}

In general, there are two kinds of string with different boundary
conditions: $closed$ and $open$ strings:
\begin{itemize}
\item Closed Strings: $X^\m(\t,\s+2\p)=X^\m(\t,\s)$. The solution
is:
\bea &&X^\m (z,\bar{z}) =X^\m_L(z)+X^\m_R(\bar{z}) ~~~~~~~~~~~~~~
\textrm{where:}\nn\\
&&X^\m_L(z)={x^\m \over 2}-i{\a' \over 2} p^\m_L \log z +
i\sqrt{\a' \over 2}\sum_{k\neq 0}{\tilde{\a}^\m_k \over k} z^{-k}~,\nn\\
&&X^\m_R(\bar{z})={x^\m \over 2}-i{\a' \over 2} p^\m_R \log
\bar{z} + i\sqrt{\a' \over 2}\sum_{k\neq 0}{\a^\m_k \over k}
\bar{z}^{-k}~, \label{Closed}\eea
where for non-compact dimensions $p^\m_L=p^\m_R=p^\m$. The
$\a^\m_k, \tilde{\a}^\m_k$ are Fourier modes where $k$ runs over
all integers. Reality conditions give relations between opposite
sign Fourier modes: $(\a^\m_k)^* =\a^\m_{-k}$ and
$(\tilde{\a}^\m_k)^* =\tilde{\a}^\m_{-k}$.
\item Open Strings: There are two different boundary conditions
that can be imposed to the ends of an open string:
\begin{itemize}
\item Neumann boundary conditions (N): $\partial_\s
X^\m|_{end}=0$~,
\item Dirichlet boundary conditions (D): $\partial_\t
X^\m|_{end}=0$~.
\end{itemize}
These two choices, eliminate an extra boundary term that appears
from the variation of the Polyakov action.
Therefore, open strings can have different boundary conditions on
their endpoints. All the possible combinations are: NN, DD, ND
with different solutions:
\bea &&X^\m_{NN}(z,\bar{z})=x^\m-i\a'p^\m\log z\bar{z} + i
\sqrt{\a'\over 2}
\sum_{k\neq 0}{\a^\m_k \over k} (z^{-k}+\bar{z}^{-k})~,\nn\\
&&X^\m_{DD}(z,\bar{z})=-{c^\m \over 2\p}\log (z/ \bar{z}) + i
\sqrt{\a'\over 2}
\sum_{k\neq 0}{\a^\m_k \over k} (z^{-k}-\bar{z}^{-k})~,\nn\\
&&X^\m_{ND,DN}(z,\bar{z})=\sqrt{\a'\over 2}\sum_{k\in
\Zint+1/2}{\a^\m_k \over k} (z^{-k}\mp\bar{z}^{-k})~.
\label{Open}\eea
The open strings have been parametrized as $\s\in [0,\p]$. The
$x^\m$ and $p^\m$ are the position and momentum of the center of
mass of the open string. Notice also that we have imposed two
different conditions in the DD case: $X^\m|_{\s=0}=0$ and
$X^\m|_{\s=\p}=c^\m$. This will be very important later on.
\end{itemize}
The physical states obey also the Virasoro constrains.
%
%
In the conformal gauge, these constrains take the form:
$T_{zz}=0$, $T_{\bar{z}\bar{z}}=0$. Defining the Fourier modes of
these elements of the stress-tensor, we have:
\begin{itemize}
\item Closed strings
\bea L_m={1\over 2\p i} \oint dz z^{n+1}T_{zz}(z)
={1\over 2} \sum_n \a^\m_{m-n}\a_{\m,n}~,\nn\\
\bar{L}_m={1\over 2\p i} \oint d\bar{z}
\bar{z}^{n+1}T_{\bar{z}\bar{z}}(\bar{z}) ={1\over 2} \sum_n
\tilde{\a}^\m_{m-n}\tilde{\a}_{\m,n}~.
\label{VirasoroOperatorsCosed}\eea
\item Open Strings
\bea L_m={1\over 2\p i} \int_{C} \left[dz z^{m+1}T_{zz}+d\bar{z}
\bar{z}^{m+1} T_{\bar{z}\bar{z}}\right] ={1\over 2} \sum_n
\a^\m_{m-n}\a_{\m,n} ~,\label{VirasoroOperatorsOpen}\eea
\end{itemize}
where we have used (\ref{Closed}, \ref{Open}).

In the Hamiltonian picture we have the equal-$\t$ Poisson brackets
for the dynamical variables and their conjugate momenta:
\bea \{ X^\m(\s,\t),\partial_\t X^\n(\s',\t)\}_{PB}=2\p\a'
\h^{\m\n}\d(\s-\s') \eea
and $\{X^\m,X^\n\}_{PB}=\{\partial_\t X^\m,\partial_\t
X^\n\}_{PB}=0$. Inserting (\ref{Closed}, \ref{Open}) in the above
we find relations for the oscillator modes:
\bea &&\{ \a^\m_m,\a^\n_n \}_{PB}= \{
\tilde{\a}^\m_m,\tilde{\a}^\n_n
\}_{PB}= -im \h^{\m\n} \d_{m+n,0}~,\nn\\
&&\{ \a^\m_m,\tilde{\a}^\n_n \}_{PB}= 0 ~,~~~~~~~\{
x^\m,p^\n\}_{PB}=\h^{\m\n}~.
\label{PBrelation}\eea
In the open string case there are no $\tilde{\a}^\m_m$. Using
these relations we find that the Virasoro constraints form the
$classical$ $Virasoro$ $algebra$:
\bea &&\{ L_m,L_n \}_{PB}= -i (m-n)L_{m+n}~~,\nn\\
&&\{ \bar{L}_m,\bar{L}_n \}_{PB}= -i (m-n)\bar{L}_{m+n}~~,~~~~~~
\{L_m,\bar{L}_n \}_{PB}=0~.\eea

\subsubsection{Quantization}

There are various different ways to quantize the classical bosonic
string. All these ways agree whenever they can be applied. We will
describe the $light$-$cone$ method that is based on first solving
the Virasoro constraints and then replacing the fields with
operators and the Poisson brackets with commutators.

However, even after we have fixed the conformal gauge, there is
some invariance leftover. 
Defining $X^{\pm}={1\over 2}(X^0 \pm X^1)$ and using this symmetry
we can eliminate all the oscillators from the ``$+$" direction.
After imposing the Virasoro constraints (\ref{Virasoro}), we can
express all the $\a^-_m$ oscillators as functions of $\a^i_m$.
%
%
%
%
%
%
%
Since we have solved the Virasoro constraints, we can now quantize
the string by the usual field $\to$ operators and $\{~,~\}_{PB}\to
-i [~,~]$ replacements in (\ref{PBrelation}). The index
$i=2,\cdots, D-1$.

The choice of the light-cone gauge, however, obscures the
Lorentz-invariance of the theory. Finding the operators $M^{\m\n}$
that generate Lorentz transformations, and varying that they will
have the correct algebra with $p^\m$, one finds that this is true
only in $D=26$ spacetime dimensions.
%

\subsubsection{Spectrum}

%
From all the $\a^\m_m, ~\tilde{\a}^\m_m$, we define as $raising$
and $lowering$ operators, modes with $m<0$ and $m>0$ respectively.
By the commutation relations we realize that $x^\m$ and $p^\m$
commute with all $\a^\m_n$ and therefore we can diagonalize one of
them. Choosing the momenta, the ground state is labeled by
$|p^\m\rangle$:
\bea \a^\n_m |p^\m\rangle=0~~~\textrm{for}~~m>0 ~.\eea
The zero modes of the Virasoro operators define the mass-shell
condition for the physical states:
\bea &&M^2={2\over \a'}\left(L_0+\bar{L}_0-2\right)
~~~~~\textrm{for closed strings}~,\nn\\
&&M^2={1\over \a'}\left(L_0-1\right)~~~~~~~~~~~~\textrm{for open
strings}~.
\label{Mass}\eea
There is an extra constraint $L_0-\bar{L}_0=0$ for the closed
strings, from the fact that there is not any special initial point
on the string.

From (\ref{Mass}) is clear that the ground state is a tachyon for
both, closed and open strings.
The massless states for the closed strings can be decomposed into:
\bea \a^i_{-1}\tilde{\a}^j_{-1}|p\rangle \to G^{ij}+B^{ij}+\F
~.\eea
which are a spin-2 particle $G^{ij}$ (graviton), an antisymmetric
tensor $B^{ij}$ and a scalar $\F$, respectively, the massless
state of the open string:
\bea \a^i_{-1}|p\rangle ~,\eea
is a vector of $SO(24)$.
The tachyon and the absence of space-time fermions make obvious
the need of another enlarged theory, this theory is the
$supersting$.

\subsection{Chan-Paton factors}\label{ChanPaton}

We can add a non-dynamical degree of freedom to the endpoints of
an open string. Considering that the endpoints can take values
$i=1,2,\ldots,N$, the ground-state is labelled, in addition to the
momentum, by the charge on the endpoints: $|p^\m;ij\rangle$.
These labels are called $Chan$-$Paton$ factors. 
In general, we can have $N^2$ different labels (for oriented
strings) that give rise to a $U(n)$ gauge group.
Originally, the motivation of this was to introduce $SU(3)$ flavor
quantum numbers: the endpoints are like quarks and antiquarks,
connected by a color-electric flux tube.

Since each open string state has $N^2$ copies, we can introduce
Hermitian matrices $\l^\a_{ij}$, normalized such that:
\bea Tr[\l^\a\l^\b]=\d^{\a\b}~.\eea
The $\l$s form a complete set of states for the two endpoints.
Therefore, they form a representation of $U(N)$. Massless vectors
will be associated with this gauge symmetry.

Interactions of open strings imply that the endpoint of one string
will be the end point of the other. Therefore, a tree-level
amplitude of $k$ external open strings will contain an extra term:
\bea Tr[\l^{\a_1}\l^{\a_2} \cdots \l^{\a_k}]~.\eea
String amplitudes have an obvious $U(N)$ global symmetry $\l^\a\to
U\l^\a U^\dag$.

Later on we will see that these labels are associated with same
extended dynamical objects, the D-branes.

\subsection{Superstrings}

As we mentioned before, the bosonic string spectrum contains a
tachyon and no spacetime fermions. A generalization of the
Polyakov action has been shown to solve these problems, which
contains fermions on the worldsheet (we leave aside the Euler
number)\footnote{First, we will explore the closed superstring
theory.}:
\bea &&S_{SP}=-{1\over 4\p a'}\int d^2\x \sqrt{g} \bigg(g^{\a\b}
\partial_\a X^\m \partial_\b X_\m + {i\over 2}\bar{\y}^\m_M
\sla{\partial} \y_{M\m} \nn\\
&&~~~~~~~~~~~~~~~~~~~~~~~~~~~~~~~~~ +{i\over 2}\bar{\c}_\a
\g^\b\g^\a\y_{M\m} \left(\partial_b X^\m-{i\over
4}\bar{\c}_b\y^\m_M \right)\bigg)~,\eea
where $\y^\m_M=\left(\ba{c}\y^\m \\
\tilde{\y}^\m \ea \right)$ are two-dimensional Majorana spinors,
the superpartners of $X^\m$ couple to two-dimensional supergravity
fields: the zweibein $e^a_\a$ \footnote{We remind that $e^\a_a
e^\b_b g_{\a\b}=\h_{ab}$, for $a,b,\a,\b=0,1$ and the Dirac
matrices are $\g^\a=e^\a_{a}\g^a$.} and the Majorana gravitino
$\c_\a$. The last term is inserted to manifest local worldsheet
supersymmetry.
This action has the following symmetries:
\begin{itemize}
\item Local worldsheet supersymmetry~.
%
%
\item Local super-Weyl invariance~.
\item Worldsheet Lorentz invariance~.
\item Worldsheet reparametrization invariance~.
\end{itemize}
Choosing a gauge (analog of the conformal gauge) we can eliminate
the gravitational fields.
%
%
%
%
Finally, the action takes the form:
\bea S_{SP}=-{1\over 4\p a'}\int d^2z \left(
\partial X^\m \bar{\partial} X^\n +\y^\m \bar{\partial}
\y^\n+\tilde{\y}^\m \partial
\tilde{\y}^\n\right)\h_{\m\n}~.\label{SuperstringAction}\eea
The EOM for the fermions denote that $\y$ and $\tilde{\y}$ are
holomorphic and antiholomorphic functions of $z, \bar{z}$.

\subsubsection{Solving the equations of motion}

As we mention in the previous section, there are two sectors
living on the worldsheet: the bosonic and the fermionic sector.
The boundary conditions for the bosonic sector are identical to
the ones in the bosonic string (\ref{Closed}) giving the same
results as above. For the fermionic sector we can make two
inequivalent choices of boundary conditions:
\bea &&\textrm{Ramond (R):}~~~~~~~~~~~ \y^\m(\s+2\p)=\y^\m(\s)\\
&&\textrm{Neveu-Schwarz (NS):}~~~ \y^\m(\s+2\p)=-\y^\m(\s) \eea
Expanding the fermions we find (in the $z, \bar{z}$ basis):
\bea \y^\m(z)=\sum_{r\in \Zint+\n}{\y^\m_r \over
z^{r+1/2}}~~,~~~~
\tilde{\y}^\m(z)=\sum_{r\in \Zint+\tilde{\n}}{\tilde{\y}^\m_r
\over \bar{z}^{r+1/2}}\label{FermionicExpansion}\eea
where $\n=0,1/2$ in the R and NS sector respectively. Since the
left and right movers of the closed string do not interact, we can
make four inequivalent choices for the periodicity conditions of
the fermions that are called RR, RNS, NSR, NSNS.

In addition to the Virasoro operators $L_m$, which come from the
Fourier expansions of the (bosonic) energy-momentum tensor, there
are also the $G_r$ operators which come from the Fourier
expansions of the fermionic energy-momentum tensor:
\bea && L_m= {1\over 2}\sum_m :\a^\m_{m-n}\a_{\m, n}:+ {1\over
2}\sum_r (r-{m\over 2}):\y^\m_{m-r}\y_{\m, r}:+\d_{m,0}\D~,\nn\\
&& G_r=\sum_n \a^\m_n \y_{\m, r-n}~,\label{SuperVirasoro}\eea
where $r$ is half-odd integer for the NS sector and integer in the
R sector. For each fermionic coordinate, the corresponding normal
ordering shift $\D$ is $-1/48$ and $1/24$ in the NS and R
respectively. Each periodic bosonic coordinate contributes
$-1/24$. As a result, in D dimensions in the light-cone basis, we
have a total $-{1\over 16}(D-2)$ from the NS sector and $0$ from
the R.

As in the bosonic case, we can go to the light-cone gauge and
solve the ``super" Virasoro constrains:
\bea G_r|physical\rangle=0~, ~~r>0~;~~~~
(L_n-\d_{n,0})|physical\rangle=0~, ~~n\geq 0~.\eea
Finally, we eliminate the ``$+$" and we express the ``$-$"
coordinates as a function of the ``$i$"s for both bosonic and
fermionic states.

The critical dimension for the supersymmetric version of the
bosonic string is $D=10$.

Next, we quantize the theory. Canonical quantization requires, in
addition to the quantum version of (\ref{PBrelation}) for the
bosonic modes, also anticommutation relations for the fermionic
modes:
\bea \{
\y^\m_r,\y^\n_s\}=\{\tilde{\y}^\m_r,\tilde{\y}^\n_s\}=\h^{\m\n}
\d_{r+s}~.\eea
(Since we are in the light-cone gauge two coordinates have been
expressed as functions of the other coordinates. Therefore, $\m\to
i=2,\cdots,9$).

\subsubsection{Spectrum}\label{Spectrum}

There are three independent left moving sectors living on the
worldsheet of the closed superstring\footnote{Same study can take
place also for the right moving modes.}: For the bosonic sector
$\a_n^i$, the annihilation, creation operators and the vacuum
state are identical to those introduced for the pure bosonic
closed string. The NS and the R fermionic modes are new sectors
and we will study them separately:
\begin{itemize}
\item The NS sector: The anticommutation relations for $\n=1/2$
show that we can define the ground state to be annihilated by all
$r>0$ modes:
\bea \y_r^i |0\rangle_{NS}=0  ~~~~~\textrm{for $r>0$}~.\eea
Obviously, all modes with $r<0$ are raising operators.
\item The R sector: In the R sector there are zero modes. For the
non-zero modes we define again:
\bea \y_r^i |vacuum\rangle_{R}=0  ~~~~~\textrm{for $r>0$}~.\eea
The $\y^i_0$ satisfy an $O(8)$ Clifford algebra:
$\{\y^i_0,\y^j_0\}=\d^{ij}$. The R vacuum is degenerate and the
fermionic zero modes change ground state. We can choose a basis:
\bea \y_i^{\pm}={1\over \sqrt{2}}(\y_0^{2i+2}\pm\y_0^{2i+3})~,\eea
where $\{\y^+_i,\y^-_j\}=\d^{ij}$. The $\y^-_i$ will be the
annihilation operators. Thus, the R vacuum is
\bea
|vacuum\rangle_{R}=|s_0,s_1,s_2,s_3\rangle_{R}~~~~~~s_i=\pm1/2~,\eea
and it is constructed by $2^{8/2}=16$ ground states. These ground
states can be decomposed into the ${\bf 8}_s$ with an even number
of $-1/2$s and the ${\bf 8}_c$ with odd number of $-1/2$s (even or
odd is clearly a convention).
\end{itemize}
The mass formula for the superstring is again provided by the
$L_0$, $\bar{L}_0$ constraint and it is:
\bea M^2={2\over \a'} \left(L_0+\bar{L}_0\right)~.
%
\eea
The NS vacuum is clearly tachyonic due to the non-vanishing of
$\D$ and $\bar{\D}$ (\ref{SuperVirasoro}).

In order to achieve spacetime supersymmetry and eliminate the
tachyon, the spectrum is projected onto states with an odd number
of fermions. This is called the $GSO$ $projection$\footnote{There
is another reason for projection out the odd or the even fermionic
modes and this is modular invariance. We will come back in this
when we will discuss the 1-loop amplitudes.}.
The GSO operators are defined as
\bea GSO_{NS}=(-1)^F~~,~~~~~GSO_{R}=(-1)^{\sum_i s_i}~,\eea
where $F$ is the worldsheet fermion number. To eliminate the
tachyon we keep the NS states that have an odd number of fermions.
However, things are not so clear in the R sector. Which states
should we project out? The ${\bf 8}_s$ or the ${\bf 8}_c$? This
question has a relative answer since spinor or conjugate-spinor is
just a matter of definition. The question is: ``What should be the
GSO projection to the left compared to the right movers?"
Since left and right movers are disconnected, we can make the same
or different choice. There are two inequivalent theories that are
called $Type~IIA$ and $Type~IIB$, where in the $A$ we choose
different and in the $B$ the same GSO projections for the two
sectors.
The massless spectrum of the two theories are (in $SO(8)$
content):
\bea &&\textrm{Type IIA:} ~~~({\bf 8}_v\oplus {\bf 8}_s)_L\otimes
({\bf 8}_v\oplus {\bf 8}_c)_R\nn\\
&&\textrm{Type IIB:} ~~~({\bf 8}_v\oplus {\bf 8}_s)_L\otimes ({\bf
8}_v\oplus {\bf 8}_s)_R \eea
The spectrum is provided below. The $G^{ij}$ is the graviton. The
$\y^i$ are gravitino with different and same chirality in $A$ and
$B$ theories respectively.
\begin{center}
\begin{tabular}{|c|c|c|c|c|}
\hline
& NSNS & RR & NSR & RNS  \\
\hline \hline
Type IIA  & $\F\oplus
B^{ij}\oplus G^{ij}$  & $A^i\oplus C^{ijk}$ & $\y^i_{\dot\a}$ &
$\tilde{\y}^j_\b$\\
Type IIB & $\F\oplus B^{ij}\oplus G^{ij}$ & $\F'\oplus B'^{ij}
\oplus D^{ijkl}$
& $\y^i_{\dot\a}$ & $\tilde{\y}^j_{\dot\b}$ \\
\hline
\end{tabular}
\end{center}
The above massless spectra are described by $10D$ supergravity
theories, the so called: $Type$ $IIA$ and $Type$ $IIB$.

\subsection{Open Superstings}

Before we describe the open strings we have to mention that a pure
open string theory cannot be consistent.
Open strings can always interact by themselves giving open and
closed strings. The complete theory is one that describes both
open+closed strings and is called $Type$ $I$. We will explore this
theory later on when we will discuss the orientifold models.

The open superstring action is again (\ref{SuperstringAction})
where $\s\in [0,\p]$ and $\t\in (-\infty, +\infty)$. The bosonic
sector has the same solutions as in the pure bosonic case
(\ref{Open}).
We have again two choices for the fermionic boundary conditions:
\bea &&
\y^\m(0,\t)=\tilde{\y}^\m(0,\t)
~~,~~~~~~~~ \y^\m(\p,\t)=\tilde{\y}^\m(\p,\t)~,\label{+}\\
&&
\y^\m(0,\t)=-\tilde{\y}^\m(0,\t) ~~,~~~~~
\y^\m(\p,\t)=\tilde{\y}^\m(\p,\t)~. \label{-}\eea
Traditionally, we want the, so called, R sector to have the same
moding as the bosonic part. Therefore, in the NN and DD open
strings the R sector is the one with boundary conditions (\ref{+})
and NS with (\ref{-}).
To visualize the connection of the open and closed R and NS
sectors, we can combine $\y^m$, $\tilde{\y}^m$ in a single field
$\Y^\m$ with the extended range $\s\in [0,2\p]$. Defining
$\Y^\m(\s)\equiv \y^\m(\s)$ and $\Y^\m(2\p- \s)\equiv
\tilde{\y}^\m (\s)$. These left moving fields are periodic in R
and antiperiodic in NS. This is called the $doubling$ $trick$ and
allows us to treat the open sector as the left moving sector of
the closed string.

Having express the open string as the left moving sector of a
closed string, we can use the quantization procedure and GSO
projection introduced in the previous chapter (\ref{Spectrum}).
The open string spectrum is the same to the left moving spectrum
of the closed string.

For the ND open strings the choice of the R and NS sector is the
opposite to the one of NN and DD ones. This interchanges the
properties between the R and NS giving the spinorial vacuum to the
NS sector. We will describe all these cases in more detail later
on.

\subsection{Compactification}

As we mentioned before, string theory lives in $10D$. Therefore,
if we want to discuss interesting phenomenological aspects we have
to somehow reduce the visible dimensions to our familiar $4D$
spacetime.
One of the most straight-forward ideas is to compactify the extra
six dimensions to a compact manifold:
\bea \Realint^{(1,3)}\times {\cal M}_6 ~,\eea
where the $1+3$ real dimensions form the Minkowski space.

The effects of compact dimensions in a theory are many. Kaluza and
Klein had shown in the beginning of the 20$^{th}$ century that in
a theory in $D=5$ dimensions with one compact $x^4=x^4+2\p {\cal
R}$, the momentum in the compact dimension is quantized such that
$p^4_n=n/{\cal R}$.
Massless scalars in $5D$ can be expanded $\f(x^N )=\sum
\f_m(x^\m)e^{im x^4/{\cal R}}$ ($M, N$ run in all and $\m, \n$ run
in the non-compact dimensions) giving a family of scalars of mass
$\textsl{m}_m=m/{\cal R}$ in $4D$:
\bea \partial^M \partial_M \f (x^N )=0 ~~~~~\to ~~~~~
\left(\partial^\m \partial_\m -{m^2\over {\cal R}^2} \right) \f_m
(x^\n)=0 ~.\eea
This family is a tower of states characterized by $m$ which are
called the $Kaluza$-$Klein$ $modes$.

$5D$ gravity with one compact dimension also has interesting
effects. Decompose $G^{MN}$ into $G^{\m\n}$, $G^{\m 4}$, $G^{44}$.
As is known, $5D$ local coordinate transformations are a symmetry
where:
\bea x^M\to x^M +\e^M(x) ~~,~~~~~ G^{MN} \to G^{MN}-\partial^M
\e^N -\partial^N \e^M ~.\eea
Local transformations of the type $\e^4(x^\m), \e^\m=0$ (rotations
of the circle) can be interpreted as gauge transformation of a
"vector" field $G_{\m 4}=A_\m$: $A^\m\to A^\m-\partial^\m \e^4$.

The effective action for the massless theory in a curved
background contains the graviton $G_{MN}$, an antisymmetric tensor
$B_{MN}$, and a dilaton $\F$. Considering that none of these
fields is $x^5$ dependant the graviton-dilaton action becomes:
\bea S_{symmetric}&=&{1\over 16 \p G^N_{(5)}}\int d^5x
\sqrt{-G_{(5)}}e^{-2\F}\big( R^{(5)} +4 \partial_M\F \partial^M\F
\big)\nn\\
&=&{1\over 16 \p G^N_{(4)}}\int d^4x
\big(\sqrt{G_{(4)}}\big)e^{-2\F_4}\times \nn\\
&&~~~~~~~~~~~~~\big( R^{(5)}-\partial_\m \f \partial^\m \f +4
\partial_\m\F_4 \partial^\m\F_4 -{1\over 4}e^{2\f}
F^{\m\n}F_{\m\n}\big) \eea
%
%
%
%
%
%
%
%
where $G_{44}=e^{2\f}$, $\F_4=\F-\f/2$,
the Newton constants in $5D$ and $4D$ are related through
$G^N_{(5)}=2\p {\cal R} G^N_{(4)}$. Therefore, $5D$ gravity plus
the dilaton in a spacetime with one compact dimension can be
interpreted as $4D$ gravity coupled to electromagnetism and two
scalar fields $\f$, $\F_4$.

Closed strings that live in spaces with compact dimensions have
another very interesting effect that does not appear in particle
physics. They can rap around the compact dimension. This gives a
topological charge the $winding$ $number$ $n$.
Solving again the EOM for the bosonic string living in compact
dimensions, we find that the momenta in compact dimensions are not
equal any more. The solution for the closed string is given again
by (\ref{Closed}) with:
\bea p_L={m\over {\cal R}} +{n {\cal R}\over \a'} ~~~~~,~~~~~
p_R={m\over {\cal R}} -{n {\cal R}\over \a'} ~,\eea
for the compact dimensions.

Notice that if we exchange $m \leftrightarrow n$ and ${\cal R}\to
1/{\cal R}$ we end up with a theory where $p_L \to p_L$ and $p_R
\to -p_R$. It is important to mention that the spectrum and the
currents also respect this property. This property is called
$T$-$duality$ and implies that conformal field theory cannot
distinguish a circle of radius ${\cal R}$ from another of radius
$1/{\cal R}$. It states that two a priori different theories are
in fact equivalent. In the next sections, we will find some more
interesting properties of this duality.

\subsection{Orbifolds}\label{Orbifolds}

There is a class of exactly soluble compactifications on spaces
known as $orbifolds$ \cite{Dixon:1985jw}.
The notion of orbifold arises when we consider a manifold $M$ that
has a discrete symmetry group $G$. We may consider a new manifold
$\tilde{M}=M/G$, which is obtained from the old one by moding out
the symmetry group $G$. If $G$ is freely acting, the manifold
$\tilde{M}$ is smooth. If the manifold $\tilde{M}$ has
fixed-points, it has conical singularities at the fixed points.

Orbifolds are interesting in the context of CFT and string theory,
since they provide spaces for string compactifications that are
richer than tori, but admit an exact CFT description. Moreover,
although their classical geometry can be singular, strings
propagate smoothly on them. In other words, the correlation
functions of the associated CFT are finite.

To be more precise, we will explore a specific example. Consider a
circle $S^1$ parametrized by $X=X+2\p {\cal R}$. The orbifold
action will be a $Z_2$ discrete symmetry where: $G=\{1,R\}$ and
$R:~X\to -X$. This identification gives rise to two sectors: the
so-called untwisted and twisted sectors.
The fact that the string wavefunction must be invariant under the
element $R$ gives rise to the $untwisted~sector$. Imposing the
reflection condition on the mode expansion (\ref{Closed}) we find:
\bea m=n=0~,~~~~ \a_k\to -\a_k~,~~~~\tilde{\a}_k\to -\tilde{\a}_k
~.\label{Z2}\eea
%

\begin{figure}
\begin{center}
\epsfig{file=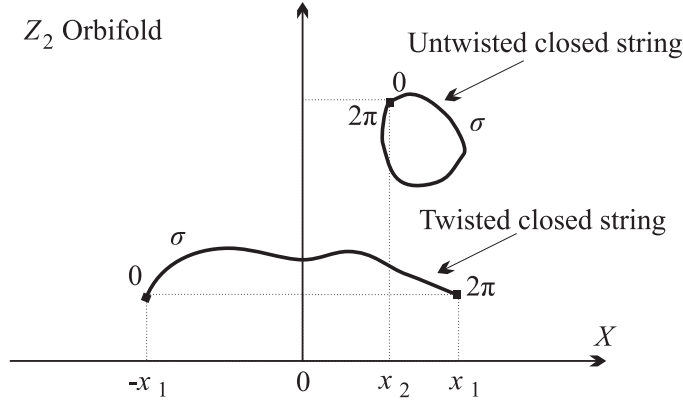,width=90mm}
\end{center}
\caption{The closed strings of the $Z_2$
orientifold.}\label{Twisted}
\end{figure}

In addition, we also have a new sector in the closed string
spectrum, in which, the boundary conditions for the bosonic part
of the string are twisted:
\bea X(\s+2\p)=-X(\s)~.\label{TwistedZ2}\eea
These strings are called $twisted~sector$ strings and they are
closed only under the identification (\ref{TwistedZ2})
(Fig.\ref{Twisted}).
Solving the EOM using the twisted boundary condition we find:
\bea X=x_{fixed~points} + i\sqrt{\a'\over 2} \sum_{r\in
\Zint+v}\left({\a_r \over r} z^{-r}+{\tilde{\a}_r \over r}
\bar{z}^{-r} \right)~.\label{ClosedTwisted}\eea
where for the $Z_2$ case $v_{R}\equiv g=1/2$\footnote{We denote by
$g$ the rotation angle of a $Z_2$ element $R$.}. Notice the
differences between the mode expansions (\ref{Closed}) and
(\ref{ClosedTwisted}) in the absence of momenta and in the modding
of the oscillators. The center of mass is localized on
$x_{fixed~points}=0,~\p {\cal R}$, the fixed points of the
manifold. Therefore, we have one ground state on each fixed poind
$|H^{0,\p {\cal R}}\rangle$ that is annihilated by the positive
moding $\a_{r}$.
The action of $R$ on the oscillator modes of the twisted sector is
again given by (\ref{Z2}).

We have also to impose boundary conditions on the fermionic
twisted sectors.
Since in the untwisted sector the R and the bosonic sector have
the same moding, we define as R twisted sector the half moded one.
Therefore, the NS sector has the zero modes.

In general, we could project with an element $\a=e^{2\p i v}$ (as
we mention above, the reflection element $R$ is a special case
with $v_{R}=1/2$). The untwisted wavefunction should be invariant
under the action of the new element and the twisted states will be
modes of the general kind: $\a_{k+v}$. Notice that right movers
will be moded as: $\tilde{\a}_{k-v}$. The fermions on the other
hand will be also twisted with $\a=e^{2\p i (v+\n)}$ where
$\n=0,1/2$ for R and NS respectively. The field expansion will be
similar to (\ref{FermionicExpansion}) where the moding will run
to: $r\in \Zint +\n +v$. Notice also that different sectors will
be localized on different fixed points.

The existence of the two sectors has its origins in a deeper
reason that is $modular$ $invariance$ of the 1-loop diagram.

\subsubsection{Partition function and modular invariance}

Consider the 1-loop  vacuum to  vacuum amplitude of an oriented
closed string which is obviously a torus diagram.
To evaluate the path-integral we have to sum over all possible
tori.
The torus is a two dimensional surface that can be seen as two
independent one-cycles, parametrized as $\s^1,\s^2\in [0,1]$.
It is completely specified by giving a flat metric and a $complex$
$structure$ $\t$ with $Im (\t)\geq 0$ that cannot be changed by
any infinitesimal diffeomorphisms or Weyl rescaling. Defining
complex coordinates $w=\s^1+\t \s^2$ and $\bar{w}=\s^1+\bar{\t}
\s^2$, the periodicity conditions become:
\bea w\to w+1~,~~~~~~w\to w+\t~.\eea
The torus can be thought as a point of the complex plane $w$
identified under two translation vectors corresponding to the
complex numbers $1$ and $\t$ .
\begin{figure}
\begin{center}
\epsfig{file=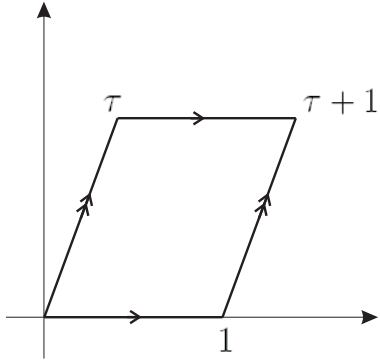,width=50mm}
\end{center}
\caption{The torus as a periodic lattice.}\label{Torus}
\end{figure}

\begin{figure}
\begin{center}
\epsfig{file=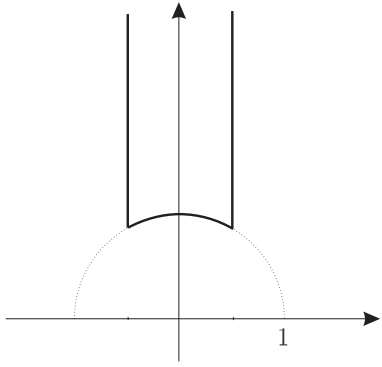,width=50mm}
\end{center}
\caption{Fundamental domain of the torus.}
\label{FundamentalDomain}
\end{figure}

Not all $\t$ describe different tori. The periodicity conditions
show that transformations of the type:
\bea \t'={a\t +b\over c\t+d} ~~,~~~~~ \textrm{with:}~~ad-bc=1
~,\eea
keep the torus invariant. This is the group SL(2,$\Zint$). The
generators of this group are:
\bea T: \t\to \t+1 ~,~~~~~~S: \t\to -1/\t~.\label{TS}\eea
It can be shown that the fundamental domain $\cal F$ of the
modular group of the torus is $|\t^1|\leq 1/2$ and $||\t||\geq 1$
(Fig.\ref{FundamentalDomain}).

The path-integral of a conformal field theory on a torus is the
1-loop vacuum energy. As we mentioned above, the Hamiltonian is
$H=L_0+\bar{L}_0$, which is the generator of translations in
worldsheet time $\t$. The generator of rotations around $\s$ is
$P=L_0-\bar{L}_0$. Putting everything together we have:
\bea {\cal T}=\int e^{-S}= \int_{\cal F} {d^2\t\over
\t_2^2}Tr[e^{-2\p \t H}e^{2\p i \s P}] =\int_{\cal F} {d^2\t\over
\t_2^2} Tr[q^{L_0-1}\bar{q}^{\bar{L}_0-1}]
~,\label{PathIntegral}\eea
where $q=e^{2\p i \t}$. This is the so-called $torus$ $partition$
$function$, since expanding in powers of $q, \bar{q}$, the powers
in the expansion refer to the mass squared level of excitations.
Notice that (\ref{PathIntegral}) does not contain divergencies
since the integration area, $\cal F$, does not touch the origin.

One very important property of (\ref{PathIntegral}) is that it is
$modular$ $invariant$ (invariant under (\ref{TS})). This property
is crucial and it means that we correctly integrate over all
inequivalent tori.

The partition function of an orbifold has to be modified since we
want to project onto states that in the $Z_2$ case have $R=+1$.
The untwisted contribution is
\bea {\cal T}_U={1\over 2}\int_{\cal F} {d^2\t\over
\t_2^2}Tr_U[(1+R)q^{L_0-1}\bar{q}^{\bar{L}_0-1}]~.\eea
The trace part of the $1$ term is modular invariant like
(\ref{PathIntegral}). However, the trace part with the insertion
of $R$ is not. We have to add some extra terms/sectors if we want
the total partition function to be modular invariant.
The twisted sector recovers this problem giving the full modular
invariant partition function:
\bea Z_{U+T}={1\over 2}\int_{\cal F} {d^2\t\over
\t_2^2}Tr_{U+T}[(1+R)q^{L_0}\bar{q}^{\bar{L}_0}]~.\eea
We will describe more precisely the orbifold construction in the
next section where we will discuss a generalized version of it:
the orientifold.

\newpage

\section{Orientifolds}

Orientifolds are generalized orbifolds, where except from the
orbifold discrete symmetry we include orientation reversal on the
worldsheet \cite{Pradisi:1988xd, Gimon:1996rq, Berkooz:1996dw,
Ibanez:1998qp, Angelantonj:2002ct}. This expansion generates a
theory of unoriented closed strings (plus open strings as we will
see in a while).

Orientation reversal, $\Omega$, means interchanging of left and
right movers. Here, we shall consider the Type IIB closed theory
since it contains a symmetry on left and right modes\footnote{We
could also take the Type IIA for a $\Omega'$ element that changes
also the chirality of the left and right moving fermions.}.
$\Omega$ interchanges $\s\to-\s$ (or $z\leftrightarrow \bar{z}$).
Looking at (\ref{Closed}, \ref{ClosedTwisted}) we realize that the
action of $\Omega$ on the oscillators is to interchange
$p_L\leftrightarrow p_R$ and $\a\leftrightarrow \tilde{\a}$.
The action of $\Omega$ on the bosonic zero modes of a compact
dimension is:
\bea \Omega |m,n\rangle= |m,-n\rangle~.\label{OmegaOnZero}\eea
The action and the quantization procedure preserve the worldsheet
parity. For all the oscillator modes we have:
\bea \Omega \a^\m_k \Omega^{-1}= \tilde{\a}^\m_k~~,~~~~~\Omega
\tilde{\a}^\m_k \Omega^{-1}= \a^\m_k~~,~~~~~\Omega \y^\m_r
\Omega^{-1}= \tilde{\y}^\m_r~~,~~~~~\Omega \tilde{\y}^\m_r
\Omega^{-1}= -\y^\m_r~,~~~\eea
for integer and half-integer $r$. The minus in the last equation
is included to give $\Omega \y\tilde{\y} \Omega^{-1} =\y
\tilde{\y}$, so that the graviton is invariant under the $\Omega$
projection.

The total orientifold group contains elements of two kinds:
internal symmetries of the worldsheet theory, forming a group $G$,
and elements of the form $\Omega\cdot g$, where $g$ is some
symmetry element that is taken from a group $G^{\prime}$. Closure
implies that $\Omega\cdot g \cdot\Omega\cdot g^{\prime} \in G$ for
$g, g^{\prime} \in G^\prime$. The full orientifold group is
$G+\Omega G^\prime$.

In our study we will concentrate on groups where $G=G'$. For
simplicity, the compact manifold will be formed by three tori
where our $10D$ space will be parametrized as:
\bea \Realint^4\times T_1^2\times T_2^2 \times T_3^2~.
\label{RxT2xT2xT2}\eea
We define complex coordinates for each torus:
$z^i=X^{2i+2}+iX^{2i+3}$ and similarly for the complex fermionic
states $\y^i=\y^{2i+2}+i\y^{2i+3}$, for $i=1,2,3$. In general, $G$
contains two kinds of elements: rotations and translations:
\begin{itemize}
\item $Rotation$ elements are a subgroup of the Poincar\'e group
and they are defined as:
\bea \a=\exp\bigg(2\p
i(v_\a^1J_{45}+v_\a^2J_{67}+v_\a^3J_{89})\bigg) ~,\eea
where $J_{mn}$ are $SO(6)$ Cartan generators. The resulting
manifold has fixed points. To preserve some of the supersymmetry,
the $v_\a^i$s should satisfy the condition $\sum_i v_\a^i=0$. This
ensures that there are gravitini in both the NSR and RNS untwisted
sectors.

Notice that a $Z_2$ element $R_i$ (to preserve some of the
supersymmetries)\footnote{$Z_2$ reflecting elements are also
denoted in the literature as $I$. In particular, the reflection
element that does not break supersymmetry acts on the coordinates
of two tori and it is denoted by $I_4$.} leaves unaffected one
torus $T^2_i$. We denote the components of the rotation vector as
$u^i_R\equiv g^i$. For example, an $R_1$ element has a shift
vector: $g_1=\{0,1/2,-1/2\}$. Such elements will play a key role
in the rest of our studies. In the next table we show the general
rotation elements of $G$ and we denote the tori in which they act:
\begin{table}[h]
\begin{center}
\begin{tabular}{cccc}
Elements of $G$ & $~~T^2_1~~$ & $~~T^2_2~~$ & $~~T^2_3~~$ \\
\hline
$\a:~ v_\a^3=0$&  $X$ & $X$ & \\
$\a :~ v_\a^3\neq 0$ & $X$ & $X$ & $X$\\
$R_1 $ &  & $X$ & $X$\\
$R_2 $ & $X$ &  & $X$ \\
$R_3 $ & $X$ & $X$ & \\
\end{tabular}
\end{center}
\end{table}

The direct action of such an element on the bosonic zero modes of
a compact dimension is:
\bea \a ~|m,n\rangle= |e^{2\p i v_\a}m,e^{2\p i v_\a}n\rangle ~,
\label{RotationsOnZERO}\eea
where $m,n$ complex momentum and winding numbers coming from the
complex parametrization of the coordinates. On the oscillation
states the action is: for the bosonic and NS sector:
\be \a ~z^i = e^{2\pi i v_\a^i} z^i~~, ~~~~~~\a ~\y^i = e^{2\pi i
v_\a^i} \y^i~. \label{RotationsB+NS}\ee
The conjugate fields $z^{-i}, \y^{-i}$ transform with the phase $
e^{-2\p i v_\a^i}$. On the R sector, the action on the non-zero
modes is similar to the one on the NS, however, the R vacuum
transform as:
\bea \a ~|s_0s_1s_2s_3\rangle = e^{2\p i v_\a \cdot
s}|s_0s_1s_2s_3\rangle~,\label{RotationsR}\eea
where we extended $v_\a=\{0,v_\a^1,v_\a^2,v_\a^3\}$. The action on
the right movers is the same to the one on the left movers.
\item $Translation$ elements $h$ are freely acting elements which
are also a subgroup of the Poincar\'e group.
The generic symmetry of a $d$-dimensional toroidal CFT contains
the $\rm U(1)^d_L\times U(1)_R^d$ chiral symmetry. The
transformations associated with it are arbitrary lattice
translations. They act on a state with  momenta $m_i$ and windings
$n_i$ as
\bea h_{\rm translation}= \exp\left[2\pi
i\sum_{i=1}^d(m_i\theta_{i}+n_i\phi_{i})\right] ~,\label{240}\eea
where $\theta_i,\phi_i$ are rational in order to obtain a discrete
group. There are also symmetries that are subgroups of the
$O(d,d)$ group not broken by the moduli $G_{ij}$ and $B_{ij}$.
These depend on the point of the moduli space.
%
%
For the rest of our study we will concentrate on translation
elements that act on one only coordinate as momentum shifts of
order $N$ ($\theta= 1/N, ~\phi_{i}=0$). Generalization to more
dimensions and to winding shifts is straight forward.

Clearly, the action $h_N$ affects only the bosonic zero modes of
the states where by acting in direct it gives an eigenvalue, and
by twisting it shifts the winding number:
\bea &&\textrm{Direct action}~~~~~~~~h_N:|m,n\rangle \to e^{2\p i
m/N}|m,n\rangle~.\nn\\
&&\textrm{Twist}~~~~~~~~~~~~~~~~~~h_N:|m,n\rangle \to
|m,n+1/N\rangle~.\label{TranslationsOnZERO}\eea

Translation elements that are accompanied by elements that treat
bosons and fermions in a different way break supersymmetry.
These type of actions are called $Sherck$-$Schwarz$ deformations
(SS) \cite{sss, Antoniadis:1998ki}.
For this work we will consider only $Z_2$ SS deformations:
\bea h=(-1)^{\cal F}~h_2 ~,\label{Sherck-Schwarz}\eea
where ${\cal F}$ is the space-time fermion number. The geometric
action of this element is to halve the radius of the corresponding
dimension that it acts onto: $X \to X + \p R$.
Notice that:
\begin{itemize}
\item The $\tilde{\a}^\m_{-1}|S^I_\a \rangle \otimes |m,n\rangle$
has the space-time quantum numbers of the gravitino. It transforms
with $(-1)^{m+1}$ sign under the $h$ action. Therefore, the
massless state $|m,n\rangle=|0,0\rangle$ is projected out
(massless gravitino) but not the $|m,n\rangle=|1,0\rangle$ state
that has mass $m_{3/2}^2=1/4R^2$ ($\to$ massive gravitino).
\item Supersymmetry is broken spontaneously and it restores in the
large radius limit $R \to \infty$. 
\end{itemize}
%

Rotation $\a$ and translation $h$ elements belong in $G$ only if
they commute $[\a,h]=0$. Therefore, in the direction where a $Z_2$
Scherk-Schwarz deformation acts, we can only consider rotations by
$Z_2$ elements $R$.

To summarize, in the supersymmetric case the most general rotation
element $\a$ has rotation vector $v_\a= (v_\a^1, v_\a^2, v_\a^3)$
with $v_\a^3=0$ or $v_\a^3\neq 0$.
On the other hand, in non-supersymmetric models where a SS
deformation acts onto one coordinate of the third torus $T^2_3$,
the most general rotation element $\a$ will act as
$v_\a=(v_\a^1,v_\a^2,0)$ or
$v_\a=(v_\a^1,v_\a^2,1/2)$\footnote{The former can be written in
the form $v_\b+g_{i=1,2}$ with $v_\b^3=0$ a rotation in the
$T_1^2\times T_2^2$ torus and $g_{i=1,2}$ a $Z_2$ element.
Therefore, without loss of generality we can take $\a$ such that
$v_\a^3=0$.}.


%
\end{itemize}

\subsection{The closed string spectrum}

The closed string spectrum is constructed combining left and right
states with the same chirality, invariant under the orbifold
action.
We will concentrate on the massless states. Untwisted states:
\begin{itemize}
\item From the NSNS states we have a graviton
($(\y_{-1/2}^{\{\m}\tilde{\y}_{-1/2}^{\n\}}-
\d^{\m\n}\y^{\r}_{-1/2}\tilde{\y}_{-1/2,\r}) |0,0\rangle$), an
axion ($\y_{-1/2}^{[\m}\tilde{\y}_{-1/2}^{\n]} |0,0\rangle$) and a
dilaton ($\y_{-1/2}^{\r}\tilde{\y}_{-1/2,\r} |0,0\rangle$). Since
parity projects onto symmetric states the axion is eliminated.

The matter states depend on the orbifold action $(v_1,v_2,v_3)$.
Consider a state:
\bea \a~\y_{-1/2}^{\pm i}\tilde{\y}_{-1/2}^{\pm j}|0;0\rangle\to
e^{\pm 2\p i (v_i+\tilde{v}_j)k}\y_{-1/2}^{\pm
i}\tilde{\y}_{-1/2}^{\pm j}|0;0\rangle~.\eea
It will not be excluded only in the case $(v_i+\tilde{v}_j)k\in
\Zint$. $\Omega$ projection will keep only left-right invariant
states.
\item Similar for the RR sector:
\bea \a~|s_i;\tilde{s}_j\rangle\to e^{\pm 2\p i (s_i\cdot
v_i+\tilde{s}_j\cdot\tilde{v}_j)k}|s_i;\tilde{s}_j\rangle~,\eea
where $i,j=0,1,2,3$. The invariant states are those where
$(s_i\cdot v_i+\tilde{s}_j\cdot\tilde{v}_j)k \in \Zint$. $\Omega$
projects onto antisymmetric combinations of left-right.
\item Finally, for the RNS and NSR the procedure is similar to the
above. Invariant states under $\Omega$ are taken by the symmetric
combination NSR+RNS.
\end{itemize}
For the twisted sector the procedure is similar. However, only
$Z_2$ twisted elements are invariant under the $\Omega$ projection
as it was mentioned in section \ref{Orbifolds}.

\subsection{Klein Bottle}\label{Klein-Bottle}

Consider the 1-loop vacuum amplitude of a theory with orientation
reversal. 
%
Consider also $\a\in G$, an element of a $N$ dimensional group
$G$. The 1-loop partition function for this generalized orbifold
will be:
\bea {1\over 2\times N}\sum_{\a} Tr[(1+\Omega)\a
q^{L_0}\bar{q}^{\bar{L}_0}]\label{KleinBottle-GENERAL}\eea
Notice that we project onto even states under $\Omega$ ($+1$
eigenvalue) since odd states do not form a closed set under
interactions.
The amplitudes that do not contain $\Omega$ describe the
propagation of oriented closed strings. It is the usual torus
amplitudes $\cal T$. The amplitude that contains $\Omega$
describes strings that propagate and flip orientation.
Geometrically, this is described by an unoriented two-dimensional
surface with Euler number zero (equal to the Euler number of the
torus): the Klein Bottle amplitude $\cal K$.

\begin{figure}
\begin{center}
\epsfig{file=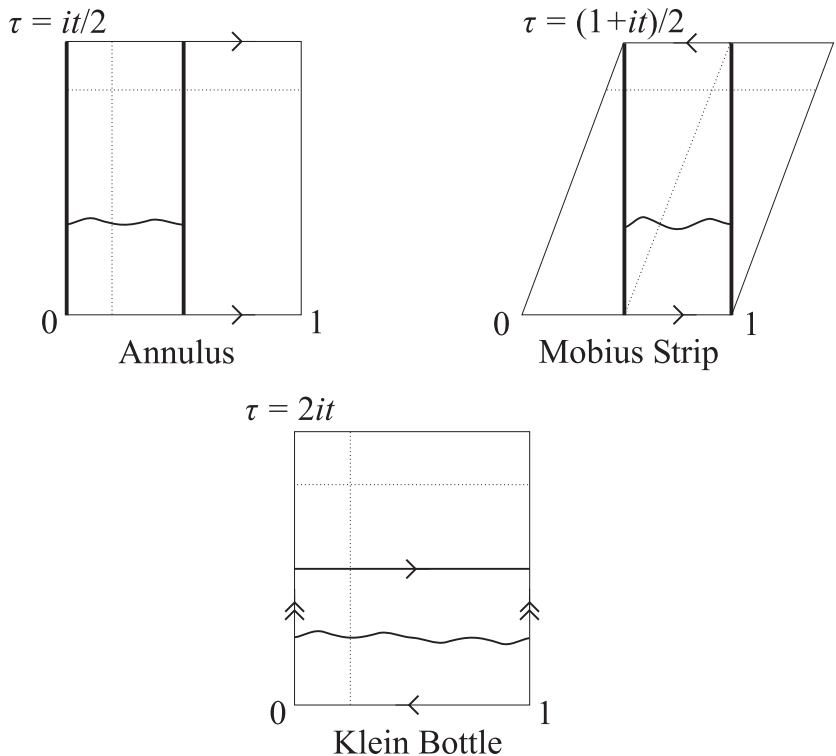,width=110mm}
\end{center}
\caption{Covering tori and fundamental cells for the three one
loop surfaces. The cycles are represented by dashed lines. The
point $M'$ is the image of $M$ under the appropriate
involutions.}\label{DoubleCover}
\end{figure}

Topologically, the Klein Bottle can be obtained from its covering
torus with complex structure $\t=2i\t_2$, if the lattice
translations are supplemented by the anticonformal involution
$w\to 1-\bar{w}+i \t_2$. This representation will be denoted as
$\cal K$.
There is a second choice of polygon that defines an inequivalent
horizontal time. It is obtained by halving the horizontal side
while doubling the vertical one and thus leaving the area
unchanged. The end result has the virtue of displaying an
equivalent representation of this surface as a tube terminating on
two crosscaps, and the horizontal side is the proper time elapsed
as a closed oriented string propagates between the crosscaps. The
change of orientation is taking place on the crosscaps. This will
be denoted as $\tilde{\cal K}$ (Fig.\ref{DoubleCover}).

To evaluate the path-integral we have to integrate over all
possible ${\cal K}$, that means integration on $\t_2\in [0,\infty
)$. We will see that this integration gives in general
ultraviolent (UV) divergences (due to tadpoles)
\cite{Polchinski:1987tu}.
%


\subsubsection{Supersymmetric Orientifolds}

In this section we will evaluate the UV limit of
(\ref{KleinBottle-GENERAL}). First, we will concentrate on
supersymmetric cases and after we will generalize to include the
Scherk-Schwarz deformation that spontaneously breaks
supersymmetry.

We can work out the contribution of an element $\a\in G$ to the
Klein Bottle amplitude by using the trace formula:
\bea {\cal K}_\a = Tr_{U+T}\left[ \Omega \a ~~q^{L_0}
\bar{q}^{\bar{L}_0}\right]~, \nonumber \eea
where the subscripts $U$ and $T$ refer to the untwisted and
twisted closed string states of the type IIB orbifold model
considered.
General twisted states have different moding between left and
right movers (we recall (\ref{ClosedTwisted}) and comments below)
which coincide only for the $Z_2$ case. Therefore, only $Z_2$
twisted sectors will survive the $\Omega$ reflection.
The contribution to the Klein Bottle amplitude of an element $\a
\in G$ can be written in the form
\bea {\cal K}_\a \sim T[^{~0}_{2v_\a}] + T[^{~g}_{2v_\a}]~,
\label{gf} \eea
where the second term exists only in case where there are $Z_2$
factors denoted by $g$ (section \ref{Orbifolds}). The form of
$T[^{u}_{v}]$ is given in the appendix (\ref{Tov}, \ref{Tgv}). In
the transverse channel, the contribution of an element $\a$
corresponds to a propagation of a closed string state projected by
$(\Omega \a)^2=\a^2$ which explains the $2v_\a$ factor in
(\ref{gf}).

As we mentioned above, the 1-loop diagram gives in general UV
divergencies, since $\t_2\in [0,\infty)$. The way to compute the
divergent contribution is to evaluate $\tilde{\cal K}$. In this
picture, the horizontal side is the proper time elapsed as a
closed oriented string propagates between two crosscaps.
%
%
%
%
%
%
%
%
\begin{figure}
\begin{center}
\epsfig{file=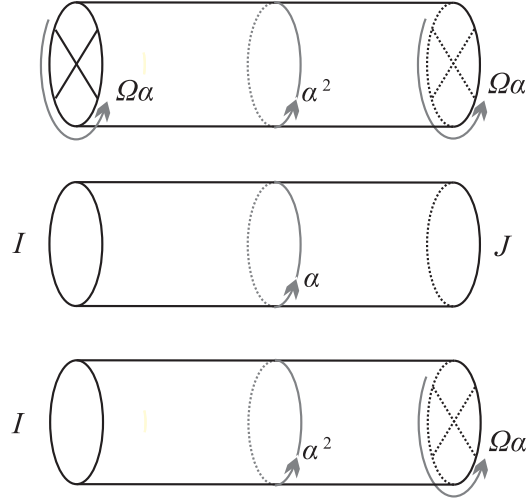,width=70mm}
\end{center}
\caption{Klein-bottle, Annulus and M\"obius strip. The one-loop
amplitudes become tree-level in the transverse picture where an
$\a^2$-twisted closed string propagates between crosscaps and
boundaries.}\label{KleinAnnulusMobius}
\end{figure}
Technically, to go from the one representation of the $\cal K$ to
the other $\tilde{\cal K}$ we need to perform a modular
transformation, $l=1/4t$, where $t$ is the loop modulus and $l$
the cylinder length \cite{Gimon:1996rq}. To extract the
divergencies we evaluate the UV limit by taking $l\to \infty$.

If the orbifold group $G$ contains $Z_2$ factors denoted by $R_i$
\footnote{We remind that $R_i$ is a $Z_2$ rotation element which
leaves unaffected the $T^2_i$ torus and acts on the other two
tori.}, then there is an extra contribution since $(\Omega R_i
\a)^2= \a^2$:
\be T[^{~~0}_{2 g_{i} v_\a }] + T[^{~~g_j}_{2 g_i v_a}]
~,\label{TSUSY}\ee
where $i,j$ denote the different $Z_2$ elements in $G$. In
general, elements $\a \in G$ can leave one torus unaffected or act
on all tori. Without loss of generality, we consider as the
unaffected torus the $T^2_3$ (\ref{RxT2xT2xT2}). Therefore, the
various orientifolds can be classified by $v_\a^3 = 0$ or $v_\a^3
\neq 0$.

Taking the UV limit $l\to \infty$ of $\tilde{\cal K}$, we can
factorize and compute the contributions to the divergences of each
of the two crosscups, between which closed strings propagate
(Fig.\ref{Factorization}). The results are provided in the
appendix where we use representative pictures for the tadpoles.
The type of the twist of the emitted closed string ($\a^2$) is
marked on the right of the tadpole.
\begin{figure}
\begin{center}
\epsfig{file=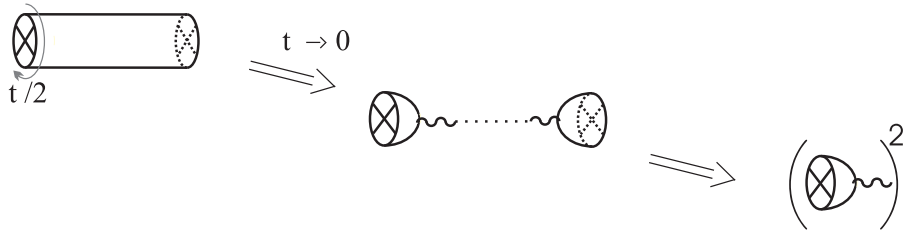,width=120mm}
\end{center}
\caption{We can factorize and compute the contributions to the
divergences of each of the two crosscups  between which closed
strings propagate.}\label{Factorization}
\end{figure}
Using this notation, we can classify all cases in a compact way:
\begin{itemize}
\item The contribution to the Klein Bottle from an element $\a$
with $v_\a=(v_\a^1,v_\a^2,0)$ will be:
\begin{itemize}
\item[-] If the orbifold group $G$ does not include $R$ factors,
the only contribution of $\a$ to the massless tadpoles will come
from the untwisted sector states (the first term in (\ref{gf})):
\bea
(1_{NS}-1_R)\bigg(\Omega \a
\raisebox{-1.2ex}[0cm][0cm]{\epsfig{file=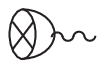,width=12mm}}
~\a^2\bigg)^2~.
%
\label{k}\eea
\item[-] In case the group $G$ contains $Z_2$ factors, $R\in G$,
that commute with $\a$, we have extra contributions from the
twisted states. We classify the contributions to the tadpoles by
the different $Z_2$ elements that are included in $G$:
\begin{itemize}
\item[i.] If $R_3\in G$ we have two sources of divergences:
\bea
(1_{NS}-1_R)\bigg(\Omega \a
\raisebox{-1.2ex}[0cm][0cm]{\epsfig{file=TadRPP_2.eps,width=12mm}}
~\a^2+\Omega R_3\a
\raisebox{-1.2ex}[0cm][0cm]{\epsfig{file=TadRPP_2.eps,width=12mm}}
~\a^2\bigg)^2~,
%
\label{ki}\eea
\item[ii.] If $R_{i}\in G$ for a given $i=1$ or $2$:
\bea
(1_{NS}-1_R)\bigg(\Omega \a
\raisebox{-1.2ex}[0cm][0cm]{\epsfig{file=TadRPP_2.eps,width=12mm}}
~\a^2+\Omega R_i\a
\raisebox{-1.2ex}[0cm][0cm]{\epsfig{file=TadRPP_2.eps,width=12mm}}
~\a^2\bigg)^2~,
%
\label{kii} \eea
\item[iii.] If $R_l\in G$ with $l=1,2,3$:
\bea
(1_{NS}-1_R)\bigg(\Omega \a
\raisebox{-1.2ex}[0cm][0cm]{\epsfig{file=TadRPP_2.eps,width=12mm}}
~\a^2+\sum_{l=1}^3\Omega R_l\a
\raisebox{-1.2ex}[0cm][0cm]{\epsfig{file=TadRPP_2.eps,width=12mm}}
~\a^2\bigg)^2~.
%
\label{kiii} \eea
\end{itemize}
\end{itemize}
%
%
%
All the amplitudes above are proportional to $(1_{NS}-1_{R})$ and
their multiplicatives appear as perfect squares
\cite{Pradisi:1988xd, Angelantonj:2002ct}. We should mention that
for this kind of orbifold action all the amplitudes are volume
depended (${\cal V}_3$ is the volume of the third torus which is
not affected by $\a$).
They are of the general form:
\be (1_{NS}-1_{R}) \left[ K_1 \sqrt{{\cal V}_3}+{K_2 \over
\sqrt{{\cal V}_3}}\right]^2~, \label{KB-general}\ee
where $K_1$ and $K_2$ are constants.
\item Next, we consider the case where $\a$ acts on all tori
($v_\a=(v_\a^1,v_\a^2,v_\a^3)$ with $v_\a^{l=1,2,3}\neq 0$).
We can classify again:
\begin{itemize}
\item[-] If $G$ contains no $R$ factors.
\bea
(1_{NS}-1_R)\bigg(\Omega \a
\raisebox{-1.2ex}[0cm][0cm]{\epsfig{file=TadRPP_2.eps,width=12mm}}
~\a^2\bigg)^2~.
%
\label{kk} \eea
\item[-] In case $G$ contains $R$ factors:
\begin{itemize}
\item[i.] If $R_i\in G$ for a given $i$.
\bea
(1_{NS}-1_R)\bigg(\Omega \a
\raisebox{-1.2ex}[0cm][0cm]{\epsfig{file=TadRPP_2.eps,width=12mm}}
~\a^2+\Omega R_i\a
\raisebox{-1.2ex}[0cm][0cm]{\epsfig{file=TadRPP_2.eps,width=12mm}}
~\a^2\bigg)^2~.
%
\label{kki} \eea
\item[ii.] If $R_l\in G$ for $l=1,2,3$.
\bea
(1_{NS}-1_R)\bigg(\Omega \a
\raisebox{-1.2ex}[0cm][0cm]{\epsfig{file=TadRPP_2.eps,width=12mm}}
~\a^2+\sum_{l=1}^3\Omega R_l\a
\raisebox{-1.2ex}[0cm][0cm]{\epsfig{file=TadRPP_2.eps,width=12mm}}
~\a^2\bigg)^2~.
%
\label{kkii} \eea
\end{itemize}
\end{itemize}
All the amplitudes are again perfect squares as in
(\ref{KB-general}) without the volume dependence.
\end{itemize}
Schematically, the above classification of tadpoles can be
visualized as:
\begin{center}
\epsfig{file=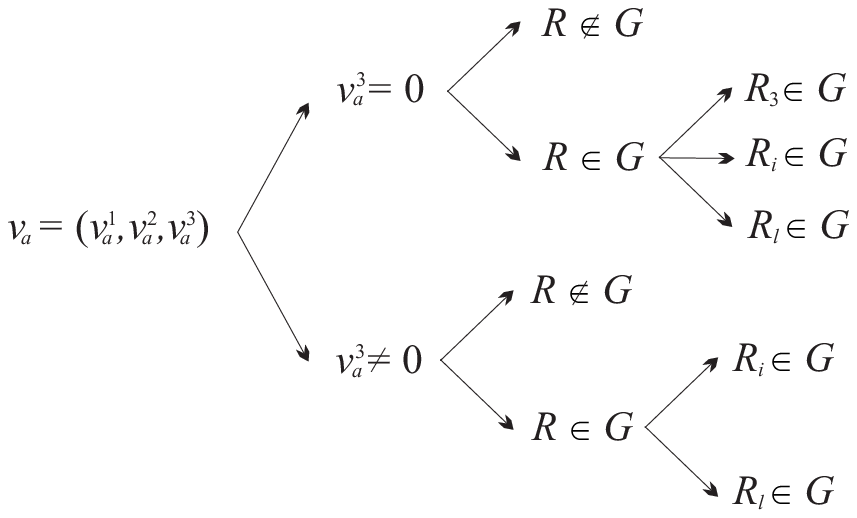,width=100mm}
\end{center}

\subsubsection{O-planes}

Tadpoles in general are amplitudes for creation of a single
particle from the vacuum. They are artifacts of the perturbation
theory and they appear in higher loops.

The tadpoles that we found in the Klein-Bottle amplitude
(\ref{Klein-Bottle}), can be interpreted as sources of massless
closed fields in space-time introduced by the orientifold
($\Omega$) projection. They couple to the massless IIB fields, in
particular the metric (so they have energy or tension), the
dilaton and the RR-forms (under which they are minimally coupled).
Such sources are localized in sub-manifolds of spacetime,
typically $p+1$ dimensional hyperplanes that are known as
orientifold planes, $O_p$.
%

Depending on the tension and charge, we define the following
notation: $O_+$ an $O$-plane with negative tension and charge,
$O_-$ an $O$-plane with positive tension and charge. The
$\bar{O}_+$ and $\bar{O}_-$ have same tension and opposite charge
to the $O_+$ and $O_-$ respectively.

The NSNS tadpoles can be seen by an analogous phenomenon in field
theory. Consider for example the action:
\bea \int d^dx\left({1\over 2}\partial_\m\f \partial^\m\f
+Q\f\right)~.\eea
The equation of motion is: $\partial_\m \partial^\m\f =Q$. If we
expand around $\f(x)=0$ we will encounter Feynman diagrams like:
\bea
Q~
\raisebox{-2.5ex}[0cm][0cm]{\unitlength=0.45mm
\begin{fmffile}{aBBa_8}
\begin{fmfgraph*}(40,25)
\fmfpen{thick} \fmfleft{i1} \fmfright{o1} \fmf{photon}{i1,o1}
\fmfv{decor.shape=circle, decor.filled=empty, decor.size=.20w}{i1}
\fmffreeze \fmfdraw \fmfv{d.sh=cross,d.size=.20w}{i1}
\fmfv{decor.shape=circle, decor.filled=empty, decor.size=.20w}{o1}
\fmffreeze \fmfdraw \fmfv{d.sh=cross,d.size=.20w}{o1}
\end{fmfgraph*}
\end{fmffile}}
~~Q ~~~\sim~~~ {1\over k^2}=\int^\infty_0 dl \exp(-k^2 l)~,
\label{QFT-Tadpole}\eea
that have divergences at vanishing momentum. From
(\ref{QFT-Tadpole}) we realize that the divergence originates as
$l\to \infty$. We could avoid this divergence if we had expanded
around the correct vacuum. The NSNS divergence in $\tilde{\cal K}$
has the same origin. If we had expanded around the correct
non-constant metric and dilaton, the amplitude would converge.

The RR tadpoles have different origin and they need to be
cancelled since they refer to not vanishing charges. In noncompact
spaces this may be acceptable since the Faraday lines can end at
infinity. However, this is not possible in compact spaces where a
non vanishing of the total charge violates the Gauss law.

\subsection{Open strings and D-branes}

Stability of the above unoriented closed string theory requires a
new ``twisted" sector under $\Omega$ (analogous to the twisted
sector of the orbifold construction). This is the $open$ $string$
sector.

Open strings have endpoints. We can always define a $p+1$
dimensional hyperplane, that is called the $Dp$-$brane$, where the
ends of the open strings attach. Open strings that are attached to
the brane can freely move in the $p$ longitudinal directions (they
obey Neumann boundary conditions) and they are fixed in the
remaining $9-p$ transverse directions (where they obey Dirichlet
boundary conditions).
\begin{center}
\epsfig{file=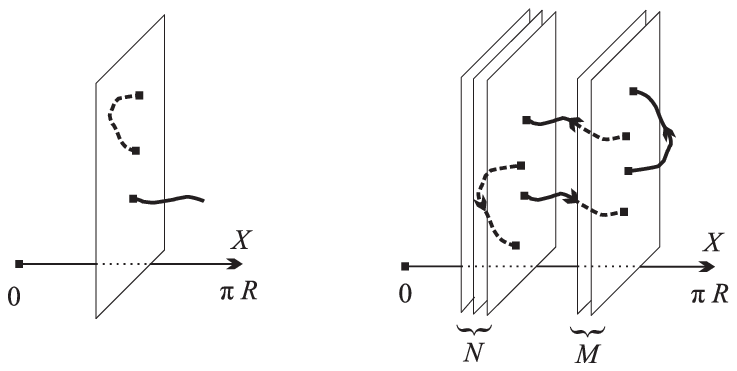,width=95mm}
\end{center}

As we mentioned before, we can introduce charges on the endpoints
of the open strings, the Chan-Paton factors (\ref{ChanPaton}). In
this picture, strings with different charges can be interpreted as
strings ending on different stacks of branes.

There is an interesting property of T-duality on D-branes. When we
T-dualize a dimension, Dirichlet boundary conditions become
Neumann and vice versa. This implies that if we T-dualize a
longitudinal or a transverse direction of a D$_p$-brane, it will
become D$_{p-1}$ or D$_{p+1}$ brane, respectively.


D-branes are dynamical objects that couple to the NSNS and RR
states. The lagrangian of a D-brane is given by:
\bea
S_{D-brane}&=&-{\t_p}\int d^{p+1}\x~ e^{-\f}
\sqrt{\det(G_{ab}+B_{ab}+2\p \a' F_{ab})}\nn\\
&& +\m_p \int_{{\cal M}_{p+1}}C_{p+1}\wedge Tr[e^{B+2\p \a'F}]
~.\eea
The first term is called the "Dirac-Born-Infeld action" and it
contains the induced metric on the brane $G_{ab}$ (that is
connected with the spacetime metric via $G_{ab}=G_{\m\n}\partial_a
X^\m \partial_b X^\n$), an antisymmetric tensor $B_{ab}$,
\footnote{$B_{ab}$ is introduced due to the T-duality that
connects D(p+1) and D(p-1) branes.} coming from the closed string
sector and the field strength of a gauge field, $F_{ab}$, that
lives in the brane. $\t_p$ is the tension of the brane.
The second term is the ``Wess-Zumino" action that describes the
coupling of the D-brane to the RR sector. $C_{p+1}$ are the RR
forms and $\m_p$ the RR charge.
%
%
%

\subsubsection{Orientifold action on open strings}\label{ActionOnCP}

In general, an open string is denoted as $|\Y, ab\rangle$, where
$\Y$ refers to the worldsheet fields and $ab$ to the Chan-Paton
indices that are associated to the string endpoints on D$p$-branes
and D$q$-branes. The Chan-Paton labels are contracted with a
hermitian matrix $\l_{ab}$. The action of a group element $\a$ of
the orientifold group $G$ is given by:
\bea \a~:~|\Y, ab\rangle\to (\g_{\a,p})_{aa'}|\a\Y, a'b'\rangle
(\g_{\a,q})^{-1}_{b'b}~,\eea
where $\g_\a$ unitary matrices associated to $\a$.
The action of the same element accompanied with $\Omega$ gives:
\bea \Omega\a~:~|\Y, ab\rangle\to (\g_{\Omega\a,p})_{aa'}|\a\Y,
b'a'\rangle (\g_{\Omega\a,q})^{-1}_{b'b}~.\eea
Since $1\in G$, acting twice with $\Omega$ we find:
\bea \g^T_\Omega =\pm\g_\Omega~.\eea
A nontrivial argument of Gimon and Polchinski shows that for
D9-branes the $\g_{\Omega}$ is symmetric and for D5-branes
antisymmetric.

The worldsheet parity $\Omega$ acts by interchanging the string
$\s\to \p-\s$ (or $z\to -\bar{z}$ in the complex plane basis):
$X(\s)\to X(\p-\s)$ and $\y(\s)\to \pm\tilde{\y}(\p-\s)$. Applying
this on (\ref{Open}) and (\ref{FermionicExpansion}) we find that:
\bea \Omega\a^\m_m \Omega^{-1}= \pm e^{i\p m}\a^\m_m
~,~~~~~~~\Omega\y^\m_m \Omega^{-1}= \e e^{i\p m}\y^\m_m~,\eea
where ``$+$" is for NN and ``$-$" is for DD strings (\ref{Open}).
These transformations are valid for integer and half-integer $m$.
The $\e=\pm 1$ leaving an irrelevant sign freedom. There is no
corresponding result for the ND sector since $\Omega$ takes it to
a different DN sector.

The action of the rotating elements $\a$ on the open strings is
the same as the action on the closed ones (\ref{RotationsB+NS},
\ref{RotationsR}).

\subsubsection*{Rotation elements on Chan-Paton factors}

The action of the orientifold elements on the Chan-Paton (CP)
factors is defined in such a way that the total open string
wavefunction $|\Y, ab\rangle \l_{ab}$ will remain invariant under
the orientifold action. As an example, we will provide the
transformation rules for the massless open spectrum of a generic
orientifold model.
\begin{itemize}
\item For Dp-Dp states, where $p=9,5_i$:
\begin{itemize}
\item The massless NS sector is $\y^M_{-1/2}|0, ab\rangle\l_{ab}$.
This includes gauge bosons for $M=\m$ and matter scalars for
$M=\pm i$, with $i=1,2,3$. For the gauge fields, the
$\y^\m_{-1/2}$ do not transform under $\a$. However, for the
scalars, the $\y^i_{-1/2}$ acquire a phase $e^{\pm2\p i v^i_\a}$.
Therefore, to construct totally invariant states, the $\l_{ab}$s
should transform in the opposite way:
\bea \y^\m_{-1/2}|0,
ab\rangle\l_{ab}^{(0)}&:&\l^{(0)}=\g_{\a,p}\l^{(0)}\g_{\a,p}^{-1}
~,~~~~~~~~~~~\l^{(0)}=-\g_{\Omega,p}\l^{(0)T}\g_{\Omega,p}^{-1}
~,~~~~~\label{ImposingOnLamda-0}\\
\y^i_{-1/2}|0, ab\rangle\l_{ab}^{(i)}&:&\l^{(i)}=e^{2\p i
v^i_\a}\g_{\a,p}\l^{(i)}\g_{\a,p}^{-1}
~,~~~~~\l^{(i)}=-\g_{\Omega,p}\l^{(i)T}\g_{\Omega,p}^{-1}
~,~~~~~\label{ImposingOnLamda-1}\eea
on the fixed points. Scalar fields $\y^j_{-1/2}|0, ab\rangle$ on
D5$_i$-branes with $i\neq j$ transform with a minus sign in the
$\Omega$ projection due to the DD boundary conditions on the $j$
directions transverse to the brane:
\bea \y^j_{-1/2}|0, ab\rangle\l_{ab}^{(j)}&:&\l^{(j)}=e^{2\p i
v^j_\a}\g_{\a,5_i}\l^{(j)}\g_{\a,5_i}^{-1}
~,~~~~~\l^{(j)}=\g_{\Omega,5_i}\l^{(j)T}\g_{\Omega,5_i}^{-1}
~~~~~~.~\label{ImposingOnLamda-2}\eea
In case we can move some D5$_i$-branes away from the fixed points,
rotation elements do not act on the fields and we should omit the
first equation in (\ref{ImposingOnLamda-0},
\ref{ImposingOnLamda-1}, \ref{ImposingOnLamda-2}).
\item The massless R sector is the vacuum:
$|s_0s_1s_2s_3,ij\rangle \l_{ij}$. GSO requires an even number of
``$-1/2$"s. Using the $\sum_i v^i_\a=0$ we find that: states with
$s_0=s_1=s_2=s_3$ do not transform and their relative CP matrix
transforms as $\l^{(0)}$. However, states that have: $s_i=s_0\neq
s_j=s_k$ transform with a phase $e^{\pm2\p i v^i_\a}$ and their
relative CP matrices transform as the $\l^{(i)}$s.
\end{itemize}
\item For D9-D5$_i$ states:
\begin{itemize}
\item The massless NS sector is $|s_js_k, ab\rangle\l_{ab}$. GSO
projection requires $s_j=s_k$. These fields transform under $\a$
acquiring a phase $e^{2\p i (v^j_\a s_j+v^k_\a s_k)}$. The
$\l_{ab}$ transform as:
\bea |s_js_k, ab\rangle\l_{ab}&:&\l_{59}=e^{2\p i (v^j_\a
s_j+v^k_\a s_k)} \g_{\a,5_i}\l_{5_i9}\g_{\a,9}^{-1}~,\eea
on the fixed points. Obviously, there is no constraint due to
$\Omega$, since it relates different states ND$\leftrightarrow$DN.
\item The massless R sector is the vacuum: $|s_0s_i,
ab\rangle\l_{ab}$. GSO projection requires $s_0=s_i$. These fields
transform under $\a$ acquiring a phase $e^{2\p i v^i_\a s_i}$. The
$\l_{ab}$ transform as:
\bea |s_0s_i, ab\rangle\l_{ab} &:& \l_{59}=e^{2\p i v^i_\a s_i}
\g_{\a,5_i}\l_{5_i9}\g_{\a,9}^{-1}~.\eea
\end{itemize}
\end{itemize}
Notice that the condition $\sum_i v^i_\a=0$ relates bosonic and
fermionic states as it was stated above.

\subsubsection*{Scherk-Schwarz deformation on Chan-Paton}

Similarly to the above we can define the action of translation
elements on the Chan-Paton factors. We will concentrate onto the
$Z_2$ Scherk-Schwarz deformation element $h$
(\ref{Sherck-Schwarz}).

The action of this element is defined in such a way that
supersymmetry will be restored if we take the decompactification
limit of the torus where it acts.
\begin{itemize}
\item Consider the D$p$-D$p$ string states with $p=9,5_i$. Bosonic
states do not transform under $h$. To ensure totally invariant
states, their relative $\l$ should also not transform:
\bea
\y^{M}_{-1/2}|0,ab\rangle \l_{ab} &~:~& \l=\g_{h,p} \l
\g^{-1}_{h,p} ~.\label{TransLambdaBos.99}\eea
Space-time fermionic states acquire a minus sign and their $\l$
should also transform in this way:
\bea
|s_0 s_1 s_2 s_3,ab\rangle \l_{ab} &~:~&\l=-\g_{h,p} \l
\g^{-1}_{h,p}~. \label{TransLambdaFer.99} \eea

\item For the mixed 95$_i$ states, space-time bosons do not
transform, giving:
\bea
|s_j s_k,ab\rangle \l_{ab} &~:~&\l=\g_{h,9} \l \g^{-1}_{h,5_i}~,
\label{TransLambdaBos.59}\eea
where $j\neq i\neq k$ and GSO projection demands $s_j=s_k$.
Space-time fermions acquire a minus sign that must be eliminated
by the transformation of their relative $\l$:
\bea
|s_0 s_i,ab\rangle \l_{ab} &~:~& \l=-\g_{h,9} \l \g^{-1}_{h,5_i}~,
\label{TransLambdaFer.59}\eea
where GSO projection demands $s_0=s_i$.
\end{itemize}
Having the transformation conditions of the Chan-Paton matrices,
we need the analytic expressions of the $\g$s to find the
representations of the massless fields. Later, we will see that
$\g$ matrices obey some consistency conditions.

\subsection{Annulus}

To evaluate the 1-loop partition function, we have to include also
the contribution of the new sector of the theory (the open
strings).
The 1-loop diagram of an open string is the $Annulus$. The annulus
can be taken from the torus with the involution $z\to -\bar{z}$
and $z\to 1-\bar{z}$ (Fig.\ref{DoubleCover}). The $\t$ is purely
imaginary and the $\t_2$ is the proper time of an open string that
sweeps the annulus $\cal A$. However, there is a distinct
horizontal choice that defines the proper time elapsed while a
closed string propagates between the two boundaries $\tilde{\cal
A}$. These boundaries are the D-branes that the open string ends
on.

The Annulus amplitudes can be computed for all kinds of D-branes
existing in the theory and the contribution of an element $\a$ is
given by the trace formula:
\bea {\cal A}_{IJ,\a} = Tr_{IJ} \left[\a ~q^{L_0} \right]~, \nn
\eea
where now the trace is over all open string states attached
between $I$ and $J$ D-branes.
When there are no reflecting elements $R$ in the theory, only
D9-branes are necessary to cancel the RR and NSNS charges.
However, when there are $R_i$-factors we need in addition
D5$_i$-branes extended along the $R^4\times T_i^2$ and sitting on
the $R_i$-fixed points of the other tori.
The contribution of an element $\a$ can be written in the form:
\bea {\cal A}_\a = \bigg(Tr[\g_{\a,9}]^2 +Tr[\g_{\a,5_i}]^2\bigg)
T[^{~0}_{v_\a}] + 2 Tr[\g_{v_\a,9}] Tr[\g_{\a,5_i}]
T[^{g_i}_{v_\a}] ~.\label{av} \eea
To extract the tadpole contributions we need to perform a modular
transformation to the transverse channel, $l=1/2t$, and then take
the limit $l\to \infty$ \cite{Gimon:1996rq}. We can perform a
similar factorization to the one that we already did for the
Klein-Bottle (Fig.\ref{Factorization}) and evaluate the tadpoles
for the different D-branes:
\begin{itemize}
\item for an element $\a$ such that: $v_\a=(v_\a^1,v_\a^2,0)$:
\begin{itemize}
\item[-] If $G$ contains no $Z_2$-factors, then the only
contribution to the tadpoles in the annulus amplitude is coming
from the 99 strings
\bea
(1_{NS}-1_R)\bigg(\textrm{D9}
\raisebox{-1.2ex}[0cm][0cm]{\epsfig{file=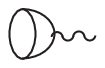,width=12mm}}
~\a\bigg)^2~.
%
\label{a} \eea
\item[-] In the case where the group $G$ contains $R$-factors,
then we have also contributions from the corresponding D5-branes.
As in the Klein Bottle case, we have the following cases:
\begin{itemize}
\item[i.] if $R_3\in G$,
\bea
(1_{NS}-1_R)\bigg(\textrm{D9}
\raisebox{-1.2ex}[0cm][0cm]{\epsfig{file=TadBoundary.eps,width=12mm}}
~\a+\textrm{D5}_3
\raisebox{-1.2ex}[0cm][0cm]{\epsfig{file=TadBoundary.eps,width=12mm}}
~\a\bigg)^2~,
%
\label{ai}\eea
\item[ii.] if $R_i\in G$, for a given $i=1$ or $2$,
\bea
(1_{NS}-1_R)\bigg(\textrm{D9}
\raisebox{-1.2ex}[0cm][0cm]{\epsfig{file=TadBoundary.eps,width=12mm}}
~\a+\textrm{D5}_i
\raisebox{-1.2ex}[0cm][0cm]{\epsfig{file=TadBoundary.eps,width=12mm}}
~\a\bigg)^2~,
%
\label{aii} \eea
\item[iii.] if $R_l\in G$, with $l=1,2,3$
\bea
(1_{NS}-1_R)\bigg(\textrm{D9}
\raisebox{-1.2ex}[0cm][0cm]{\epsfig{file=TadBoundary.eps,width=12mm}}
~\a+\sum_{l=1}^3 \textrm{D5}_l
\raisebox{-1.2ex}[0cm][0cm]{\epsfig{file=TadBoundary.eps,width=12mm}}
~\a\bigg)^2~.
%
\label{aiii} \eea
\end{itemize}
\end{itemize}
In all the above cases the general structure is again proportional
to zero $(1_{NS}-1_{R})$ and the multiplicative is a function of
the volume of the unaffected torus:
\be (1_{NS}-1_{R}) \left[ A_1 \sqrt{\cal V}_3+{A_2 \over
\sqrt{\cal V}_3}\right]^2 ~.\label{A-general}\ee
The $A_1$ and $A_2$ are functions of the traces of the Chan-Paton
factors, $Tr[\g_{\a, I}]$. The $A_1$ is the contribution of
strings that are longitudinal to the torus which is unaffected by
$v_\a$ (they have Neumann boundary conditions in this torus).
Therefore, it is proportional to $Tr[\g_{\a,9}]$ and
$Tr[\g_{\a,5_3}]$. The $A_2$ is the contribution of the strings
that are transverse to ${\cal V}_3$ and it is a function of
$Tr[\g_{\a,5_i}]$ for $i=1,2$.

\item If now $v_\a=(v_\a^1,v_\a^2,v_\a^3)$, then the
classification is similar to the Klein Bottle one:
\begin{itemize}
\item[-] If the orbifold group $G$ has no $Z_2$ factors, we have
just the contribution of the 99 strings.
\bea
(1_{NS}-1_R)\bigg(\textrm{D9}
\raisebox{-1.2ex}[0cm][0cm]{\epsfig{file=TadBoundary.eps,width=12mm}}
~\a\bigg)^2~.
%
\label{aa} \eea
\item[-] If the group $G$ contains $R$ factors, then:
\begin{itemize}
\item[i.] if $R_i\in G$ for a given $i$, we should include its
corresponding $D5_i$-branes as well
\bea
(1_{NS}-1_R)\bigg(\textrm{D9}
\raisebox{-1.2ex}[0cm][0cm]{\epsfig{file=TadBoundary.eps,width=12mm}}
~\a+ \textrm{D5}_i
\raisebox{-1.2ex}[0cm][0cm]{\epsfig{file=TadBoundary.eps,width=12mm}}
~\a\bigg)^2~,
%
\label{aai} \eea
\item [ii.] if $R_l\in G$ with $l=1,2,3$, we should include its
corresponding $D5_l$-branes
\bea
(1_{NS}-1_R)\bigg(\textrm{D9}
\raisebox{-1.2ex}[0cm][0cm]{\epsfig{file=TadBoundary.eps,width=12mm}}
~\a+\sum_{l=1}^3 \textrm{D5}_l
\raisebox{-1.2ex}[0cm][0cm]{\epsfig{file=TadBoundary.eps,width=12mm}}
~\a\bigg)^2~.
%
\label{aaii} \eea
\end{itemize}
\end{itemize}
The structure of these amplitudes is similar to (\ref{A-general})
without the volume dependance.
\end{itemize}

\subsection{M\"obius Strip}

In general, we can expect that there are closed strings which
propagate between the two sources of closed strings: the
$O$-planes and the D-branes.
The amplitude that describes such transmission is topologically a
1-loop amplitude and contributes at the same level in string
perturbation theory as the Klein-bottle and the Annulus. It is the
so called \textit{M\"{o}bius Strip}.

The M\"obius Strip can be taken from a double covered torus by the
involution: $z\to 1-\bar{\t}_2+i\t_2$ (Fig.\ref{DoubleCover}). The
parameter $\t_2$ describes the proper time elapsed while an open
string sweeps the M\"obius Strip ${\cal M}$. There is again an
alternative choice where the horizontal parameter describes a
closed string propagating between a boundary and a crosscap
$\tilde{\cal M}$.

The contribution of an element $\a$ accompanied by $\Omega$ is:
\bea {\cal M}_{I,\a} = Tr_I \left[\Omega \a ~q^{L_0}\right]~,
\label{MaI}\eea
where the trace is over open strings attached to a $DI$-brane.
Finally, this contribution has the form:
\bea {\cal M}_\a = &-&\Big(Tr[\gamma^T_{\Omega
\a,9}\gamma^{-1}_{\Omega \a,9}] T[^{~0}_{v_\a}]
+Tr[\gamma^T_{\Omega R_i\a,9}\gamma^{-1}_{\Omega R_i\a,9}]
T[^{~0}_{g_iv_\a}]
\nn\\
&&+Tr[\gamma^T_{\Omega \a,5_i}\gamma^{-1}_{\Omega \a,5_i}]
T[^{~0}_{g_iv_\a}] +Tr[\gamma^T_{\Omega
g_iv,5_i}\gamma^{-1}_{\Omega g_iv,5_i}] T[^{~0}_{v_\a}]\Big)~, \nn
\eea
the overall minus sign is conventional. However, we should make
the same choice of sign as for the identity element of $G$.
To extract the tadpole conditions we must perform a modular
transformation to the transverse channel of the form $P=T S T^2 S
T$ where, $T:\t \to \t+1$ and $S:\t \to - 1/\t$, where in this
case $l=1/8t$. Finally, we take the UV limit $l \to\infty$
(Fig.\ref{DoubleCover}).
%
%

The M\"obius strip transverse channel amplitude is the mean value
between the transverse channel Klein Bottle and Annulus amplitudes
\cite{Pradisi:1988xd, Angelantonj:2002ct}.
Therefore, comparing the UV limit of the M\"obius strip amplitude
(\ref{MaI}) with the mean value of the UV limits of the Klein
Bottle and Annulus amplitudes, we obtain the following constraints
on the matrices $\g_{\a,I}$ and $\g_{\Omega.\a,I}$:
\bea &&Tr[\gamma^T_{\Omega \a,9}\gamma^{-1}_{\Omega
\a,9}]=Tr[\gamma_{\a^2,9}]~~,~~~~~~~~~
%
Tr[\gamma^T_{\Omega R_i \a,9}\gamma^{-1}_{\Omega R_i \a,9}]=
-Tr[\gamma_{\a^2,9}]~~, \nn\\&&
Tr[\gamma^T_{\Omega \a,5_i}\gamma^{-1}_{\Omega \a,5_i}]=
-Tr[\gamma_{\a^2,5_i}]~~,~~~~
%
Tr[\gamma^T_{\Omega R_i \a,5_i}\gamma^{-1}_{\Omega R_i \a,5_i}]=
Tr[\gamma_{\a^2,5_i}]~~,
\nn\\
&&Tr[\gamma^T_{\Omega R_j \a,5_i}\gamma^{-1}_{\Omega R_j \a,5_i}]=
-Tr[\gamma_{\a^2,5_i}] ~,\label{consta} \eea
where in the last equation $i\neq j$ and $i,j= 1,2,3$.
These constraints appear for either $v_\a=(v_\a^1,v_\a^2,0)$ or
$v_\a=(v_\a^1,v_\a^2,v_\a^3)$.

\subsection{Tadpole conditions}\label{Tadpoles-NoSS}

The massless part of the transverse channel amplitudes
$\tilde{\cal K}_\a +\tilde{\cal A}_\a +\tilde{\cal M}_\a$ provide
the tadpole conditions. Let us examine all the different cases for
an element $\a^2$ where:
\begin{itemize}
\item $\a$ is such that $v_\a=(v_\a^1,v_\a^2,0)$:
\begin{itemize}
\item[-] If $G$ contains no $Z_2$ factors:
\bea Tr[\gamma_{\a^2,9}]= 32 \prod_l \cos \pi v_\a^l
~.\label{tvk1} \eea
\item[-] If $G$ contains $Z_2$ factors, then we have the following
cases:
\begin{itemize}
\item[i.] if $R_3\in G$,
\bea Tr[\gamma_{\a^2,9}]+ 4\prod_{l} \sin2\pi
v_\a^l~Tr[\gamma_{\a^2,5_{3}}]= 32~ \bigg(\prod_{l} \cos\pi v_\a^l
+\prod_{l} \sin\pi v_\a^l\bigg) ~,~~~~~
\label{tvk2i} \eea
\item[ii.] if $R_{i}\in G$ for a given $i=1$ or $2$,
\bea
Tr[\gamma_{\a^2,9}]&=& 32~\prod_{l} \cos\pi v_\a^l~,\nonumber\\
2\sin2\pi v_\a^j~Tr[\gamma_{\a^2,5_{i}}]&=& 32~\cos\pi v_\a^i
\sin\pi v_\a^j~. \label{tvk2ii} \eea
\item[iii.] if $R_l\in G$ with $l=1,2,3$,
\bea Tr[\gamma_{\a^2,9}]+4 \prod_{l} \sin2\pi
v_\a^lTr[\gamma_{\a^2,5_3}] &=& 32~\bigg(\prod_{l} \cos\pi v_\a^l
+ \prod_{l} \sin\pi v_\a^l\bigg)~,
\nonumber\\
\sum_{i\neq j =1,2} 2\sin2\pi v_\a^j Tr[\gamma_{\a^2,5_i}]&=&
32~\sum_{i\neq j =1,2} \epsilon_{ij} \cos\pi v_\a^i \sin\pi
v_\a^j~.
~~~~~~~~\label{tvk2iii} \eea
\end{itemize}
\end{itemize}
\item $\a$ is such that $v_\a=(v_\a^1,v_\a^2, v_\a^3)$:
\begin{itemize}
\item[-] If $G$ does not contain any $Z_2$ factors:
\bea Tr[\gamma_{\a^2,9}]= 32~\prod_l \cos \pi v_\a^l
~.\label{tvl1} \eea
\item[-] If $G$ does contain $Z_2$ factors, then:
\begin{itemize}
\item[i.] if $R_i\in G$ for a given $i$:
\bea Tr[\gamma_{\a^2,9}]&+& 4\prod_{l\neq i} \sin2\pi
v_\a^l~Tr[\gamma_{\a^2,5_{i}}]\nn\\
&=& 32~ \bigg(\prod_l \cos\pi v_\a^l + \prod_l \sin\pi
v_\a^l\bigg)~,
\label{tvl2i}\eea
\item[ii.] if $R_l\in G$ with $l=1,2,3$:
\bea Tr[\gamma_{\a^2,9}]&+& 4\sum_{i=1}^3 \prod_{l\neq i} \sin2\pi
v_\a^l Tr[\gamma_{\a^2,5_i}]\nonumber\\
&=& 32~ \bigg(\prod_l \cos\pi v_\a^l + \sum_i \cos\pi v_\a^i
\prod_{l\neq i} \sin\pi v_\a^l\bigg)~.
\label{tvl2ii} \eea
\end{itemize}
\end{itemize}
\end{itemize}
Notice that in all these cases, the tadpole conditions hold for
both NS and R sectors due to supersymmetry.

The tadpole condition for an element $\a \in G$ that is not the
square of any other element of $G$ (there is no element $\b \in G$
such that $\a = \b^2$), will receive contribution only from the
Annulus amplitude. If this element is such that
$v_\a=(v_\a^1,v_\a^2,0)$ or $g_3 v_\a$, tadpole conditions will be
the same as before with zeros in the right hand side of
(\ref{tvk1}-\ref{tvk2iii}). It is not difficult to work out the
tadpole conditions for elements $g_i v_\a$:
\bea Tr[\gamma_{R_i\a,9}]&+& 4\sin\p v_\a^i \cos\p v_\a^j
Tr[\gamma_{R_i\a,5_3}]
\nn\\
&+&2\cos\pi v_\a^j Tr[\gamma_{R_i\a,5_i}]+2\sin\p v_\a^i
Tr[\gamma_{R_i\a,5_j}]=0~, \label{tgiv} \eea
where it is understood that $i\neq j =1,2$ and the different terms
exist only if the corresponding $R$s do. If $v_\a= (v_\a^1,
v_\a^2, v_\a^3)$, the tadpole conditions are the same as
(\ref{tvl1}-\ref{tvl2ii}) without the right hand side (\ie the
right hand side is zero).

In the next chapters, we will give some applications of the
formulae we have obtained in this section and compare with the
supersymmetric orientifolds already studied in the literature.

\subsection{Breaking Supersymmetry with Scherk-Schwarz
deformation}

In this section we include a $Z_2$ Scherk-Schwarz (SS) deformation
in order to break supersymmetry.
Without loss of generality we consider that the translation $h_2$
of (\ref{Sherck-Schwarz}) acts on a direction of the third torus
$T^2_3$. This deformation is compatible only with an orbifold
action that commutes with it, therefore, we will restrict
ourselves to elements $a$ with rotation angles of the form
$v_\a=(v^1_\a,v^2_\a,0)$ or $v_\a g_i$ where $g_i$ is the rotation
angle of $R_i$, a $Z_2$ element which leaves the coordinates of
the $T^2_i$ torus invariant and gives a minus sign to the others,
$i=1,2$.

\subsubsection{Klein Bottle}\label{KleinBottle+SS}

The trace in $\cal K$, is taken over all states, and gives rise to
a term coming from the zero modes:
\bea \sum_{m,n} q^{\a'p_L^2/4}\bar{q}^{\a'p_R^2/4}\langle
m,n|\Omega \a| m,n\rangle~.\label{KB-ZeroModes}\eea
By (\ref{OmegaOnZero}), we realize that the $h$-twisted sector
does not survive the $\Omega$ projection since it has shifted
windings and $\Omega$ keeps only $n=0$ states.
This sector can survive iff there is $R$, a $Z_2$ element in $G$
where, acting with $\Omega$ keeps the $n\neq 0$ states:
$\Omega R~|m,n\rangle \rightarrow ~|-m,n\rangle $.
In this case, the invariant states are those with vanishing
momenta, $m=0$. Therefore, the $h$-twisted sector will survive
this projection if $h$ and $R$ act in the same direction.
It is easy to realize that $R$ and $Rh$ twisted fields generate
$O_5$ and $\bar{O}_5$-planes sitting on the corresponding
%
%
fixed points 
%
\cite{Antoniadis:1998ki, Scrucca:2001ni}.

To extract the massless tadpole contribution we need to perform a
modular transformation, $l=1/4t$, and then take $l\to \infty$, as
in the previous section. In addition to (\ref{TSUSY}), we will
have extra contributions from the $h$ and $hR_i$ twisted sector:
\bea T[^{~~h}_{2g_{i} v_\a}], ~~~T[^{~~~h}_{2g_{i}h v_\a}] , ~~~
T[^{~hg_j}_{2g_{i} v_\a}], ~~~T[^{~~hg_j}_{2g_{i}h v_\a}]~,
\label{kghv1} \eea
where $i\neq j =1,2$ ($T[^a_b]$ are provided in the appendix).
These sectors contribute as $(1_{NS}+1_R)$ to the tadpoles.
\begin{itemize}
\item[-] If the orbifold group $G$ does not contain a $Z_2$
element, the contribution to the Klein Bottle will come only from
the untwisted sector as for the case without SS (\ref{k}). There
is no contribution from ${\cal K}_{h\a}$ due to the shift (it
gives rise only to massive states).
%
%
\item[-] If the group $G$ contains an $R$ factor, we have also
contributions from the twisted states.
\begin{itemize}
\item[i.] If $R_3\in G$,
%
%
the contribution is exactly as before (without SS (\ref{ki}))
because the Scherk-Schwarz deformation is acting transverse to the
$R_3$ factor and so the twisted states by $R_3 h$ do not
contribute.
\item[ii.] if $R_{i}\in G$ for a given $i=1$ or $2$
\bea
&&1_{NS}\bigg(\Omega \a
\raisebox{-1.2ex}[0cm][0cm]{\epsfig{file=TadRPP_2.eps,width=12mm}}
~\a^2+\Omega R_i\a
\raisebox{-1.2ex}[0cm][0cm]{\epsfig{file=TadRPP_2.eps,width=12mm}}
~\a^2
+\Omega R_i h\a 
\raisebox{-1.2ex}[0cm][0cm]{\epsfig{file=TadRPP_2.eps,width=12mm}}
~\a^2 \bigg)^2\nn\\
&-&1_R~\bigg(\Omega \a
\raisebox{-1.2ex}[0cm][0cm]{\epsfig{file=TadRPP_2.eps,width=12mm}}
~\a^2+\Omega R_i\a
\raisebox{-1.2ex}[0cm][0cm]{\epsfig{file=TadRPP_2.eps,width=12mm}}
~\a^2
-\Omega R_i h\a 
\raisebox{-1.2ex}[0cm][0cm]{\epsfig{file=TadRPP_2.eps,width=12mm}}
~\a^2 \bigg)^2
%
%
\label{khii} \eea
\item[iii.] if $R_l\in G$ with $l=1,2,3$.
\bea
1_{NS}\bigg(\Omega \a
\raisebox{-1.2ex}[0cm][0cm]{\epsfig{file=TadRPP_2.eps,width=12mm}}
~\a^2
&+&\Omega R_3\a
\raisebox{-1.2ex}[0cm][0cm]{\epsfig{file=TadRPP_2.eps,width=12mm}}
~\a^2\nn\\
&+&\sum_{i=1}^2\left(\Omega R_i\a
\raisebox{-1.2ex}[0cm][0cm]{\epsfig{file=TadRPP_2.eps,width=12mm}}
~\a^2
+\Omega R_i h\a
\raisebox{-1.2ex}[0cm][0cm]{\epsfig{file=TadRPP_2.eps,width=12mm}}
~\a^2 \right)\bigg)^2~~~~~~~~\nn\\
-1_R\bigg(\Omega \a
\raisebox{-1.2ex}[0cm][0cm]{\epsfig{file=TadRPP_2.eps,width=12mm}}
~\a^2
&+&\Omega R_3\a
\raisebox{-1.2ex}[0cm][0cm]{\epsfig{file=TadRPP_2.eps,width=12mm}}
~\a^2\nn\\
&+&\sum_{i=1}^2\left(\Omega R_i\a
\raisebox{-1.2ex}[0cm][0cm]{\epsfig{file=TadRPP_2.eps,width=12mm}}
~\a^2
-\Omega R_i h\a
\raisebox{-1.2ex}[0cm][0cm]{\epsfig{file=TadRPP_2.eps,width=12mm}}
~\a^2 \right)\bigg)^2~~~~~~~~
%
%
%
%
\label{khiii} \eea
%
%
%
\end{itemize}
\end{itemize}
All these amplitudes are perfect squares as they should be.
However, the cases (ii.) and (iii.) do not appear as
$(1_{NS}-1_R)$ any more. This dissimilarity of the coefficients of
the NS and R oscillators is due to the effect of SS deformation
and the breaking of supersymmetry via the term $\Omega R_i \a h$.
All the amplitudes have the general form:
\bea 1_{NS}\left[ K_{NS,1} \sqrt{\cal V}_3+{K_{NS,2} \over
\sqrt{\cal V}_3}\right]^2 -1_{R}\left[ K_{R,1} \sqrt{\cal
V}_3+{K_{R,2} \over \sqrt{\cal
V}_3}\right]^2~,\label{K-general+SS}\eea
where $K_{NS,2} \sim (1+1)f(v_\a)$, $K_{R,2}\sim (1-1)f(v_\a)=0$
and $f(v_\a)$ is a function of the vector $v_\a$. This explains
the appearance of the factor of 2 in the NS sector in (\ref{khii})
and (\ref{khiii}) and the absence of the factor proportional to
$1/\sqrt{\cal V}_3$ in the R sector.

\subsubsection{Annulus}

To cancel these tadpoles one needs to add D9, D5$_3$ and
D5$_i$-branes as well as D5$_i$-antibranes in the case $R_i \in
G$, with $i=1,2$, where the Scherk-Schwarz element $h$ acts in the
$T^2_3$ torus. The anti D5$_i$-branes sit on the $R_ih$ fixed
points \cite{Scrucca:2001ni}.
\begin{center}
\epsfig{file=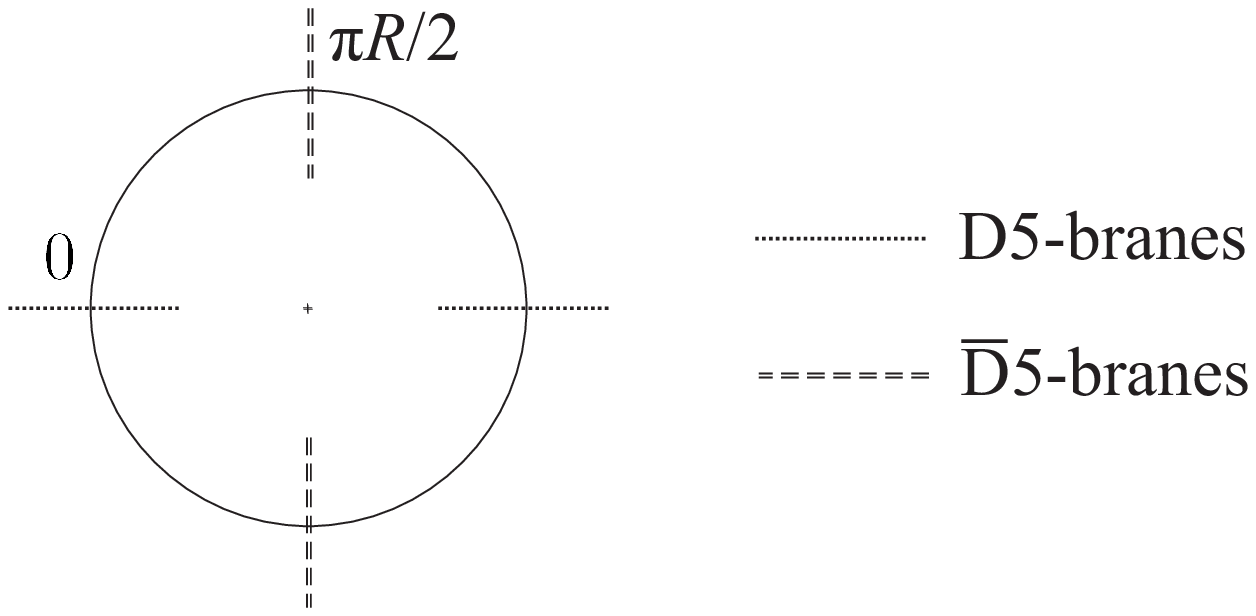,width=80mm}
\end{center}
The contribution from the annulus amplitudes to the tadpole
conditions are the same as for the case without SS deformation,
with in addition the anti-D$5_{i\neq 3}$-brane sector when $R_i\in
G$. Note that the annulus amplitudes between the same type of
branes contribute as $(1_{NS}-1_R)$ whereas, the ones between a
D-brane and an anti-D-brane give $(1_{NS}+1_R)$. The contribution
of the element $\a$ in the Annulus amplitudes is:
\begin{itemize}
\item[-] When $G$ does not contain $Z_2$ factors\footnote{what we
mean by $Z_2$ factors are those other than the Scherk-Schwarz
deformation $h$.}, the contribution is the same as (\ref{a}).
%
%
\item[-] When $G$ contains $Z_2$ factors, then:
\begin{itemize}
\item[i.] if $R_3\in G$, the contribution is the same as
(\ref{ai}).
%
%
\item[ii.] if $R_{i}\in G$ for a given $i=1$ or $2$:
\bea
&&1_{NS}\bigg(\textrm{D9}
\raisebox{-1.2ex}[0cm][0cm]{\epsfig{file=TadBoundary.eps,width=12mm}}
~\a+\textrm{D5}_i
\raisebox{-1.2ex}[0cm][0cm]{\epsfig{file=TadBoundary.eps,width=12mm}}
~\a
+\overline{\textrm{D5}}_i
\raisebox{-1.2ex}[0cm][0cm]{\epsfig{file=TadBoundary.eps,width=12mm}}
~\a \bigg)^2\nn\\
&-&1_R~\bigg(\textrm{D9}
\raisebox{-1.2ex}[0cm][0cm]{\epsfig{file=TadBoundary.eps,width=12mm}}
~\a+\textrm{D5}_i
\raisebox{-1.2ex}[0cm][0cm]{\epsfig{file=TadBoundary.eps,width=12mm}}
~\a
-\overline{\textrm{D5}}_i
\raisebox{-1.2ex}[0cm][0cm]{\epsfig{file=TadBoundary.eps,width=12mm}}
~\a \bigg)^2~
%
%
%
%
%
\label{ahii} \eea
\item[iii.] if $R_l\in G$ with $l=1,2,3$:
\bea
1_{NS}\bigg(\textrm{D9}
\raisebox{-1.2ex}[0cm][0cm]{\epsfig{file=TadBoundary.eps,width=12mm}}
~\a &+& \textrm{D5}_3
\raisebox{-1.2ex}[0cm][0cm]{\epsfig{file=TadBoundary.eps,width=12mm}}
~\a\nn\\
&+&
\sum_{i=1}^2 \left(\textrm{D5}_i
\raisebox{-1.2ex}[0cm][0cm]{\epsfig{file=TadBoundary.eps,width=12mm}}
~\a
+\overline{\textrm{D5}}_i
\raisebox{-1.2ex}[0cm][0cm]{\epsfig{file=TadBoundary.eps,width=12mm}}
~\a \right)\bigg)^2~\nn\\
-1_R~\bigg(\textrm{D9}
\raisebox{-1.2ex}[0cm][0cm]{\epsfig{file=TadBoundary.eps,width=12mm}}
~\a &+& \textrm{D5}_3
\raisebox{-1.2ex}[0cm][0cm]{\epsfig{file=TadBoundary.eps,width=12mm}}
~\a\nn\\
&+&
\sum_{i=1}^2 \left(\textrm{D5}_i
\raisebox{-1.2ex}[0cm][0cm]{\epsfig{file=TadBoundary.eps,width=12mm}}
~\a
-\overline{\textrm{D5}}_i
\raisebox{-1.2ex}[0cm][0cm]{\epsfig{file=TadBoundary.eps,width=12mm}}
~\a \right)\bigg)^2~
%
%
%
%
%
%
%
\label{ahiii} \eea
\end{itemize}
\end{itemize}
The massless contribution from the above amplitudes is not
proportional to $(1_{NS}-1_{R})$ as in the supersymmetric case.
The general form is a function of the volume of the unaffected
torus:
\be 1_{NS}\left[ A_{NS,1} \sqrt{\cal V}_3+{A_{NS,2} \over
\sqrt{\cal V}_3}\right]^2 -1_{R}\left[ A_{R,1} \sqrt{\cal
V}_3+{A_{R,2} \over \sqrt{\cal
V}_3}\right]^2~.\label{A-general+SS}\ee
The general $A_i$s are again functions of the traces of the
Chan-Paton factors, $Tr[\g_{\a, I}]$. The $A_{NS,1}$ and $A_{R,1}$
are proportional to $Tr[\g_{\a,9}]$ and $Tr[\g_{\a,5_3}]$. The
$A_{NS,2}$ and $A_{R,2}$ are proportional to $Tr[\g_{\a,5_i}]$ and
$Tr[\g_{\a,\bar{5}_i}]$, for $i=1,2$. This is precisely the effect
of the anti $\overline{\textrm{D}}$-branes.

\subsubsection{M\"obius Strip}\label{Mobius+SS}

Finally, the M\"obius strip amplitude that is derived in two
inequivalent ways (as a direct amplitude and as the mean value of
the Klein Bottle and the Annulus amplitudes) leads to the same
constraints as before (\ref{consta}), where in addition:
\begin{itemize}
\item[-] if $R_{i}\notin G$, then:
\begin{itemize}
\item[i.] if $R_3\notin G$:
\bea Tr[\gamma^{T}_{\Omega h \a,9}\gamma^{-1}_{\Omega h \a,9}]=\pm
Tr[\gamma_{\a^2,9}]~, \eea
\item[ii.] if $R_3\in G$:
\bea Tr[\gamma^{T}_{\Omega h \a,5_3}\gamma^{-1}_{\Omega h
\a,5_3}]= \pm Tr[\gamma_{\a^2,5_3}] ~.\eea
\end{itemize}
The signs are the same for the $D9$ and $D5_3$ sectors due to
T-duality. Examples of this cases have been discussed in
\cite{Anastasopoulos:2003ha}.
\item[-] If $R_{i}\in G$, for a given $i=1$ or $2$:
\bea Tr[\gamma^{T}_{\Omega h \a,I}\gamma^{-1}_{\Omega h
\a,I}]=Tr[\gamma_{\a^2,I}], ~~~I=9,5_3, 5_i,\bar{5}_i ~.\eea
In all these cases $\gamma^2_{R,I}=-1$, with $I=9,5_l,\bar{5}_i$
for all $R$s.
\end{itemize}


\subsubsection{Tadpole conditions}\label{Tadpoles-WithSS}

The tadpole conditions for an element $\a$ such that
$v_\a=(v_\a^1,v_\a^2,0)$ are classified as:
\begin{itemize}
\item[-] $G$ contains no $Z_2$ factors:
\bea Tr[\gamma_{\a^2,9}]= 32 \prod_l \cos \pi v_\a^l
~,\label{tvkh1} \eea
the tadpole condition is the same as in the case without SS
deformation (\ref{tvk1}).
\item[-] $G$ contains $Z_2$ factors:
\begin{itemize}
\item[i.] if $R_3\in G$:
\bea Tr[\gamma_{\a^2,9}]&+& 4\prod_{l} \sin2\pi
v_\a^l~Tr[\gamma_{\a^2,5_{3}}]
\nn\\
&=& 32~ \bigg(\prod_{l} \cos\pi v_\a^l + \prod_{l} \sin\pi
v_\a^l\bigg) ~,\label{tvkh2i} \eea
\item[ii.] if $R_{i}\in G$ for a given $i=1$ or $2$:
\bea
&& ~~~~~~~~~~~~~Tr[\gamma_{\a^2,9}]=32~\prod_{l} \cos\pi v_\a^l\nn\\
&& 1_{NS}: ~~~~2\sin2\p v_\a^j~\Big(Tr[\gamma_{\a^2,5_i}]+
Tr[\gamma_{\a^2,\bar{5}_i}]\Big)
= 32~\cos\pi v_\a^i~\sin\pi v_\a^j~,\nn\\
&& 1_R:~~~~~2\sin2\pi v_\a^j~\Big(Tr[\gamma_{\a^2,5_{i}}]-
Tr[\gamma_{\a^2,\bar{5}_{i}}]\Big)= 0 ~,\label{tvkh2ii} \eea
\item[iii.] if $R_l\in G$ with $l=1,2,3$:
\bea &&Tr[\gamma_{\a^2,9}]+4 \prod_{l} \sin2\pi v_\a^l
Tr[\gamma_{\a^2,5_3}]
=32~\Big(\prod_{l} \cos\pi v_\a^l + \prod_{l} \sin\pi v_\a^l\Big)
\nonumber\\
&& 1_{NS}:~~~~~~~~2\sum_{i\neq j =1,2} \sin2\pi v_\a^j
\Big(Tr[\gamma_{\a^2,5_i}]+ Tr[\gamma_{\a^2,\bar{5}_{i}}]\Big) \nn\\
&&~~~~~~~~~~~~~~~~~~~~~~~~~~~~~~~~~~~~ =32~\sum_{i\neq j =1,2}
\epsilon_{ij}~\cos\pi v_\a^i \sin\pi v_\a^j~,
\nn\\
&& 1_{R}:~~~~~~~~~2\sum_{i\neq j =1,2} \sin2\pi v_\a^j
\Big(Tr[\gamma_{\a^2,5_i}]-Tr[\gamma_{\a^2,\bar{5}_{i}}]\Big)=0~.
\label{tvkh2iii} \eea
\end{itemize}
\end{itemize}

Finally, there could be elements that cannot be expressed as the
square of any other element in $G$, these elements will not
receive contribution from the Klein Bottle amplitude. For such
elements the tadpole conditions are the same as in
(\ref{tvkh1}-\ref{tvkh2iii}) with zero on the right hand side. For
the elements $h\a$ and $R_3h\a$ the tadpole conditions are as in
(\ref{tvkh1}) and (\ref{tvkh2i}) and because $D5_{i}$ and
$D\bar{5}_{i}$ are transverse to the direction where $h$ acts,
there are no conditions on $Tr[\g_{h\a,5_{i}}]$ and
$Tr[\g_{h\a,\bar{5}_{i}}]$. For the element $R_i\a$ the tadpole
condition is:
\bea Tr[\gamma_{R_i\a,9}]&+& 4\sin\pi v_\a^i \cos\pi v_\a^j
Tr[\gamma_{R_i\a,5_3}]
\nonumber\\
&+&2\cos\pi v_\a^j Tr[\gamma_{R_i\a,5_i}]+2\sin\pi v_\a^i
Tr[\gamma_{R_i\a,5_j}]=0~, \eea
where the D$5_i$-antibranes do not contribute because they do not
sit on the fixed points of this element. This condition is valid
for both NS and R sectors (the contribution is proportional to
$(1_{NS}-1_R)$). For the element $R_ih\a$ we find:
\bea 1_{NS}:~~Tr[\gamma_{R_ih\a,9}]&+& 4\sin\pi v_\a^i \cos\pi
v_\a^j Tr[\gamma_{R_ih\a,5_3}]
\nn\\
&+&2\cos\pi v_\a^j Tr[\gamma_{R_i\a,\bar{5}_i}]
+2\sin\pi v_\a^i Tr[\gamma_{R_ih\a,\bar{5}_j}]=0~,\nn\\
1_{R}:~~Tr[\gamma_{R_ih\a,9}]&+& 4\sin\pi v_\a^i \cos\pi v_\a^j
Tr[\gamma_{R_ih\a,5_3}]
\nn\\
&-&2\cos\pi v_\a^j Tr[\gamma_{R_ih\a,\bar{5}_i}] -2\sin\pi v_\a^i
Tr[\gamma_{R_ih\a,\bar{5}_j}]=0~, \eea
where the $D5_i$-branes do not contribute because they do not sit
on the fixed points of $R_ih\a$.

\subsection{Solving the tadpole conditions}

%
%

A simple way to impose the tadpole conditions on the Chan-Paton
matrices $\l$ is to recast them in a Cartan-Weyl basis. In this
case, constraints on the $\l$s will emerge as restrictions on
weight vectors \cite{Ibanez:1998qp}.

Suppose that we imply some constraints on the $\l$s and we find
that they are constrained to be generators of a specific
Lie-algebra. Therefore, they can be organized into charged
generators: $\l_a=E_a$ and Cartan generators, $\l_I=H_I$ such
that:
\bea [H_I,E_a]=\r_I^a E_a~,\label{Cartan}\eea
where $\r_I^a$ the roots associated to the generators $E_a$ of the
Lie-algebra. The matrix $\g_{\a}$ and its powers represent the
action of the orientifold on the Chan-Paton factors, and they
correspond to elements of a discrete subgroup of the abelian group
spanned by the Cartans. Hence, we can write:
\bea \g_{\a}=e^{-2\p i V_\a\cdot H}~,\label{ShiftVector}\eea
where the $shift$ vector $V_\a$ has the dimension of the number of
the Cartan matrices of the initial Lie-group.
Different elements have different shift vectors that are defined
by the relevant tadpole conditions.

Recalling the formula
$e^{-B} A e^B=\sum_{n=0}^\infty [A,B]_n$,
with $[A,B]_{n+1}=[[A,B]_{n},B]$ and $[A,B]_0=A$, and using
(\ref{Cartan}), it is easy to show that 
%
$\g_\a E_a \g_\a^{-1}=e^{-2\p i \r_a\cdot V_\a} E_a$. 
All the equations that provide the massless spectrum can be
expressed in the following way:
\bea \r_a \cdot V^\a_{pq}= f_{pq} ~,
%
\eea
where ``$f_{pq}$" is a number associated with the transformation
of the various strings that are stretched between Dp-Dq branes.
Notice the difference between 99, 5$_i$5$_i$ and 95$_i$ strings:
%
%
%
\bea f_{99,5_i5_i}= \left\{ \ba{lcl} 0+k && \textrm{for vectors}\\
                            v_\a^i+k && \textrm{for scalars $\y_i$}\\
                            s\cdot v_\a+k && \textrm{for fermions}
                            \ea \right. 
%
~,~
%
f_{95_i}= \left\{ \ba{lcl} s_jv_j+s_lv_l+k && \textrm{for scalars}\\
                            s_i v_i+k && \textrm{for fermions}
                            \ea \right. ~~~\label{Shifts}\eea
%
where $k\in \Zint$. In the next section, we will be more precise
by exploring the: $SO(N)$, $U(N)$ and $USp(N)$ algebras which
always appear in orientifold constructions.

\subsubsection*{Orientifolds with commuting $\g$s}

Consider the action of $\Omega$ on 99 states. The parity
transformation is represented by a symmetric $\g_{\Omega,9}$
matrix. The constraint, $\l=-\g_{\Omega,9}\l\g_{\Omega,9}^{-1}$,
restricts the original 32$\times$32 matrices $\l$ to be generators
of the $SO(32)$ algebra. Therefore, the Cartan will be
$H^I_{ij}=\d_{i,2I}\d_{j,2I}-\d_{i,2I+1}\d_{j,2I+1}$\footnote{We
have normalize the Cartans as $Tr[H_I H_J]=2\d_{IJ}$.}. The roots
$\r_\a$ have the form $(\underline{\pm 1, \pm 1,0,\ldots,0})$,
where the underlining indicates that all possible permutations
must be considered.

Every rotation element that commutes with $\g_{\Omega,9}$ can be
written in the form (\ref{ShiftVector}). We have mentioned already
that $\g_{\a}^N=\pm 1$ \cite{Ibanez:1998qp}. For even elements
$\g_{\a}$ only the minus sign is allowed, and the shift vector can
have the general form:
\bea V_\a={1\over
2N}\Big(1,\ldots,1,3,\ldots,3,\ldots\ldots,N-1,\ldots,N-1\Big)~.
\label{Veven}\eea
The number of the entries is determined by the tadpole conditions.
In the case where there are two commuting rotation elements that
commute also with $\g_{\Omega,9}$, they can both be expressed in
the form of (\ref{Veven}). However, we should be careful that the
mixed tadpole conditions between the commuting elements are
satisfied.

Consider now the action of $\Omega$ on 55 states. The parity
transformation is represented by an antisymmetric $\g_{\Omega,5}$
matrix. The constraint on $\l$ restricts the original 32$\times$32
matrices $\l$ to be generators of the $USp(32)$ algebra. The roots
$\r_\a$ have the same form as the ones above: $(\underline{\pm 1,
\pm 1,0,\ldots,0})$, however, we have to add some extra long ones:
$(\underline{\pm 2,0,\ldots,0})$.
Whenever, the D5-branes are on top of the fixed points, the long
roots are projected out.
If all D5-branes sit at the same fixed point, we can take
$V_{55}=V_{99}$, giving the same spectrum for both cases.

The 95 sector is handled using an auxiliary $SO(64)\supset
SO(32)_{(99)}\otimes SO(32)_{(55)}$ algebra. Since we have
generators acting simultaneously on both D9 and D5 branes, only
roots of the form:
\bea \r_{(95)}=\r_{(9)}\otimes \r_{(5)}=(\underline{\pm 1,
0,\ldots,0};\underline{\pm 1, 0,\ldots,0})~,\eea
must be considered. The shift vector is defined as
$W_{(95)}=V_{(9)}\otimes V_{(5)}$.

Each commuting element gives an extra contribution to the
spectrum.
%
Consider an orientifold of the type $Z_{2N}'=Z_2\times Z_N$, where
$Z_N$ commutes with the $Z_2$. We can consider the direct shift
vector $V_{2N}$ and evaluate the massless spectrum. However, we
will separate and study the action of the $Z_N$ on the spectrum
created by $Z_2$:
\begin{itemize}
\item[-] We will first evaluate the spectrum of a single $Z_2$
element that is acting as $v_2=(1/2,-1/2,0)$ and $V={1\over
4}(1,1,\ldots,1)$. Using the technique that we describe above, we
find that the $Z_2$ orientifold has gauge group $U(16)$. The
Cartans of $U(16)$ are the same as the ones of the $SO(32)$,
however, the roots of the $SO(32)$ that give the adjoint of the
$U(16)$ are only the: $(\underline{+1,-1,0,\ldots})$.

There are scalar fields, $\y^1_{1/2}|0\rangle$,
$\y^2_{1/2}|0\rangle$ in the $\Yasymm$ and $\bYasymm$, associated
to the $(\underline{+1,+1,0,\ldots})$ and $(\underline{-1,
-1,0,\ldots})$ roots of the initial $SO(32)$, respectively. There
are also scalars, $\y^3_{1/2}|0\rangle$ in the adjoint.

\item[-] On top of that, we have to act with an extra $Z_N$
element that acts as $v_N=(0,-1/N,1/N)$ (without loss of
generality).
The new gauge group will be given by the condition $\r
V_N=0\textrm{mod} \Zint$ where $\r=(\underline{+1,-1,0,\ldots})$.
Similarly, scalars $\y^1_{1/2}|0\rangle$, $\y^2_{1/2}|0\rangle$
will have $\r V_N=0\textrm{ mod } \Zint$, $\r V_N=-1/N\textrm{ mod
} \Zint$, where the roots are $(\underline{+1,+1,0,\ldots})$ and
$(\underline{-1,-1,0,\ldots})$ that gave the antisymmetric reps in
the $Z_2$ case. Finally, the $\y^3_{1/2}|0\rangle$, will have $\r
V_N=1/N\textrm{ mod } \Zint$ where again
$\r=(\underline{+1,-1,0,\ldots})$.
Similarly for the 55 and 59 sectors.
\end{itemize}
Rotation elements that commute with $\g_{\Omega}$ and have
$\g_{\a}^N=+1$, have shift vector in the general form:
\bea V_\a={1\over N}\Big( 0,\ldots,0,1,\ldots,1,\ldots
\ldots,(N-1)/2, \ldots,(N-1)/2\Big)~. \label{Vodd}\eea
The number of the entries is again determined by the tadpole
conditions.

\subsubsection*{Orientifolds with non-commuting $\g$s}

In the previous section, we studied the action of various
commuting elements on the $\l$s. We showed that we can use the
shift vectors to evaluate the spectrum. Any extra condition breaks
the representations further.

When we have non commuting $\g$s we cannot apply directly the
above method since we cannot diagonalize all $\g$ matrices
together. Models with non-commuting elements contain $Z_2 \times
Z_2$ as a subgroup.
The $Z_2 \times Z_2$ orientifold contains three $R_i$ reflecting
elements (where $i=1,2,3$) that each generate different
D5$_i$-branes. The $\g_{R_i}$s do not commute since all of them
should have $\g_{R_i}^2=-1$. After some tedious calculations we
find that the gauge group is $USp(16)$ for all branes. There are
also scalars in the antisymmetric rep of $USp(16)$
\cite{Berkooz:1996dw}. 95$_i$ and 5$_i$5$_j$ states transform in
bifundamental representations.

Having the spectrum of $Z_2 \times Z_2$, we can apply extra shift
elements on it. We will use as a basis the Cartan and the roots of
the $USp(16)$. As an example, we will consider the $Z_2\times Z_6$
(which is equivalent to $Z_2\times Z_2\times Z_3$) orientifold.
We will act just with the shift vector of $Z_3$ and we will use
the proper roots for each field. The shift vector in this case
will be:
\bea V_\a={1\over N}\Big( 0,\ldots,1,\ldots \ldots,(N-1)/2,
\ldots,(N-1)\ldots\Big) ~,\label{Vodd1}\eea
where $N=3$ \footnote{Notice that this shift vector is the same
with the one provided in (\ref{Vodd}) upon rotation. For example
the "1"s and "N-1"s can be identified upon rotation. The reason
for this choice is that tadpole conditions between different
elements cannot be satisfied with all elements in the form
(\ref{Veven}, \ref{Vodd}).}. To find the gauge group of $Z_2\times
Z_6$, we will use the roots of $USp(16)$ and for the scalars we
will remove the long roots.

\subsubsection*{Scherk-Schwarz deformation}

The action of Scherk-Schwarz deformations on open strings is
similar to the action of rotation elements.

The $\g_h$ can in general be $\g_h^2=\pm 1$. A generic choice for
these two cases is:
\bea V_h={1\over 4} \left\{ \ba {lcl}
(1_a,-1_b) & \quad & \textrm{for $\g^2_h=-1$,}\\
&&\\
(2_a,0_b)& \quad & \textrm{for $\g^2_h=+1$,} \ea \right.
\label{V-h}\eea
where the index refers to the number of the same components in the
vector. In the case where $\g_h^2=-1$, we have $a=b$, however,
there is no constraint for $\g_h^2=+1$.
The related $f_h$ for the Scherk-Schwarz deformation is just:
\bea f_{h}= \left\{\ba{lcl} 0+k && \textrm{for spacetime bosons,}\\
                            1/2+k && \textrm{for spacetime fermions,}
                            \ea \right.
\label{Shifts-h}\eea
where again $k\in \Zint$.

\subsection{Applications}


As we mention above, we can simplify the initial problem of
finding the reps of the orientifold group by using the proper
shift vector $V=\{V_i\}$ with number of identical entries $n_i$
(\ref{Veven},\ref{Vodd}).

\subsubsection*{Even elements}

Consider a shift vector of an even element where $\g^N=-1$. By the
definition (\ref{Veven}), we have: $V_{i}=(2i-1)/2N$. Therefore,
the massless spectrum will be in general:
\begin{itemize}
\item Vectors in: $\prod_{i}U(n_i)$.
\item Scalars $\y_{-1/2}^I|0\rangle$ in:
%
$(n_i,n_{1-i+f_{99} N})$, $(n_i,\overline{n}_{i-f_{99} N})$,
$(\overline{n}_i,\overline{n}_{1-i-f_{99} N})$.

\end{itemize}
Notice that there will be antisymmetric reps iff $2i-1=f_{99} N$.
Similarly for the fermions.

According to (\ref{Shifts}), 55 states states form similar reps.
The $95_i$ states have:
\begin{itemize}
%
%
\item Scalars $|s_j,s_k\rangle$ in: $(n_i,\tilde{n}_{1-i+f_{95}
N})$, $(n_i,\overline{\tilde{n}}_{i-f_{95} N})$,
$(\overline{n}_i,\tilde{n}_{i+f_{95} N})$,
$(\overline{n}_i,\overline{\tilde{n}}_{1-i-f_{95} N})$.
\item Fermions $|s_0,s_i\rangle$ in: $(n_i,\tilde{n}_{1-i+s_iv_i
N})$, $(n_i,\overline{\tilde{n}}_{i-s_iv_i N})$,
$(\overline{n}_i,\tilde{n}_{i+s_iv_i N})$,
$(\overline{n}_i,\overline{\tilde{n}}_{1-i-s_iv_i N})$.
\end{itemize}

\subsubsection*{Odd elements}

Consider a shift vector of an odd element where $\g^N=1$. By the
definition (\ref{Vodd}), we have: $V_{i}=(i-1)/N$. Therefore, the
massless spectrum will be in general:
\begin{itemize}
\item Vectors in: $A\times \prod^{(N+1)/2}_{i\neq 1}U(n_i)$. Where
$A=\{SO(n_1),USp(n_1)\}$ depending on the existence of commuting
or non-commuting $Z_2$ elements.
\item Scalars $\y_{-1/2}^I|0\rangle$ in: $(n_i,n_{2-i+f_{99} N})$,
$(n_i,\overline{n}_{i-f_{99} N})$,
$(\overline{n}_i,\overline{n}_{2-i-f_{99} N})$.
\end{itemize}
Representations of $SO(n_1)$ or $USp(n_1)$ appear as
$n_1+\bar{n}_1$ that represent the vector $n_{1,v}$.
There will be antisymmetric reps in the $U(n_i)$ iff $2i-2=v_I N$.
Similarly for the fermions.

In case there are D5-branes, 55 states form similar reps to the
above. The $95_i$ states have:
\begin{itemize}
%
%
\item Scalars $|s_j,s_k\rangle$ in: $(n_i,\tilde{n}_{2-i+f_{95}
N})$, $(n_i,\overline{\tilde{n}}_{i-f_{95} N})$,
$(\overline{n}_i,\tilde{n}_{i+f_{95} N})$,
$(\overline{n}_i,\overline{\tilde{n}}_{2-i-f_{95} N})$.
\item Fermions $|s_0,s_i\rangle$ in: $(n_i,\tilde{n}_{2-i+s_iv_i
N})$, $(n_i,\overline{\tilde{n}}_{i-s_iv_i N})$,
$(\overline{n}_i,\tilde{n}_{-i+s_iv_i N})$,
$(\overline{n}_i,\overline{\tilde{n}}_{2-i-s_iv_i N})$.
\end{itemize}

\subsubsection*{Scherk-Schwarz deformation}

Scherk-Schwarz deformation commutes with each rotation element.
Therefore, we can represent the $\g_h$s with a shift vector
(\ref{Shifts-h}). In general, each component of the rotating shift
vector can have different components of the SS deforming vector.
For example, consider $V_h$ where $\g_h^2=-1$. Components $V_i$
will split $V_i\to V_i^1+V_i^2$ with $n_i=n_i^1+n_i^2$. The
components of the SS deformation will be: $V^h_i\to
V_i^{h,1}+V_i^{h,2}$ where $V_i^{h,1}=-V_i^{h,2}=1$. Following the
same spirit, we realize that the action of the SS deformation
breaks the representations. We can summarize by considering a
representation:
\bea(m,n)_+ \to\left\{\ba{lcl} (m_1,n_1)+(m_2,n_2) && \textrm{bosons,}\\
(m_1,n_2)+(m_2,n_1) && \textrm{fermions,}\ea \right.\\
%
(n,m)_- \to\left\{\ba{lcl} (m_1,n_2)+(m_2,n_1) && \textrm{bosons,}\\
                            (m_1,n_1)+(m_2,n_2) && \textrm{fermions,}
                            \ea \right.\eea
where the index $+,-$ denote the $\g_h^2=\pm 1$ and $m,n$ are both
in fundamental or antifundamental reps. The bifundamental reps
split for both $\g_h^2=\pm 1$, as follows:
\bea(m,\overline{n})_\pm \to\left\{\ba{lcl}
(m_1,\overline{n}_1)+(m_2,\overline{n}_2)
&& \textrm{bosons,}\\
(m_1,\overline{n}_2)+(m_2,\overline{n}_1) && \textrm{fermions,}\ea
\right.\eea

Therefore, the effect of the SS deformation on the open strings in
a given supersymmetric model is to break the gauge group for
$\g^{2}_h=-1$ as
\be U(N)\rightarrow U(n)\times U(N-n)~,\hskip 1cm
SO(2N)\rightarrow U(N)~,\ee
whereas for $\g^{2}_h=+1$ as
\be U(N)\rightarrow U(n)\times U(N-n)~,\hskip 1cm SO(N)\rightarrow
SO(n)\times SO(N-n)~. \ee

\subsubsection{Some specific examples}


\subsubsection*{Supersymmetric $T^2\times K3$}

The first example of groups are supersymmetric models with $G=Z_N$
for $N=2,3,4,6$ acting on $T^4$ \cite{Gimon:1996rq}. The tadpole
conditions are given by (\ref{tvkh1}-\ref{tvkh2i}) with $v_\a^1=
-v_\a^2= k/N,~v_\a^3=0$ leading for odd $N$:
\bea Tr[\g_{2k,9}]= 32 \cos^2\frac{\p k}{N} ~,\nn \eea
whereas for even $N$:
\bea &&Tr[\g_{2k,9}]-4~\sin^2\frac{2k\p}{N}~Tr[\g_{2k,5_{3}}]= 32~
\cos\frac{2k\p}{N}~,
\nn\\
&&Tr[\g_{2k-1,9}]-4~\sin^2\frac{(2k-1)\p}{N}~Tr[\g_{2k-1,5_{3}}]=
0~. \nn \eea
Solving these equations, we find $\g$s and by the
(\ref{ActionOnCP}) we find the gauge group and massless spectrum
of these models which are provided in the appendix
\ref{MasslessSpectrums}.

\subsubsection*{Non-supersymmetric $T^2\times K3$}

Next, consider an orientifold of the type $G=Z_N\times
Z^\prime_2$. The extra $Z_2'$ is a freely acting SS deformation
$h$ which acts in a transverse circle of $T^4/Z_N$ and breaks
supersymmetry spontaneously.

Upon projecting this orbifold by the world sheet parity $\Omega$,
the massless limit of the tree channel Klein Bottle amplitude has
non-vanishing RR tadpoles and thus reveals the presence of
orientifold planes in the background. Besides the $O9$-plane that
extends in the non-compact directions, wraps the $T^2\times T^4$
and it is present for any $N$, for even $N$ the model contains
also $O5$-planes that extend along the non-compact directions,
wrap around the $T^2$ and sit at the $\a^k$-fixed points of the
transverse $T^4$. In order to cancel the associated to the
orientifold planes massless tadpoles one has to introduce D9 and
D5-branes. The contribution of the D-branes to the tadpoles is
encoded in the massless limit of the transverse channel Annulus
and M\"obius strip amplitudes.

The matrices $\g_{1,9}$ and $\g_{1,5}$ that correspond to the
identity element of $Z_N\times Z'_2$ can be chosen to be the
$32\times 32$ identity matrices, so that
$Tr[\g_{1,9}]=Tr[\g_{1,5}]=32$. This is a constraint on the number
of D-branes that originates from tadpole cancellation in the
untwisted sector. The twisted tadpole conditions on the other hand
in the $\a^k$ twisted sector, for $N$ even are given by
\cite{Gimon:1996rq}
\bea &&Tr[\g_{\a^{2k-1},9}]-4\sin^2 {(2k-1)\pi \over N}~
Tr[\g_{\a^{2k-1},5}]=0~,
\label{tad1}\\
&&Tr[\g_{\a^{2k},9}]-4\sin^2 {2\pi k\over N}~
Tr[\g_{\a^{2k},5}]-32\cos \frac{2\pi k}{N} =0~, \label{tad2}\eea
whereas for $N$ odd they read
\bea Tr[\g_{\a^{2k},9}]-32\cos^2 \frac{\pi k}{N}=0~.
\label{tad3}\eea
From the $\a^k h$ and $h$ twisted sectors we do not get further
constraints on $Tr[\g_{\a^{k} h,9}]$, $Tr[\g_{\a^{k}h,5}]$,
$Tr[\g_{h,9}]$ and $Tr[\g_{h,5}]$. Notice that for N even, the
tadpole conditions are consistent with T-duality transformations
along the $T^4$ torus that exchanges the $D9$ and $D5$-branes. On
the other hand, for the circle along which the shift is performed,
we have a freedom in taking $\g^{2}_{h,9}=\pm 1$ and also
$\g^{2}_{h,5}=\pm 1$, however T-duality constrains them to have
the same sign. In summary, we will obtain two open string spectra
for each $N$, related by Wilson lines.

Let us describe the massless spectrum starting from the closed
string sector. The closed string spectra of the supersymmetric
$T^4/Z_N$ orientifolds have been computed in \cite{Gimon:1996rq}.
Sectors twisted by $h$ do not contribute to the massless part of
the Torus and the Klein-Bottle since they correspond to half
integer winding \cite{Angelantonj:2002ct}.
Every other massless sector in the Torus is the same as in the
corresponding supersymmetric model \footnote{By corresponding
supersymmetric model we simply mean the model obtained by
eliminating the SS part, which is supersymmetric for all values of
$N$ discussed here.} plus an identical sector where the sign of
the fermions is reversed.
This simply means that $h$ projects out the fermions altogether
from the closed string sector. The bosons remain multiplied by a
factor of two which is cancelled by the 1/2 of the $h$-projector
$(1+h)/2$ in the trace. The Klein-Bottle on the other hand remains
the same as in the corresponding supersymmetric model. The extra
1/2 from the $h$-projector is now cancelled by a factor of two
coming from the doubling of the surviving the $\Omega$ projection
states, since any sector and its projected by $h$ counterpart give
the same contribution to the Klein-Bottle.
The closed string spectrum therefore for any $N$ is just the
bosonic part of the corresponding supersymmetric model
compactified on a $T^2$ torus.
The full open string spectrum will be presented in the appendix
(\ref{MasslessSpectrums}) for each value of $N$ considered here.
It is easy to check that the spectrum do not suffer from
irreducible gauge anomalies. This is due to the fact that all
fermions are in vector like representations. Alternatively, the
models we have considered are effectively five dimensional and
therefore do not have anomalies.

\subsection{Chapter Summary}

In this chapter we give an introduction to the orientifolds and we
explore the breaking of supersymmetry by the Scherk-Schwarz
deformation. We give general formulae for the tadpole conditions
and we provide the general form of the massless spectrum.

\newpage

\section{D-brane realization of the Standard Model and anomalies}

One of the important motivations in favor of string theory is the
fact that it seems to include in principle all the ingredients
required to embed the Standard Model (SM) inside a full unified
theory with gravity. A standard approach that tries to embed the
SM into string theory is the so called $top$-$down$ $approach$.
One starts by a string theory and tries to reduce the number of
dimensions, supersymmetries and the gauge group by an appropriate
orientifold compactification leading to a massless spectrum as
similar as possible to the SM.

Lately, the string theories that are analyzed are open string
theories (orientifolds) where the SM gauge group can be obtained
from the D-branes.
A low string scale compatible with the known value of the Planck
scale can be easily accommodated in ground states of unoriented
open and closed strings. Solvable vacua of this type are
orientifolds of closed strings. Such vacua include various type of
D-branes stretching their worldvolumes in the four non-compact
dimensions while wrapping additional worldvolume dimensions around
cycles of the compact six torus. Moreover, they include
non-dynamical orientifold planes that cancel the charges of the
D-branes, implementing the (un)orientability condition and
stabilizing the vacuum (cancellation of tadpoles).


Since masses of open strings are proportional to their lengths, it
is obvious that the branes that give rise to the SM fields must be
very close together in the internal space. Thus, we can talk about
the local group of SM D-branes and we may focus our discussion on
this. The presence of other branes further away may affect global
rather than local properties of the model (but can be important
for the overall stability of the configuration).

As we mention before, the standard relation between the string
scale and the Planck scale, namely $M_P^2={V_6\over g_s^2}M_s^2$
implies that the internal volume must be very large in string
units. The hierarchy problem in this context is the question of
what stabilizes the value of $V_6\gg 1$. No compelling answer
exists to this question so we will bypass it and move on.
However, the possibility of low string scale $M_s$ \cite{low} in
these theories and supersymmetry breaking at that scale without
suffering directly from the ordinary hierarchy problem of the
scalar masses makes these theories particularly interesting. If
the string scale is around a few TeV, observation of novel effects
at the near future experiments becomes a realistic possibility.



The minimal D-brane configuration that can successfully
accommodate the SM gauge group consists of three sets of
branes\footnote{As we mentioned above, $N$ coincident D-branes
typically generate a Unitary group $U(N)$.} with gauge symmetry
$U(3)\times U(2)\times U(1)$\footnote{Bottom to top model building
shows that we have to introduce another single D-brane which
provides an extra $U(1)'$ gauge boson \cite{Antoniadis:2000en}.
However, we can omit this extra brane for the rest of our studies
since it does not participate to the hypercharge}.
%
%
The SM particles are considered as open string states attached on
different stacks of D-branes. Therefore, in these models the SM
fields are open strings that are stretched onto a stack of 3, a
stack of 2 and one brane (at least):
\bea U(3)\times U(2)\times U(1)\to SU(3)\times SU(2)\times U(1)_3
\times U(1)_2 \times U(1) ~.\eea
Notice that every stack of branes supplies the model with extra
abelian gauge fields.
Such $U(1)$ fields have generically four-dimensional anomalies. In
the rest of this chapter we will discuss about anomalies and the
Green-Schwarz mechanism that cancels them.

\subsection{Anomalies}

Anomalies are generated when classical symmetries are broken at
the quantum level \cite{Weinberg:mt, Peskin:ev, Kaku:ym}.
There are $Global$ and $Local$ (Gauge) anomalies. Global anomalies
contribute finitely to physical processes. As an example, the
decay rate $\p^0\to \g\g$ that receive contribution from the
anomalies providing the correct experimental number for three
colored quarks.

Gauge anomalies afflict symmetries necessary to normalize the
theory and they must be avoided. The longitudinal polarization of
a gauge field related to them does not decouple. The axial
Ward-identities contain an anomaly (axial current is not
conserved) leading to inconsistences.
Anomalies arise in Parity violating (chiral) theories. This means
that left and right handed fermions do not transform in the same
way under the gauge symmetry.
%
%
%
%
%
%
%

Consider an effective action of Dirac fermions coupled to gauge
fields $\G(A_\m, \y)$, that
\bea e^{i\G[A_\m]} \sim \int [D\bar{\y}][D\y] ~ \exp\left\{ -\int
d^dx \bar{\y} \left({1+\g^{d+1} \over 2} \right) \sla{D} \y
\right\} ~.\label{AnomalousAction}\eea
In general, we can evaluate the Anomalies by:
\begin{itemize}
\item Functional integration (which is the so-called ``Fujikawa's
method"):

In this case, the anomalies appear as a phase factor due to the
variation of the fermion measure $[D\bar{\y}][D\y]$. Therefore,
the variation of the lagrangian of (\ref{AnomalousAction}) does
not vanish:
\bea \d{\cal L} ={{\cal A}_{a\ldots b\ldots}\over 32\p^2}
\e_{\m\n\ldots \r\s\ldots} F^{\m\n}_a \ldots F^{\r\s}_b \ldots
~~,\label{Anomaly}\eea
where ${\cal A}_{a\ldots b\ldots}$ the anomaly.

\item Directly from the Feynman-diagrams:

In gauge theories, the longitudinal components of the associated
external gauge field in physical processes should decouple to
ensure unitarity. Therefore, one can take a diagram with on-shell
external gauge fields and check whether the matrix elements with
one polarization vector longitudinal and the rest transverse and
physical vanishes or not.

Anomalies arise when some of these diagrams do not vanish. It has
been shown that anomalies arise from parity-violation amplitudes
since they contain an $\e^{\m \ldots \r}$ tensor, which is coming
from the trace of chiral fermions in the loops.


%
%
%
\vspace{.7cm}
\begin{center}
\begin{tabular}{c c}
\unitlength=0.8mm
\begin{fmffile}{Anomalous_M_bosons_12}
\begin{fmfgraph*}(40,40)
\fmfpen{thick} \fmfleft{i1} \fmfright{o1} \fmftop{v1,v2,v7,v8,v9}
\fmfbottom{v4,v5,v10,v11,v12}
\fmf{fermion,label=$p_1$,l.side=left}{j1,j2}
\fmf{fermion,label=$p_2$,l.side=left}{j2,j3}
\fmf{fermion,label=$p_3$,l.side=left}{j3,j4}
\fmf{dots,label=$\ldots$,l.side=left}{j4,j5}
\fmf{dots,label=$\ldots$,l.side=left}{j5,j6}
\fmf{fermion,label=$p_M$,l.side=left}{j6,j1}
\fmf{photon}{i1,j1} \fmf{photon}{v2,j2} \fmf{photon}{v8,j3}
\fmf{photon}{o1,j4} \fmf{photon}{v11,j5} \fmf{photon}{v5,j6}
\fmffreeze
\fmflabel{$k^1$}{i1} \fmflabel{$k^2$}{v2} \fmflabel{$k^3$}{v8}
\fmflabel{$k^4$}{o1} \fmflabel{$\ldots$}{v11} \fmflabel{$k^M$}{v5}
\end{fmfgraph*}
\end{fmffile}
&~~ \raisebox{5.8ex}[0cm][0cm]{$ \sim {\cal A} \int d^dx ~ Tr
\left[ {i \over \sla{p}_1} \sla{\z}_1 {i \over \sla{p}_2}
\sla{\z}_2 \ldots {i \over \sla{p}_M} \sla{\z}_M \left({1+\g^{d+1}
\over 2} \right) \right]~,$}
\end{tabular}
\end{center}
\vspace{.5cm}
where
%
$p_i$ the momenta of the internal fermionic propagators,
$\z(k_i)$ the polarization vectors of each external gauge boson.
The parity matrix is projecting out all the right-fermions. ${\cal
A}=Tr[t^\a_1 t^\a_2 \ldots t^\a_M]$ is the group theory factor,
where $t^\a_i$ the generators of the gauge group in the
representation of the internal fermions. The emitted bosons are
physical and on-shell ($k \cdot \z=k \cdot k =0$).
An $\e^{\m \ldots \r}$ is arising from the trace with $\g^{d+1}$.
The number of the external bosons in the anomalous amplitudes is
$1+d/2$.

This diagram is divergent and has to be regulated. Pauli-Villars
method for example supply with masses $m$ the internal fermions
and at the end of the computation we take the limit $m\to \infty$.
Careful evaluation shows that taking one of the polarization
vectors longitudinal, the matrix element does not vanish. The form
of the anomaly is proportional to (\ref{Anomaly}).


\end{itemize}
In $4D$, the anomalous diagram is a triangle with three external
bosons.
In a theory of gauge group $U(N)$, the group theory factor implies
that the possible anomalous diagrams can be:
\bea SU(N)^3~,~~~ U(1)\times SU(N)^2~,~~~U(1)^3~ . \eea
In general, there can also be gravitational anomalies. However, we
will not discuss them in the present study. The two last diagrams
introduce the concept of the $anomalous$ $U(1)$s. The Feynman
diagrams which contribute to the $U(1)$ anomalies are:
\vspace{0.5cm}
\bea
\unitlength=0.4mm
\begin{fmffile}{QQQ_4d_2}
\begin{fmfgraph*}(40,30)
\fmfpen{thick} \fmfleft{i1} \fmfright{o1,o2} \fmftop{v1}
\fmfbottom{v2} \fmf{plain}{i1,v1,v2,i1} \fmf{photon}{v1,o2}
\fmf{photon}{v2,o1} \fmffreeze \fmfv{decor.shape=circle,
decor.filled=empty, decor.size=.20w}{i1} \fmffreeze \fmfdraw
\fmfv{d.sh=cross,d.size=.20w}{i1} \fmffreeze
\fmflabel{$U(1)_i~$}{i1} \fmflabel{$U(1)_k$}{o1}
\fmflabel{$U(1)_j$}{o2}
\end{fmfgraph*}
\end{fmffile} ~~~~~~~~~~~~~~~~
\unitlength=0.4mm
\begin{fmffile}{QGG_4d_2}
\begin{fmfgraph*}(40,30)
\fmfpen{thick} \fmfleft{i1} \fmfright{o1,o2} \fmftop{v1}
\fmfbottom{v2} \fmf{plain}{i1,v1,v2,i1} \fmf{gluon}{v1,o2}
\fmf{gluon}{v2,o1} \fmffreeze \fmfv{decor.shape=circle,
decor.filled=empty, decor.size=.20w}{i1} \fmffreeze \fmfdraw
\fmfv{d.sh=cross,d.size=.20w}{i1} \fmffreeze
\fmflabel{$U(1)_i~$}{i1} \fmflabel{$G^\a$}{o1}
\fmflabel{$G^\a$}{o2}
\end{fmfgraph*}
\end{fmffile} ~~~~~~~~~~~~~~~~
\unitlength=0.4mm
\begin{fmffile}{QRR_4d_2}
\begin{fmfgraph*}(40,30)
\fmfpen{thick} \fmfleft{i1} \fmfright{o1,o2} \fmftop{v1}
\fmfbottom{v2} \fmf{plain}{i1,v1,v2,i1} \fmf{zigzag}{v1,o2}
\fmf{zigzag}{v2,o1} \fmffreeze \fmfv{decor.shape=circle,
decor.filled=empty, decor.size=.20w}{i1} \fmffreeze \fmfdraw
\fmfv{d.sh=cross,d.size=.20w}{i1} \fmffreeze
\fmflabel{$U(1)_i~$}{i1} \fmflabel{$g_{\m\n}$}{o1}
\fmflabel{$g_{\m\n}$}{o2}

\end{fmfgraph*}
\end{fmffile}
\eea
%
%
Consider for simplicity only one anomalous $U(1)$.
In terms of a gauge transformation $A_\m^{(0)}\to
A_\m^{(0)}+\partial_\m \e$ of the effective action, the anomalies
are:
\bea \d_{\e} {S} =\int d^4x \bigg\{\e~\left( A_1~F\wedge F +
A_2~Tr[G_a\wedge G_a] + A_3~R\wedge R \right) \bigg\} ~,\eea
where $A_1=Tr[Q^3]$, $A_2=Tr[Q T^a T^a]$ and $A_3=Tr[Q]$ the group
theory factors. We suppress the indexes for simplicity.
We will concentrate our study in the mixed (second) anomalous
diagram and we will describe the Green-Schwarz mechanism that
cancels the anomaly \cite{Green:sg, Sagnotti:1992qw,
Ibanez:1998qp}. Generalization of this mechanism is strait forward
for the rest of the anomalies.

\subsection{Green-Schwarz mechanism}

In this section we will explore the Green-Schwarz mechanism in
$4D$.
The fields that contribute to the anomaly cancellation are
antisymmetric tensors $B^k_{\m\n}$ and they are coming from the
$k$th twisted closed string spectrum (they are RR fields).
We will consider one anomalous $U(1)$ and one antisymmetric
$B_{\m\n}$. The generalization is straightforward.
The lagrangian in terms of the RR 2-form is
\bea {\cal L}_B&=& -{1\over 4 g^2_0}F^{(0)}
F^{(0)}-{1\over 4 g^2_a}Tr[F^a F^a]\nn\\
&&+ {1\over 2}c_1 \tilde{H} \tilde{H} + c_3 \tilde{F}^{(0)}_{\m\n}
B^{\m\n} + 2 c_3 c_2 \tilde{\Omega}^{(\a)\m} A^{(0)}_{\m}~,
\label{LagrangianB}\eea
where $F^{(0)}$, $F^{a}$ the field strengths of the anomalous
$U(1)$ ($A^{(0)}$) and the non-abelian $SU(N)$ ($G^{\m}_a$) gauge
fields. The field strength of the RR field $H_{\m\n\r}$ is
modified by a Chern-Simons term $\Omega^{(\a)}=Tr\left[\g_k \left(
G^{\a}d G^{\a}-{2i \over 3} G^{\a}\wedge G^{\a}\wedge G^{\a}
\right)\right]$:
\bea H_{\m\n\r}&=&\partial_{[\m} B_{\n\r]}+c\sum_\a
\Omega^{(\a)}_{\m\n\r} ~, \eea
Notice the twist $\g_k$ matrix that represents the action of the
orbifold group $\a_k$. All $c_i$s are constants. $c$ is of order
of $[mass]^2$.
The third term in (\ref{LagrangianB}) is provided by the way that
the RR-forms couple to gauge field strength \cite{Douglas:1995bn,
Morales:1998ux, Klein:1999im}, :
\bea Tr[\g e^{iF}]\wedge C ~~~\to ~~~ c_3 Tr[\g_k \l]
\tilde{F}^{(0)} \wedge B ~, \eea
where $C$ is a sum over RR forms of various degrees (terms of the
correct degree of total form are kept).

It is more convenient to express the lagrangian
(\ref{LagrangianB}) using the Poincar\'e dual of $B_{\m\n}$ scalar
field $\a$ (axion):
\bea {\cal L}_{\a}&=& -{1\over 4 g^2_0}F^{(0)}
F^{(0)}-{1\over 4 g^2_a}Tr[F^a F^a] \nn\\
&&- {1\over 2} \left(d\a-2 c_1 A^{(0)}\right)^2 - {1\over 2} c_2
\a Tr[F^a \tilde{F}^a] ~.\label{LagrangianAxion}\eea
Notice that the third term in the lagrangian is not invariant
under a $U(1)$ gauge transformation unless the axion $\a$ also
transforms like:
\bea A^{(0)}_{\m} \to A^{(0)}_{\m} +\partial_\m \e~~,~~~~~~~~~\a
\to \a+2 c_1 \e~.\eea
However, this transformation of the axion generates a
non-invariance coming from the fourth term in
(\ref{LagrangianAxion}). This term will annihilate the anomalous
term that is generated by the fermionic transformation, giving an
anomaly free gauge theory.
The total variation of the lagrangian under the above gauge
transformation is:
\bea \d_{\e}{\cal L}_{total}=- \left( c_1 c_2 -{{\cal A} \over 32
\p^2} \right)~\e ~ Tr[F^a \tilde{F}^a] ~,\eea
where the first term is coming from the variation of ${\cal
L}_{\a}$ and the second are the mixed anomalies from the variation
of the measure of the chiral fermions.
The anomaly is cancelled for: ${\cal A}=32 \p^2 c_1 c_2$.

The NSNS-twisted moduli $m$ (SUSY partner of the $\a$ that they
form together a complex scalar field $\f=m+i a$) couple to the
vector fields generating Fayet-Iliopoulos D-terms:
\bea {S}_{FI} = \int d^4x {1\over g_0^2}\bigg( m + \sum_i
q_i|\F_i|^2 \bigg)^2~. \eea
where $\F_i$ denote various open strings with charge $q_i$ under
the anomalous $U(1)$s. More details are provided in appendix
(\ref{LagrangianBySUSY}).
%
%

On the fixed points we have: $\langle m \rangle= 0$. The global
$U(1)_0$ remains unbroken despite the fact that the gauge boson
became massive \cite{Poppitz:1998dj}.
Away from the fixed points we have: $\langle m \rangle \neq 0$.
Restoration of SUSY (that is more economical state for the system)
implies that the charged scalars will acquire a non-vanishing VEV.
This breaks the global $U(1)_0$ symmetry.

\subsection{Calculation of the bare mass of the anomalous $U(1)$s}

In this section we will evaluate the contribution to the anomalous
$U(1)$ mass for supersymmetric orientifolds.

Closer look to (\ref{LagrangianAxion}) shows that these terms are
coming from different orders in string perturbation theory. The
$(\partial \a^i)^2$ is a tree-level (sphere) term, the
$A^i\partial \a^i$ comes in the disk and the quadratic term in the
gauge fields is a one-loop contribution. To clarify this, we
mention that $g_i^2$ is proportional to $g_s=e^\f$ and every power
of the axion absorbs a dilaton factor $e^{-\f}$ because it is a RR
filed. The string perturbation series are weighted by $g_s^{-\c}$
where $\c=2-2h-c-b$ is the Euler character and $h$, $c$ and $b$
denote the handle, the cross-cups and the boundaries of a closed
orientable Riemann surface respectively.

The diagrams at one-loop that contribute to terms quadratic in the
gauge bosons (anomalous $U(1)$s) are the genus-one surfaces with
boundaries: the annulus and the M\"obius strip.
In the infrared (IR) region they diverge logarithmically and give
the logarithmic running of the couplings. In the ultraviolet (UV)
region the tadpoles of the annulus with both gauge bosons inserted
in the same boundary and the M\"obius strip vanish due to the
tadpole cancellation.
\begin{center}
\epsfig{file=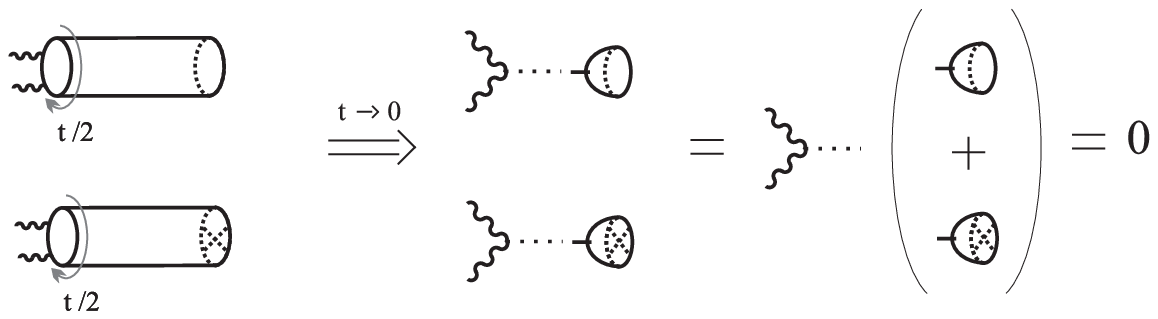,width=120mm}
\end{center}
However, in this UV limit the annulus amplitude with the gauge
bosons inserted in opposite boundaries provides the mass-term of
the anomalous $U(1)$ \cite{Antoniadis:2002cs}. Since we are
interested in the anomalous gauge boson mass, we concentrate on
the latter diagram.
The gauge boson vertex operator is
\be \tilde{V}^a=\l^a \e_\m (\partial X^\m +i(p\cdot \y)\y^\m)
e^{ip\cdot X} ~,\ee
where $\l$ is the Chan-Paton matrix and $\e^\m$ is the
polarization vector. The 2-point annulus amplitude is given by
\be {\cal A}^{ab}=-{1\over 4G}\int [d\t][dz] \int {d^dp \over
(2\p)^d}\sum_k \langle \tilde{V}^a(\e_1,p_1,z) \tilde{V}
^b(\e_2,p_2,z_0) \rangle_k ~,\label{Annulus11} \ee
where we keep undetermined the number of non-compact dimensions
$d$ and $G$ denotes the order of the orientifold group. The
fundamental polygon of the annulus is $[0,t/2]\otimes[0,1/2]$
(Fig.\ref{DoubleCover}). The index $k$ denotes the various
orbifold sectors that we may have. Using the translation symmetry
of the annulus, we fix the position of one VO to $z_0=1/2$. The
other VO is placed on the imaginary axis with $z\in [0,t/2]$.

The leading term of (\ref{Annulus11}) is
\be {\cal A}^{ab}=\int {d^dp \over (2\p)^d} \Big(\z_1\cdot
\z_2~p_1\cdot p_2-\z_1\cdot p_2~p_1\cdot \z_2\Big) \sum_k Tr[\g_k
\l^\a] Tr[\g_k \l^\b] {\cal A}^{ab}_k~, \label{Annulus11-2} \ee
where
\be {\cal A}^{ab}_k=-{1\over 2G}\int [d\t][dz] e^{-p_1\cdot p_2
\langle X(z) X(1/2)\rangle} \Big( \langle \y(z)\y(1/2)\rangle^2-
\langle X(z)\partial X(1/2)\rangle^2 \Big)Z^{ab}_k~,
\label{Annulus11ab} \ee
since the $p$-independent terms vanish due to supersymmetry. The
bosonic and fermionic correlation functions are given in the
appendix (\ref{PartialBosonicAC}, \ref{FIdentityTapp}).

It appears that the amplitude (\ref{Annulus11}) has a kinimatical
multiplicative factor that is ${\cal O}(p^2)$, thus would seem to
provide a leading  correction only to the anomalous gauge boson
coupling. We will see however, that after integration over the
position $z$ and the annulus modulus $\t_2$, a term proportional
to $1/ p_1\cdot p_2$ appears from the ultraviolet (UV) region (as
a result of the quadratic UV divergence in the presence of
anomalous $U(1)$s) that will provide the mass-term.

Strictly speaking, the amplitude above is zero on-shell if we
enforce the physical state conditions $\z\cdot p=p^2=0$ and
momentum conservation $p_1+p_2=0$. There is however a consistent
off-shell extension, without imposing momentum conservation, that
has given consistent results in other cases (see
\cite{Kiritsis:1997hj} for a discussion) and we adopt it here. We
will thus impose momentum conservation only at the end of the
calculation.

Spin structure summation of the partition function $Z^{ab}_k$,
gives zero due to space-time supersymmetry\footnote{Later we will
evaluate the bare mass of the anomalous $U(1)$s also for the
non-supersymmetric case.}. Therefore, terms in the correlation
functions which are spin-structure independent vanish. The only
spin-dependant term lies in the fermionic correlation function:
\be \langle \y(z-1/2)\y(0)\rangle^2[^\a_\b]_{annulus} = -2\p
i\partial_\t \log \vartheta[^\a_\b](0|\t)~. \label{y(z)y(o)} \ee
Equ. (\ref{y(z)y(o)}) is independent of $z$, the position of the
second VO. Thus, we can easily integrate on $dz$. Using the
modular transformations of the theta functions and keeping the
leading order of $\d$, we have:
\bea \int_0^{\t_2} dz ~e^{-\d \langle X(z)X(0)\rangle}=
\int_0^{\t_2} dz ~\t_2^{\d/2} {(2\p\h^3(\t))^\d \over
\vartheta[^0_1](z/\t|-1/\t)} =\t_2^{1+\d/2}[2\p \h^3(\t)]^\d+...
\eea
Following the procedure of \cite{Antoniadis:2002cs} we rewrite
(\ref{Annulus11ab}) as:
\be {\cal A}^{ab}_k=-{1\over 2G}\int [d\t] \t_2^{1+\d/2}[2\p
\h^3(\t)]^\d F^{ab}_k~. \label{Annulus11ab-1} \ee
defining $F^{ab}_k$ as a term which contains all the
spin-structure and the orbifold information:
\be F^{ab}_k= \sum_{\a\b} \h^{\a\b} \left[-2\p i \partial
\log\vartheta[^\a_\b]\right] \left[ {1\over (2 \p
\t)^3}{\vartheta[^\a_\b]\over \h^3} \right]
Z_{int,k}^{ab}[^\a_\b]~, \label{Fab-k}\ee
where $\h^{\a\b}={1\over 2}(-1)^{\a+\b+\a\b}$. The first bracket
is denoting the VO insertion in the annulus diagram. The second is
the six-dimensional partition function.

The integral on $t$ has a logarithmic divergence in $\d$ in the IR
and a pole in the UV.
\bea {\cal A}^{ab}_k={2C^{ab}_k\over \p \d |G|}+{\cal O}(\log
\d)~. \label{Cab}\eea
The on-shell limit
$[(\z_1\cdot \z_2)(p_1\cdot p_2)-(\z_1\cdot p_2)(p_1\cdot \z_2)]/
p_1 \cdot p_2 \to \z_1\cdot \z_2 = \z^2$
provides the un-normalized mass matrix.

\subsubsection{${\cal N}$=1 supersymmetric sectors}

In orientifold models, any element $\a\in G$ which acts onto all
tori ($v^i_\a \neq 0$ for any $i=1,2,3$) provides an ${\cal N}=1$
supersymmetric sector. In that case, it is straight-forward to
evaluate the various $C_k^{pq,UV}$ and the mass formulae:
\bea &&\left. {1\over 2}M^2_{pp,ab}\right|_{{\cal N}=1}
=\sum_{k\atop {{\cal N}=1~\rm sectors}} -{1\over \pi^3 |G|}
\prod_{i=1}^{3}|\sin[\pi kv_j]|~
Tr[\g_k\l_p^a]Tr[\g_k\l_p^b]~, \label{n199} \\
&&\left.{1\over 2}M^2_{95,ab}\right|_{{\cal N}=1}=\sum_{k\atop
{{\cal N}=1~\rm sectors}} {\sin(\pi kv_1)\over 2\pi^3|G|}
~{1\over 2}\prod_{j=1}^{3}{\sin[\pi kv_j] \over |\sin[\pi kv_j]|}
~ Tr[\g_k\l_9^a]Tr[\g_k\l_5^b]~, ~~~\label{n195} \eea
where $p=9,5$ and we have divided the 59 contribution by two, to
avoid overcounting.

\subsubsection{${\cal N}$=2 supersymmetric sectors}

${\cal N}=2$ supersymmetric sectors are present when a two-torus
remains invariant under the action of the appropriate orientifold
element.
Only massless states and their KK descendants survive the
(\ref{Fab-k}). In this case, the function $F^{ab}_k(t)=C^{ab}_k
P_2(t)$ where $C^{ab}_k$ is still given by (\ref{Cab}). $P_2(t)$
is either the appropriate momentum lattice when these directions
are NN (Neumann boundary conditions), or the winding lattice when
these directions are DD (Dirichlet boundary conditions)
\cite{Ibanez:1998qp}. No lattice sum can appear along ND
directions.
The open string momentum sum relevant in the NN case:
\bea P_2(t)=
\sum_{m\in \Zint} e^{-4\p \t_2{\a'\over 4}\left({m\over
R}\right)^2}=
{R\over \sqrt{\a' \t_2}} \sum_{w\in \Zint} e^{-{4\p
\over\t_2}{\a'\over 4}\left({wR\over \a'}\right)^2} \, ,
\label{lattice}\eea
while the open string (DD) winding sum is:
\bea W_2(t)=
\sum_{w\in \Zint} e^{-4\p \t_2{\a'\over 4}\left({wR\over
\a'}\right)^2} =
{1\over R} \sqrt{\a' \over \t_2} \sum_{m\in \Zint} e^{-{4\p
\over\t_2}{\a'\over 4}\left({m\over R}\right)^2}\,
.\label{lattice1}\eea
%

The pole contribution of (\ref{Annulus11ab-1}) has been evaluated
in \cite{Antoniadis:2002cs}:
\be {4{\cal V}_2~C^{ab,IR}_k\over \pi\delta}+{\cal
O}(\log\delta)\, . \ee
We now proceed to evaluate the contributions to the mass coming
from ${\cal N}=2$ sectors of abelian orientifolds. For such
sectors, one of the $kv_i$ is integer. We will choose without loss
of generality $k v_3=$ integer.
We compute:
\bea &&\left.{1\over 2}M^2_{ab,NN}\right|_{{\cal
N}=2}=\sum_{k\atop {{\cal N}=2~\rm sectors}} -{2{\cal V}_2\over
\pi^3|G|}\prod_{j=1}^2{|\sin[\pi kv_j]|}
Tr[\gamma_k\lambda^a]Tr[\gamma_k\lambda^b] ~,\\
&&\left.{1\over 2}M^2_{ab,DD}\right|_{{\cal N}=2}=\sum_{k\atop
{{\cal N}=2~\rm sectors}} -{1\over 2{\cal
V}_2\pi^3|G|}\prod_{j=1}^2{|\sin[\pi kv_j]|}
Tr[\gamma_k\lambda^a]Tr[\gamma_k\lambda^b]~ . \eea

Finally, for the 59 case, the relevant ${\cal N}=2$ sector is when
the longitudinal torus is untwisted. In this case, we evaluate:
\be \left.{1\over 2}M^2_{ab,DN}\right|_{{\cal N}=2}=\sum_{k\atop
{{\cal N}=2~\rm sectors}} (-1)^{kv_1}{{\cal V}_2\over
2\pi^3|G|}Tr[\gamma_k\lambda^a]Tr[\gamma_k\lambda^b]~ . \ee
We have divided the 59 contribution by an additional factor of
two. In the case where the two-torus corresponds to DD boundary
conditions (in a D7-D3 configuration for instance), one should
replace ${\cal V}_2\to 1/4{\cal V}_2$.

We should mention, that the above masses are unormalized. To
obtain the normalized mass matrix, we must also take into account
the kinetic terms of the $U(1)$ gauge bosons which are
\be S_{\rm kinetic}=-{1\over 4g_s}\left[{\cal V}_1{\cal V}_2 {\cal
V}_3 \sum_i F_i^2 +{\cal V}_3\sum_j \tilde F_j^2\right]~
.\label{SkineticU1} \ee
where $i$ and $j$ denote the gauge groups that are coming from
different stacks of D9 and D5-branes. This implies $M^2_{99}\to
M^2_{99}/({\cal V}_1{\cal V}_2{\cal V}_3)$, $M^2_{55}\to
M^2_{55}/{\cal V}_3$ and $M^2_{95}\to M^2_{95}/({\cal
V}_3\sqrt{{\cal V}_1{\cal V}_2})$.

\subsection{Applications on ${\cal N}$=1 orientifolds}

We are going to apply our formulae on various orientifolds. First,
we will compute how many anomalous $U(1)$s appear in the various
orientifold models by evaluating the mixed-anomaly traces that
give the anomalous $U(1)$s. Our normalization of the square
casimir, cubic casimir and the $U(1)$ charge of the $SU(N)$
representations are given in the following table:
%
%
\begin{center}
\begin{tabular}{clll}
$SU(N)$ Reps & Square Casimir & Cubic Casimir & $U(1)$ Charge \\
\hline
$\fund $ & $T(\fund)=1 $ & $A(\fund)=~~1 $ & $Q(\fund)=~~1$ \\
$\antifund$ & $T(\antifund)=1 $ & $A(\antifund)=-1 $ & $Q(\antifund)=-1$ \\
$\Yasymm $ & $T(\Yasymm)=N-2 $ & $A(\Yasymm)=~~N-4 $ & $Q(\Yasymm)=~~2$ \\
$\bYasymm$ & $T(\bYasymm)=N-2 $ & $A(\bYasymm)=-N+4 $ &
$Q(\bYasymm)=-2$
\end{tabular}\end{center}

For the evaluation of the mass matrix of the anomalous $U(1)$s,
the normalized generators of the anomalous $U(1)_i$ are defined
as:
\be \l^\a_i = {1\over 2\sqrt{n_i}}\sum Q^\a_i \cdot H~,
\label{lambdaGen}\ee
where $\a$ denotes the type of brane. The $Q^\a_i=
(0,\ldots,0,1,\ldots,1,0,\ldots,0)$ is a 16-dimensional vector
with $n_i$ entries of 1s where the $SU(n_i-1)$ lives. We normalize
the $\l$ matrices with $Tr[\l^2]=1/2$. Thus, the relevant trace
is:
\be Tr[\g^\a_k \l^\a_i]=Tr[e^{-2\p i k V^\a \cdot H} Q^\a_i \cdot
H] =- {i\over \sqrt{n_i}} \sin[2\p k V^\a_i] ~,
\label{Trace[gl]}\ee
where $V^\a_i$ are the overlapping components of $V^\a$ and $Q^\a$
\cite{Ibanez:1998qp}.

\subsubsection{The four-dimensional $Z'_6$ orientifold}

The orbifold rotation vector is $(v_1,v_2,v_3)=(1,-3,2)/6$. Since
there is an order two twist ($k=3$), we have one set of D5-branes.
Tadpole cancellation implies the existence of 32 D9-branes and 32
D5-branes that we put together at one of the fixed points of the
$Z_2$ action (namely the origin). The Chan-Paton 'shift' vectors
are
\be V_{5,9}={1\over 12}(1,1,1,1,5,5,5,5,3,3,3,3,3,3,3,3)~.
\label{Z6'vectors}\ee
The gauge group has a factor of $U(4)\times U(4)\times U(8)$
coming from the D9-branes and an isomorphic factor coming from the
D5-branes. The massless spectrum is provided in
Table.\ref{SpectrumD=4}. Different sectors preserve different
supersymmetries. The ${\cal N}=1$ sectors correspond to $k=1,5$,
while for $k=2,3,4$ we have ${\cal N}=2$ sectors.

The four-dimensional anomalies of the $U(1)$s have been computed
in \cite{Ibanez:1998qp} and the anomaly matrix is
\be A_{QTT}[Z_6'] \sim \left( \ba {cccccc}
 2 &  2 &  4\sqrt{2} & -2 &  0 & -2\sqrt{2} \\
-2 & -2 & -4\sqrt{2} &  0 &  2 &  2\sqrt{2} \\
 0 &  0 &      0     &  2 & -2 &     0 \\
-2 &  0 & -2\sqrt{2} &  2 &  2 &  4\sqrt{2} \\
 0 &  2 &  2\sqrt{2} & -2 & -2 & -4\sqrt{2} \\
 2 & -2 &      0     &  0 &  0 &     0    \ea \right) ~,\ee
there are two linear combinations that are free of
four-dimensional anomalies: $\sqrt{2}(A_1+A_2)-A_3$ and
$\sqrt{2}(\tilde{A}_1+\tilde{A}_2)-\tilde{A}_3$.

The contribution to the mass matrix \cite{Antoniadis:2002cs} is
\bea {1\over 2}M^2_{aa,ij}&=&-{\sqrt{3}\over
24\pi^3}\left(Tr[\g_1\l^a_i]
Tr[\g_1\l^a_j]+Tr[\g_5\l^a_i]Tr[\g_5\l^a_j]\right) \nonumber\\
&-&{1\over 4\pi^3}\left( {\cal V}_2\d_{a,9}+{1\over 4{\cal
V}_2}\d_{a,5}\right)\left(Tr[\g_2\l^a_i] Tr[\g_2\l^a_j]
+Tr[\g_4\l^a_i]Tr[\g_4\l^a_j]\right)\nn\\
&-&{{\cal V}_3\over
3\pi^3}Tr[\g_3\l^a_i]Tr[\g_3\l^a_j]~,\label{Z'6.99.D4-to-D6}\eea
for $a=5,9$ where $\d_{a,b}$ is the Kronecker delta. The mixed 59
annulus diagrams give a contribution to the mass
\bea {1\over 2}M^2_{95,ij} =&-&{\sqrt{3}\over
48\pi^3}\left(Tr[\g_1\l^9_i] Tr[\g_1\l^5_j]+Tr[\g_5\l^9_i]
Tr[\g_5\l^5_j]\right. \nonumber\\
&+& \left. Tr[\g_2\l^9_i]Tr[\g_2\l^5_j]-
Tr[\g_4\l^9_i]Tr[\g_4\l^5_j]\right) \nonumber \\
&-& {{\cal V}_3\over 12\pi^3}Tr[\g_3\l^9_i]Tr[\g_3\l^5_j]
~.\label{Z'6.59.D4-to-D6}\eea

The unormalized mass matrix \cite{Antoniadis:2002cs} has
eigenvalues and eigenvectors:
\bea m^2_1=6{\cal V}_2 &~~,~~ & -A_1+A_2\, \label{MZ6'1}\\
m^2_2={3\over 2{\cal V}_2} & ~~,~~ & -\tilde A_1+\tilde A_2 \\
m^2_{3,4}={5\sqrt{3}+48{\cal V}_3\pm\sqrt{3}\a\over 12} & ~~,~~ &
{-3\pm \a\over 4\sqrt{2} (4\sqrt{3}{\cal V}_3-1)}
(A_1+A_2-\tilde A_1-\tilde A_2)-A_3+\tilde A_3;\nonumber\\
&&\\
m^2_{5,6}={15\sqrt{3}+80{\cal V}_3\pm\b \over 12} & ~~,~~ &
{9\sqrt{3}\mp\b\over 4\sqrt{2}(20{\cal
V}_3-3\sqrt{3})}(A_1+A_2+\tilde A_1+\tilde A_2)+A_3+\tilde A_3;
\nonumber\\
&& \label{Z6'masses}\eea
with $\a=\sqrt{25-128\sqrt{3}{\cal V}_3+768{\cal V}_3^2}$ and
$\b=\sqrt{5(135-384\sqrt{3}{\cal V}_3+1280{\cal V}_3^2)}$. Note
that the eigenvalues are invariant under the T-duality symmetry of
the theory ${\cal V}_2\to 1/4{\cal V}_2$. Thus, all $U(1)$s become
massive, including the two anomaly free combinations. This result
is unexpected since there is no obvious mechanism that provides a
mass to non-anomalous $U(1)$s.

\subsubsection{The four-dimensional $Z_6$ orientifold}

The orbifold rotation vector is $(v_1,v_2,v_3)=(1,1,-2)/6$. Since
there is an order two twist ($k=3$), we have one set of
$D5$-branes that are stretched in the 4D Minkowski and wrap the
third torus $T^2_3$. Tadpole cancellation implies the existence of
32 $D9$-branes and 32 $D5$-branes that we put together at one of
the fixed points of the $Z_2$ action (namely the origin). The
Chan-Paton 'shift' vectors are
\be V_{5,9}={1\over 12}(1,1,1,1,1,1,5,5,5,5,5,5,3,3,3,3)~.
\label{Z6vectors}\ee
The gauge group has a factor of $U(6)\times U(6)\times U(4)$
coming from the $D9$-branes and an isomorphic factor coming from
the $D5$-branes. The massless spectrum is provided in the appendix
Table.\ref{SpectrumD=4}. This orientifold has different
supersymmetries in different sectors. The ${\cal N}=1$ sectors
correspond to $k=1,2,4,5$, while for $k=3$ we have ${\cal N}=2$
sectors.

The four-dimensional anomalies of the $U(1)$s have been computed
in \cite{Ibanez:1998qp} and the anomaly matrix is
\be A_{QTT}[Z_6] \sim \left( \ba {cccccc}
 6 & -3 &  \sqrt{6} &  3 &  0 &  \sqrt{6} \\
 3 & -6 & -\sqrt{6} &  0 & -3 & -\sqrt{6} \\
-9 &  9 &      0    & -3 &  3 &     0     \\
 3 &  0 &  \sqrt{6} &  6 & -3 &  \sqrt{6} \\
 0 & -3 & -\sqrt{6} &  3 & -6 & -\sqrt{6} \\
-3 &  3 &      0    & -9 &  9 &     0     \ea \right) ~,\ee
there are three linear combinations that are free of anomalies:
$A_1+A_2-\sqrt{3\over 2}A_3$,
$\tilde{A}_1+\tilde{A}_2-\sqrt{3\over 2}\tilde{A}_3$ and
$A_3-\tilde{A}_3$.

The contributions to the mass matrix \cite{Antoniadis:2002cs} are:
\bea {1\over 2}M^2_{aa,ij}&=&-{\sqrt{3}\over 48\pi^3}
\bigg(Tr[\g_1\l^a_i] Tr[\g_1\l^a_j]
+Tr[\g_5\l^a_i]Tr[\g_5\l^a_j]\nonumber\\
&& +3(Tr[\g_2\l^a_i] Tr[\g_2\l^a_j] +Tr[\g_4\l^a_i]Tr[\g_4\l^a_j])
\bigg)\nonumber\\
&&- {{\cal V}_3\over 3\p^3} Tr[\g_3\l^a_i] Tr[\g_3\l^a_j]~,
\label{Z6.aa.D4-to-D6}\eea
for $a=5,9$, while
\bea {1\over 2}M^2_{59,ij}&=&-{\sqrt{3}\over 48\pi^3}
\bigg(Tr[\g_1\l^5_i] Tr[\g_1\l^9_j]
+Tr[\g_5\l^5_i]Tr[\g_5\l^9_j]\nonumber\\
&& +3(Tr[\g_2\l^5_i] Tr[\g_2\l^9_j] +Tr[\g_4\l^5_i]Tr[\g_4\l^9_j])
\bigg)\nonumber\\
&&- {{\cal V}_3\over 12\p^3} Tr[\g_3\l^5_i] Tr[\g_3\l^9_j]~.
\label{Z6.59.D4-to-D6}\eea
Notice that the ${\cal N}=2$ sector contributes with a term
proportional to ${\cal V}_3$.
The mass matrix of the anomalous $U(1)$s has the following
eigenvalues and eigenstates \cite{Antoniadis:2002cs}:
\bea m^2_1=0 &~~~,~~~ & A_1+A_2-\tilde A_1-\tilde
A_2+\sqrt{6}(A_3-\tilde A_3); \\
m^2_2={3\sqrt{3}\over 2} & ~~~,~~~ & A_1-A_2-\tilde A_1+\tilde
A_2; \\
m^2_3=3\sqrt{3} & ~~~,~~~ & A_1-A_2+\tilde A_1-\tilde
A_2; \\
m^2_4=8{\cal V}_3 & ~~~,~~~ & -\sqrt{3\over 2}( A_1+A_2-\tilde
A_1-\tilde A_2)-A_3+\tilde A_3; \\
m^2_{5,6}={7\sqrt{3}+80{\cal V}_3 \pm \b \over 12} & ~~~,~~~ &
{40{\cal V}_3-\sqrt{3}\pm \b \over 12\sqrt{2}-40\sqrt{6}{\cal
V}_3}(A_1+A_2+\tilde
A_1+\tilde A_2)+A_3+\tilde A_3; \nonumber\\
&& \label{Z6masses}\eea
where $\b=\sqrt{147 -1040\sqrt{3}{\cal V}_3+6400{\cal V}_3^2}$.
Again, two non-anomalous $U(1)$s acquire masses.
%



As we have seen in the two last examples of $Z'_6$ and $Z_6$
orientifold models, $U(1)$ gauge fields that are free of
four-dimensional anomalies can still be massive. This is
unexpected and we should study the contribution of higher
anomalies in the mass-generation of the $U(1)$s.
We will especially study the six-dimensional anomalies since we
cannot have eight-dimensional anomalies in supersymmetric
orientifold models (which obey the condition: $\sum_i v_i=0$). We
will show that if there are decompactification limits in the
theory, six-dimensional anomalies affect four-dimensional masses.


\subsection{The structure of six-dimensional mixed gauge
anomalies}\label{6DAnomalies}

In the previous section we computed the bare masses of the
anomalous $U(1)$s by evaluating the ultraviolet tadpole of the
one-loop open string diagram with the insertion of two gauge
bosons on different boundaries.
It turns out that $U(1)$ gauge fields that are free of
four-dimensional anomalies can still be massive. This is
unexpected and we should study the contribution of higher
anomalies in the mass-generation of the $U(1)$s.
We will especially study the six-dimensional anomalies since we
cannot have eight-dimensional anomalies in orientifold models that
obey the condition $\sum_i v_i=0$. We will show that if there are
decompactification limits in the theory, six-dimensional anomalies
affect four-dimensional masses.

In six dimensions, the leading diagram that can give a
contribution to anomalies is the square diagram
\cite{Sagnotti:1992qw}.
The mixed group theory factors that do not identically vanish are
these with two or three external non-abelian gauge bosons. The
Feynman diagrams that eventually contain anomalies are:
\vspace{.6cm}
\begin{center}
\unitlength=0.6mm
\begin{fmffile}{Q2G2_6D_1}
\begin{fmfgraph*}(40,25)
\fmfpen{thick} \fmfleft{i1} \fmfright{o1} \fmftop{v1,v2,v7,v8,v9}
\fmfbottom{v4,v5,v10,v11,v12} \fmf{plain}{i1,v2,v3,v5,i1}
\fmf{gluon}{v3,o1} \fmffreeze \fmfdraw
\fmf{photon}{v2,v8} \fmf{gluon}{v5,v11} \fmfv{decor.shape=circle,
decor.filled=empty, decor.size=.20w}{i1} \fmffreeze \fmfdraw
\fmfv{d.sh=cross,d.size=.20w}{i1} \fmffreeze
\fmflabel{$U(1)_i~$}{i1} \fmflabel{$U(1)_j$}{v8}
\fmflabel{$G^\a$}{v11} \fmflabel{$G^\a$}{o1}
\end{fmfgraph*}
\end{fmffile}
~~~~~~~~~~~~~~~~~~~~~~
\unitlength=0.6mm
\begin{fmffile}{QG3_6D_1}
\begin{fmfgraph*}(40,25)
\fmfpen{thick} \fmfleft{i1} \fmfright{o1}
\fmftop{v1,v2,v7,v8,v9} \fmfbottom{v4,v5,v10,v11,v12}
\fmf{plain}{i1,v2,v3,v5,i1} \fmf{gluon}{v3,o1} \fmffreeze \fmfdraw
\fmf{gluon}{v2,v8} \fmf{gluon}{v5,v11} \fmfv{decor.shape=circle,
decor.filled=empty, decor.size=.20w}{i1} \fmffreeze \fmfdraw
\fmfv{d.sh=cross,d.size=.20w}{i1} \fmffreeze
\fmflabel{$U(1)_i~$}{i1} \fmflabel{$G^\a$}{v8}
\fmflabel{$G^\g$}{v11} \fmflabel{$G^\b$}{o1}
\end{fmfgraph*}
\end{fmffile}
\end{center}
Therefore, in the presence of an anomalous $U(1)$ field, the
effective action is not invariant under a transformation $\d A^i =
d \e^i$:
\be \d_{\e^i} {S}|_{gauge} = \int d^6x \bigg\{\e^i~\left(A_{QQTT}
~ F^j\wedge Tr[G^2] + A_{QTTT} ~ Tr[G^3] \right)
\bigg\}~,\label{dS}\ee
where $A_{QQTT}=Tr[Q_i Q_j T^\a T^\a],~ A_{QTTT}=Tr[Q_i T^\a
\{T^\b T^\g\}]$ the group theory factors. Powers of forms are
understood as wedge products. We denote by $G_{\m\n}$ the field
strength of a non-abelian gauge field $W_\m$.

Gauge invariance is preserved by the six-dimensional Green-Schwarz
mechanism. However, two inequivalent fields should contribute to
this cancellation.
The cancellation of the first anomalous term is arranged by a
2-form $B^i$ (RR twisted field) which transform under the $U(1)$
transformation like $\d B^i=-\e^i F^i$. The lagrangian of this
field is:
\be S_{QQTT} =\int d^6x \left[-{1 \over 4 g_i^2} F^{i2}_{\m\n}
-{1\over 12} \left[d B^i + \W_{A^i} \right]^2 + A_{QQTT}~B^i
\wedge Tr[G^2] \right] ~,\label{S_QQTT}\ee
where the last term is proportional to the anomaly of the first
diagram. The 3-form $\W_{A^i}=A^idA^i$ is the Chern-Simons term of
the abelian gauge field $A^i_\m$. This part of the action does not
generate a mass for the gauge boson.

By the (\ref{S_QQTT}), we can evaluate the action in terms of the
dual 2-form $\l$ of $B$ \cite{Ghilencea:2002da}. Using $Tr[G_i
\tilde{G}_i]=d \W_{W_i}$, where $\W_{W_i}=Tr[W_i dW_i+{2\over
3}W_i^3]$ is the Chern-Simons term for the non-abelian gauge field
$W^i$, we finally find:
\be \tilde{S}_{QQTT} =\int d^6x \left[-{1 \over 4 g_i^2}
F^{i2}_{\m\n} - {1\over 12}\left[d \l^i - 6 A_{QQTT}\W_{W^i}
\right]^2 - {1\over 6}\W_{A^i}\wedge (d\l^i - 6A_{QQTT}\W_{W^i})
\right]~. \label{dual_S_QQTT}\ee
The $\l^i$ are invariant under $U(1)$ gauge transformations and
transform like $\d\l^i=6C_1~Tr[G \e^i]$ under a non-abelian gauge
transformation $\d W_i^\m=D^\m \e_i$ so that the action is gauge
invariant.
Thus, under a $U(1)$ gauge transformation the variation of
$\W_{A^i}\wedge d\l^i$ (since $\d\W_{A^i}=d\e F$) vanishes due to
integration by parts and the term $\W_{A^i} \wedge \W_{W^i}$
cancels the first anomaly in (\ref{dS}).

The second anomaly is cancelled by a pseudoscalar axion that
transforms under the $U(1)$ transformation as $\d\a^i=-\e^i$:
\be S_{QTTT} =\int d^6x \left[-{1 \over 4 g_i^2} F^{i2}_{\m\n} +
{M^2\over 2} (A^i +d \a^i)^2 + A_{QTTT} \a^i~ Tr[G^3]\right]~.
\label{S_QTTT}\ee
This action supplies a mass term for the $U(1)$ gauge field and
breaks the gauge symmetry in six dimensions.

\subsection{Six-dimensional mass formulae}

The general mass formulae for the anomalous $U(1)$ gauge fields in
six-dimensional orientifolds can be easily evaluated in the same
way that we did for the four-dimensional cases. {\cal N}=1
six-dimensional orientifolds are created as $T^4/Z_N$ where
$N=2,3,4,6$.
The results for strings attached on the same kind of branes
(untwisted states) are (\ref{Annulus11ab-UV-A})
\be {1\over 2}M^2_{aa}= -{4 \over \p^2 N} \sum_k \sin^2 {\p k\over
N} ~ Tr[\g_k \l^{a}] Tr[\g_k \l^{a}]~, \label{Mass_aa}\ee
where $a=5,9$ denotes the kind of D-branes on which the open
string is attached. In the case where strings have one end on a
$D5$ and the other on a $D9$-brane (twisted states) we have:
\be {1\over 2}M^2_{59}= -{1 \over \p^2 N} \sum_k Tr[\g_k \l^{5}]
Tr[\g_k \l^{9}]~. \label{Mass_59}\ee
We should mention, that the above masses are again unormalized. To
obtain the normalized mass matrix, we must also take into account
the kinetic terms of the $U(1)$ gauge bosons which are again
(\ref{SkineticU1}), however, the volume of the torus that the
D5-branes is longitudinal to, should be normalized to identity.
This implies $M^2_{99}\to M^2_{99}/({\cal V}_1{\cal V}_2)$,
$M^2_{55}\to M^2_{55}$ and $M^2_{95}\to M^2_{95}/(\sqrt{{\cal
V}_1{\cal V}_2})$.

\subsection{Six-dimensional ${\cal N}$=1 orientifolds examples}

Usual six-dimensional decompactification limits of
four-dimensional supersymmetric orientifolds are the ${\cal N}$=1
orientifolds of Type IIB string theory, $\Realint^6\times K3/Z_N$
where the only possible choices are $N=2,3,4,6$. Thus, we will
apply the above general formulae on these orientifolds.
Tadpole cancellation guaranties that the models are free of
irreducible non-Abelian anomalies \cite{Ibanez:1998qp,
Sagnotti:1992qw}.
%

\subsubsection{$Z_2$ orientifold}\label{Z26D}

For the $Z_2$, the tadpole condition gives 32 $D9$ and 32
$D5$-branes \cite{Pradisi:1988xd, Gimon:1996rq}. The
characteristic vectors are:
\be V_{5,9}={1\over 4}(1,1,1,1,1,1,1,1,1,1,1,1,1,1,1,1)~.
\label{V-Z2}\ee
The gauge group is $U(16)_9 \times U(16)_5$. The massless states
are given in Table.\ref{SpectrumD=6}. We are interested in
anomalous diagrams with one abelian and three non-abelian gauge
bosons $U(1)\times SU(N)^3$ since their cancellation provides the
six-dimensional mass-term. We find:
\be A_{QTTT} =
32 \cdot \left( \ba {cc}
 4 & -1 \\
-1 &  4 \ea \right)~, \ee
where the columns label the U(1)s, while the rows label the
non-abelian factors. The matrix has two non-zero eigenvalues and
both anomalous $U(1)$s are expected to become massive
\cite{Scrucca:1999pp}. The unormalized mass matrix for the
anomalous $U(1)$s is calculated by the use of (\ref{Mass_aa}),
(\ref{Mass_59}) and (\ref{Trace[gl]}):
\be {1\over 2}M^2=
-{1 \over 2\p^2}\left( \ba {cc}
4~Tr[\g_1 \l^{9}] Tr[\g_1 \l^{9}] & Tr[\g_1 \l^{9}] Tr[\g_1\l^{5}] \\
Tr[\g_1 \l^{5}] Tr[\g_1 \l^{9}] & 4~Tr[\g_1 \l^{5}] Tr[\g_1\l^{5}]
\\ \ea \right)
={8\over \p^2}\left( \ba {cc}
4 & 1\\
1 & 4\\\ea \right). \label{Z2massOfU(1)}\ee
As it was expected from the effective field theory computation of
the anomalies, there are two massive eigenstates: $\pm A
+\tilde{A}$ with masses $24/\p^2$, $40/\p^2$ (we denote with $A$
the gauge boson that is coming from the D9-branes and with
$\tilde{A}$ the one that is coming from the D5).

\subsubsection{$Z_3$ orientifold}\label{Z3}

The $Z_3$ orientifold does not contain a $Z_2$ reflection element.
Thus, there are no D5-branes. The characteristic vector is:
\be V_9={1\over 3}(1,1,1,1,1,1,1,1,0,0,0,0,0,0,0,0) ~,\ee
and the gauge group $U(8)\times SO(16)$. From the massless
spectrum which is provided in Table.\ref{SpectrumD=6} we find that
the single gauge boson suffers from mixed non-abelian anomalies
\cite{Scrucca:1999pp}.
\be A_{QTTT} = 48 . \ee
Using (\ref{Trace[gl]}) we find the mass of this gauge boson:
\be {1\over 2}M^2=~ {32 \over 3\p^2} \sum_{k=1}^2 \sin^2 {\p k
\over 3} \sin^2{2\p k \over 3} =~ {12 \over \p^2}~.
\label{Z_3massOfU(1)}\ee

\subsubsection{$Z_4$ orientifold}

The $Z_4$ orientifold contains 32 D9-branes and 32 D5-branes. The
characteristic vectors are:
\be V_{5,9}={1\over 8}(1,1,1,1,1,1,1,1,3,3,3,3,3,3,3,3) ~,\ee
and the gauge group is $U(8)_9\times U(8)_9\times U(8)_5\times
U(8)_5$. The massless spectrum is provided in
Table.\ref{SpectrumD=6}. The $U(1)\times SU(N)^3$ anomalies are:
\be A_{QTTT} =
16 \cdot\left( \ba {cccc}
 3 & -1 & -1 &  0   \\
-1 &  3 &  0 & -1   \\
-1 &  0 &  3 & -1   \\
 0 & -1 & -1 &  3   \ea \right)~. \ee
where again the columns label the U(1)s and the rows the
non-abelian factors $SU(8)_9^2\times SU(8)_5^2$. Notice that we
have two equal matrices in the diagonal blocks and two other ones
equal in the off-diagonal blocks. This is a consequence of the
fact that the D9 and D5 branes are related by T-duality and split
in isomorphic groups. All those models are T-selfdual
. The anomaly matrix has four non-zero eigenvalues
\cite{Scrucca:1999pp}.

The mass matrix of the anomalous U(1) masses is
\be {1\over 2}M^2={4\over \p^2}\left( \ba {cccc}
 3 & -1 &  1 &  0   \\
-1 &  3 &  0 &  1   \\
 1 &  0 &  3 & -1   \\
 0 &  1 & -1 &  3   \ea \right)~.
\ee
Diagonalizing this matrix, we find four massive $U(1)$ fields that
are in accordance with the anomalies. The massive $U(1)$ fields
are $-A_1-A_2+\tilde{A}_1+\tilde{A}_2$, $A_1+\tilde{A}_2$,
$A_2+\tilde{A}_1$, $-A_1+A_2-\tilde{A}_1+\tilde{A}_2$ with masses
$4/ \p^2$, $12/ \p^2$, $12/ \p^2$, $20/ \p^2$ respectively.

\subsubsection{$Z_6$ orientifold}

The $Z_6$ orientifold contains 32 D9-branes and 32 D5-branes. The
characteristic vectors are:
\be V_{5,9}={1\over 12}(1,1,1,1,5,5,5,5,3,3,3,3,3,3,3,3) \ee
and the gauge group $U(4)_9\times U(4)_9\times U(8)_9\times
U(4)_5\times U(4)_5\times U(8)_5$. The massless spectrum is
provided in Table.\ref{SpectrumD=6}. The $U(1)\times SU(N)^3$
anomalies are:
\be A_{QTTT} =
8 \cdot \left( \ba {cccccc}
 3 &  0 & -2 & -1 &  0 &  0 \\
 0 &  3 & -2 &  0 & -1 &  0 \\
-1 & -1 &  4 &  0 &  0 & -2 \\
-1 &  0 &  0 &  3 &  0 & -2 \\
 0 & -1 &  0 &  0 &  3 & -2 \\
 0 &  0 & -2 & -1 & -1 &  4 \ea \right)~.
\label{AnomalyQTTT_Z4}\ee
The columns are the $U(1)$s and the rows the non-abelian factors,
always in the ordered form of Table.\ref{SpectrumD=6}. The
(\ref{AnomalyQTTT_Z4}) has five non-zero and one zero eigenvalue
which corresponds to
$A_1+A_2+A_3+\tilde{A}_1+\tilde{A}_2+\tilde{A}_3$. Our result is
in accordance with \cite{Scrucca:1999pp} where it had been shown
that one of the six $U(1)$ factor remains unbroken. The
independent axions that participate in the cancellation of the
anomaly and the mass generation are only five.

The mass matrix for the anomalous $U(1)$s is
\be {1\over 2} M^2={2\over \p^2}\left( \ba {cccccc}
3 & 0&  -\sqrt{2} &  1 &  0&  0 \\
0 & 3&  -\sqrt{2} &  0 &  1&  0 \\
-\sqrt{2} & -\sqrt{2}& 4 & 0 & 0 & 2 \\
1 & 0&  0 & 3 & 0 & -\sqrt{2} \\
0 & 1&  0 & 0 & 3 & -\sqrt{2} \\
0 & 0&  2 & -\sqrt{2}& -\sqrt{2}& 4 \ea \right)~. \ee
Diagonalizing the mass matrix, we find that five $U(1)$ fields
become massive and one remains massless. The effective field
theory computation agrees with the result above.

\subsection{4D Anomalous $U(1)$s and the relation to 6D anomalies upon
            decompactifications}

\subsubsection{Decompactification of the $Z'_6$ orientifold}

The axions that cancel the anomalies, being twisted RR fields, are
localized on the fixed points of the internal dimensions. Since
there are various orbifold sectors $k$, there are also various
axions $\a^i_k$ localized on the fixed points of the internal tori
where the $k$-th orbifold element acts \cite{Klein:1999im}. Thus,
in the $Z'_6$ orientifold, the $\a^i_1,\a^i_5$ axions are living
in the 4D Minkowski space, the $\a^i_2,\a^i_4$ in 4D Minkowski
space plus the second torus $T_2^2$ and the $\a^i_3$ in 4D
Minkowski space plus the third torus $T_3^2$.

The decompactification limit of the first torus (${\cal V}_1\to
\infty$) does not have any special interest since none of the
fields become six-dimensional.

\subsubsection*{Decompactification of the second torus (${\cal V}_2\to
\infty$)}

If we decompactify the second torus (${\cal V}_2\to \infty$) the
99 states that are coming from the $k=2,4$ sectors and the
$\a^i_2, \a^i_4$ axions become 6 dimensional fields. The gauge
group is enhanced and can be found by the action of $\g_2$, $\g_4$
on the Chan-Paton factors. The fields of the other sectors remain
four-dimensional and do not contribute to six-dimensional
anomalies. The 'shift' vector will be $2V_9$, where $V_9$ is given
in (\ref{Z6'vectors}). Following the known procedure we find that
the four-dimensional $U(4)\times U(4)\times U(8)$ gauge group is
enhanced in $U(8)\times SO(16)$. The generators of the
$U(4)_1\times U(4)_2$ are enhanced in the generators of the $U(8)$
as $T_{U(8)}\sim T_{U(4)_1} \oplus \overline{T}_{U(4)_2}$ and the
generators of the $U(8)$ in the generators of the $SO(16)$.

The rest of the matter fields are combined with some Kaluza-Klein
states, that now become massless, to give the representations of
the greater gauge group. The $(4,4,1)$, $(\bar{4},\bar{4},1)$ are
now contained in the adjoint of the $U(8)$ as the $(1,1,28)$,
$(1,1,\overline{28})$ are contained in the adjoint of the
$SO(16)$. The $(6,1,1)$, $(1,\bar{6},1)$ form the antisymmetric
$(28,1)$. The $(\bar{4},4,1)$ form the $(\overline{28},1)$.
Finally, the $(4,1,8)$, $(1,\bar{4},\bar{8})$, $(4,1,\bar{8})$ and
$(1,\bar{4},8)$ form the bi-fundamental $(8,16)$.
Thus, the effective gauge group is the one that it was taken from
the $Z_3$ six-dimensional orientifold (Table.\ref{SpectrumD=6}).

The spectrum of the $Z_3$ six-dimensional orientifold contains an
anomalous gauge boson (chapter \ref{Z3}). By the way that the
$U(4)\times U(4) \times U(8)$ gauge group is enhanced in
$U(8)\times SO(16)$, we find that the anomalous gauge boson is
$A_1-A_2$ and becomes massive due to the six-dimensional
Green-Schwarz mechanism.
This mass can be evaluated by the six dimensional formulae and it
is given in (\ref{Z_3massOfU(1)}). The $A_1+A_2$ and $A_3$ are
enhanced in the non-Abelian factors and they have no anomalies.

The contribution of the six-dimensional masses to the
four-dimensional ones can be found by taking the ${\cal V}_2\to
\infty$ limit of (\ref{Z'6.99.D4-to-D6}):
\bea {1\over 2}M^2_{99,ij}=&-&{1\over 4\pi^3}\left(Tr[\g_2\l^9_i]
Tr[\g_2\l^9_j] +Tr[\g_4\l^9_i]Tr[\g_4\l^9_j]\right)~, \eea
which is the same as the formula of the masses in the
six-dimensional $Z_3$ orientifold (\ref{Z_3massOfU(1)}) upon
normalization. The sectors $k=2,4$ of the four-dimensional $Z_6'$
orientifold in this limit are the $k=1,2$ sectors of the
six-dimensional $Z_3$ orientifold. Using (\ref{lambdaGen}) and
(\ref{Z6'vectors}), we evaluate the mass-matrix of the anomalous
$U(1)$s. The mass-matrix has two zero eigenvalues, with
eigenvectors: $A_3$, $A_1+A_2$ and a massive state with
eigenvalue:
\be -A_1+A_2~, ~~~~~~~m^2= {3\over \p^3} ~,\ee
as it was expected by the way that the initial $U(4)\times U(4)
\times U(8)$ gauge group is enhanced in $U(8)\times SO(16)$. This
six-dimensional contribution affects the four-dimensional mass
(\ref{MZ6'1}).

The results confirm that anomalous gauge bosons in six-dimensions
that become massive through the six-dimensional Green-Schwarz
mechanism, contribute to the four-dimensional mass generation by a
normalized term.

\subsubsection*{Decompactification of the third torus (${\cal V}_3\to
               \infty$)}

If we decompactify the third torus (${\cal V}_3\to \infty$), all
the string states from the $k=3$ sector and the $a^i_3$ axions
become six-dimensional. The new gauge group can be found by the
action of the $\g_3$ on the Chan-Paton. The orbifold rotation
$3(v_1,v_2)=(1,-1)/2$ shows that D5-branes survive in this limit.
The 'shift' vector is now $3V_a$ where $V_a$ is given
in(\ref{Z6'vectors}). The four-dimensional $U(4)_\a\times
U(4)_\a\times U(8)_\a$ gauge group (where $\a=5,9$) is enhanced to
$U(16)_\a$ that is the gauge group of the $Z_2$ six-dimensional
orientifold.
The generators are $T_{U(16)}\sim T_{U(4)_1} \oplus T_{U(4)_2}
\oplus \overline{T}_{U(8)}$. Therefore, $(1,\bar{4}, \bar{8})_a$,
$(4,1,8)_a$, $(\bar{4},4,1)_a$ are enhanced in the adjoint of the
$U(16)_a$. The $(6,1,1)_a$, $(1,4,\bar{8})_a$,
$(1,1,\overline{28})_a$, $(4,4,1)_a$ form the antisymmetric
$120_a$. The $(\bar{4},1,8)_a$, $(\bar{4},\bar{4},1)_a$,
$(1,1,28)_a$, $(1,\bar{6},1)_a$ are enhanced in the
$\overline{120}_a$.

From the way that the generators are formed we can expect that the
abelian factor of $U(16)_9$, $A\sim A_1+A_2-\sqrt{2} A_3$ where
the coefficients are coming from the normalization of the
generators of different rank. Similarly for the abelian factor of
$U(16)_5$, $\tilde{A}\sim
\tilde{A}_1+\tilde{A}_2-\sqrt{2}\tilde{A}_3$. As we have seen in
section \ref{Z26D}, the new gauge group contains two anomalous
bosons in six dimensions which are linear combinations of the $A$
and $\tilde{A}$. The other mass eigenstates are embedded in the
non-abelian factors.
The masses of the six-dimensional gauge bosons have been found in
(\ref{Z2massOfU(1)}). The contribution of the six-dimensional
mass-terms to the four-dimensional mass generation can be found by
taking the ${\cal V}_3\to \infty$ limit in
(\ref{Z'6.99.D4-to-D6}), (\ref{Z'6.59.D4-to-D6}) and these are
($a=5,9$):
\bea {1\over 2}M^2_{aa,ij}=&-&{1\over 3\pi^3}Tr[\g_3\l^a_i]
Tr[\g_3\l^a_j]~, \label{Mass.Z6'.aa}\eea
and for 59 states:
\bea {1\over 2}M^2_{59,ij}=&-&{1\over 12\pi^3}Tr[\g_3\l^5_i]
Tr[\g_3\l^9_j]~, \label{Mass.Z6'.59}\eea
which are the same  (upon normalization) with the contributions of
the six-dimensional generation of the $Z_2$ orientifold (section
\ref{Z26D}). In this limit, the $k=3$ sector of the
six-dimensional $Z_6'$ orientifold is the $k=1$ sector of the
six-dimensional $Z_2$ one. The mass-matrix has four zero
eigenvalues, with eigenvectors: $\sqrt{2}
\tilde{A}_1+\tilde{A}_3$, $-\tilde{A}_1+\tilde{A}_2$, $\sqrt{2}
A_1+A_3$, $-A_1+A_2$ and two massive states with eigenvalues:
\bea A_1+A_2-\sqrt{2}A_3
-\tilde{A}_1-\tilde{A}_2+\sqrt{2}\tilde{A}_3 ~&,&~~~~~~~m_3^2=
{4\over \p^3}~,
\nn\\
-A_1-A_2+\sqrt{2}A_3 -\tilde{A}_1-\tilde{A}_2 +\sqrt{2}\tilde{A}_3
~&,&~~~~~~~m_5^2= {20\over 3\p^3}~. \eea
The two massive states are the anomalous $U(1)$ which have been
found in the spectrum of the original six-dimensional $Z_2$
orientifold. The indices are taken from the four-dimensional
counting and denote which masses are affected by six-dimensional
anomalies. Notice that the linear combinations agree with our
expectations.

Another interesting limit of the $Z'_6$ orientifold is ${\cal
V}_3\to 0$. In this limit, the two linear combinations that are
free of four-dimensional anomalies become massless. This is
consistent with the fact that the six-dimensional anomalies which
are responsible for their masses cancel locally in this limit.

\subsubsection{Decompactification of the $Z_6$ orientifold}

In the $Z_6$ orientifold, the $\a^i_1,a^i_2,a^i_4,\a^i_5$ axions
are living in the 4D Minkowski space, and the $\a^i_3$ in 4D
Minkowski space plus the third torus $T_3$.

The decompactification limits of the first and second tori (${\cal
V}_1, {\cal V}_2\to \infty$) do not have any special interest
since none of the fields become six-dimensional and there are no
six-dimensional anomalies.

\subsubsection*{Decompactification of the third torus (${\cal V}_3\to
               \infty$)}

If we decompactify the third torus (${\cal V}_3\to \infty$), all
the string states from the $k=3$ sector and the $a^i_3$ axions
become six-dimensional. The rest of the sectors and axions remain
four-dimensional and do not contribute to six-dimensional
anomalies. The new gauge group can be found by the action of the
$\g_3$ on the Chan-Paton. The orbifold rotation
$3(v_1,v_2)=(1,-1)/2$ shows that D5-branes survive in this limit.
The 'shift' vector is now $3V_a$ where $V_a$ is given
in(\ref{Z6vectors}). The old $U(6)\times U(6)\times U(4)$ gauge
group is enhanced to $U(16)$, which is the gauge group of the
$Z_2$ six-dimensional orientifold (Table 1). The generators are
combined as $T_{U(16)}\sim T_{U(6)_1} \oplus T_{U(6)_2} \oplus
\overline{T}_{U(4)}$. Therefore, we can determine how the old
spectrum is enhanced to the new one. The $(\bar{6},1,\bar{4})$,
$(1,6,4)$ and $(6,\bar{6},1)$ combine in the adjoint of $U(16)$.
The $(15,1,1)$, $(1,6,\bar{4})$ are in the antisymmetric $120$ and
$(1,\overline{15},1)$, $(\bar{6},1,4)$ in the $\overline{120}$.

By the way that the generators of the $U(6)^2\times U(4)$ are
enhanced to the $U(16)$ we can expect that the six-dimensional
$U(1)$ gauge boson of the $U(16)$ will be a linear combination
$A_1+A_2-\sqrt{2 \over 3}A_3$ where the normalization coefficient
in front of $A_3$ takes into account the difference of the rank.
Similarly for the tilde.

The contributions of the six-dimensional anomalies to the
four-dimensional mass generation are given by the ${\cal V}_3\to
\infty$ limit in (\ref{Z6.aa.D4-to-D6}), (\ref{Z6.59.D4-to-D6}).
We find (for $a=5,9$):
\be {1\over 2}M^2_{aa,ij}=- {1\over 3\p^3} Tr[\g_3\l^a_i]
Tr[\g_3\l^a_j]~,\ee
while, for twisted open strings:
\be {1\over 2}M^2_{59,ij}=- {1\over 12\p^3} Tr[\g_3\l^5_i]
Tr[\g_3\l^9_j]~,\ee
which are the same (upon normalization) as the contributions of
the six-dimensional generation of the $Z_2$ orientifold (section
\ref{Z26D}).
The mass-matrix has four zero eigenvalues, with eigenvectors:
$\sqrt{2\over 3} \tilde{A}_1+\tilde{A}_3$,
$-\tilde{A}_1+\tilde{A}_2$, $\sqrt{2\over 3} A_1+A_3$, $-A_1+A_2$
and two massive states with eigenvalue:
\bea A_1+A_2-\sqrt{2\over 3}A_3
-\tilde{A}_1-\tilde{A}_2+\sqrt{2\over 3}\tilde{A}_3
~&,&~~~~~~~m_4^2= {4\over \p^3}~,
\nn\\
A_1+A_2-\sqrt{2\over 3}A_3 +\tilde{A}_1+\tilde{A}_2-\sqrt{2\over
3}\tilde{A}_3 ~&,&~~~~~~~m_5^2= {20\over 3\p^3}~. \eea
The two massive states are the anomalous $U(1)$s which have been
found in the spectrum of the original six-dimensional $Z_2$
orientifold. It is easy to verify that the four-dimensional
massless state $A_1+A_2-\tilde A_1-\tilde A_2+\sqrt{6}(A_3-\tilde
A_3)$ (\ref{Z6masses}) is still massless in six dimensions.

\subsection{Chapter Conclusions}

In this chapter we have shown that four-dimensional non-anomalous
$U(1)$s can become massive if in decompactification limits they
suffer from six-dimensional anomalies.

We have studied several four-dimensional orientifolds. In the
decompactification limit, there are sectors in such orientifolds
that become six dimensional.
The original four-dimensional massless spectrum, combined with
Kaluza-Klein states that become massless in this limit, enhanced
to the massless spectrum of six-dimensional orientifolds.
Some RR axions also become six-dimensional fields.

In the $6D$ orientifolds, we have calculated the stringy anomalous
$U(1)$ masses that are in accordance with six-dimensional
anomalies. The six-dimensional RR axions contribute to the
mass-generation of the anomalous $U(1)$s through the Green-Schwarz
mechanism.

We verified that the six-dimensional mass-matrix is the same as
the volume dependant contribution to the four-dimensional matrix.
Thus, six-dimensional anomalies play indirectly a role in
four-dimensional masses and explain why some non-anomalous $U(1)$
gauge bosons have a non-zero mass.

\newpage

%

\section{Anomalous $U(1)$s masses in non-supersymmetric open string vacua}


In this section, we are interested in the masses of the anomalous
$U(1)$s in non-supersymmetric models since such are the models
that will eventually represent the low energy physics of the
Standard Model. In particular, intersecting-brane realizations of
the Standard Model are generically non-supersymmetric.
We calculate the mass formulae using the ``background field
method" and find that they are the same as the supersymmetric ones
when we have cancellation of all tadpoles. In cases where NSNS
tadpoles do not vanish, there are extra contributions proportional
to the non-vanishing tadpole terms.

The formulae are valid even if we add Wilson lines that move the
branes away from the fixed points. The Wilson lines generically
break the gauge group and they will affect the masses of the
anomalous $U(1)$s through the traces of the model dependent $\g$
matrices.

\subsection{Computing with the background-field method}

Our purpose is to evaluate the bare masses of the anomalous $U(1)$
which appear in the one-loop amplitudes with boundaries where two
gauge fields are inserted \cite{Antoniadis:2002cs}.
Here we will use another technique which is based on turning on a
magnetic field on the D-branes and pick out the second order terms
to this magnetic field. This method is called ``the
background-field method" \cite{Bachas:bh}. We turn on different
magnetic fields $B_a$ in every stack of branes, longitudinal to
$x^1$, a non-compact dimension,
\be F^a_{23} = B_a Q_a \ , \label{MagneticField} \ee
where $Q_a$ are the $U(1)_a$ generators from every stack of
branes. The effect of the magnetic field on the open-string
spectrum is to shift the oscillator frequencies of the string
non-compact $x^2+ix^3$ coordinate by an amount $\e_a$:
\be \e_a = {1\over \p} [\arctan (\p q^a_i B_a) +  \arctan (\p
q^a_j B_a)] \ , \label{Shift} \ee
where $q^a_i$, $q^a_j$ are the $U(1)_a$ charges of the $i,j$
endpoints. The Chan-Paton states $\l_{ij}$ that describe the
endpoint $i,j$ of the open string, are the generators of gauge
group that remains after the orientifold construction.
Diagonalizing these matrixes, we can replace the $Q_i$ with
$\l_{ii}$.

The expansion of the one-loop vacuum energy is:
\bea \L(B) &=& {1 \over 2} \left( {\cal T}+{\cal K} + {\cal A}(B)
+ {\cal M}(B) \right) =
\L_0 
+ {1 \over 2} \left({B\over 2\p}\right)^2 \L_2 + \ldots \ ,
\label{oneloopvacuum} \eea
where $B$ one of the different magnetic fields.
%
%
%
Generically, it appears a linear to $B$ term that is a pour
tadpole and it is coming from the RR sector. This term vanishes
when we have tadpole cancellation.
The quadratic term in the background field contains a lot of
information. In the IR limit, we have a logarithmic divergence
whose coefficient is the $\b$-function. The UV limit provides the
mass-term of the anomalous gauge bosons. The finite part of this
term is the threshold correction in the gauge couplings
\cite{Bachas:bh}. The annulus amplitude in the $Z_N$ type I
orientifolds (without the magnetic field) can be written as:
\be {\cal A}^{ab} =-{1\over 2N} \sum_{k=1}^{N-1} \int_0^\infty \,
{dt\over t} {\cal A}^{ab}_k(q) \ , \label{BGNDanm} \ee
where $a,b$ the different kind of $D$-branes at the ends of the
open strings. The ${\cal A}^{ab}_k$ is the contribution of the
$k$th sector:
\bea {\cal A}^{ab}_k &=& {1 \over 4 \p^4 t^2} {\rm Tr} [\g^k_a]
{\rm Tr} [\g^k_b] \sum_{\a,\b=0,1} \h^{\a\b}
{\vartheta[^\a_\b]\over \h^3}
Z^{ab}_{int,k}\left[^{\a}_{\b}\right]|_{\cal A}~.
\label{BGNDannulus} \eea
Similarly, we can exchange $\cal A$ with $\cal M$ in
(\ref{BGNDanm}) to have an analogous expression for the M\"obius
strip. The ${\cal M}^{a}_k$ is given by:
\bea
{\cal M}^{a}_k &=& -{1 \over 4 \p^4 t^2} {\rm Tr}[\g_a^{2k}]
\sum_{\a,\b=0,1} \h^{\a\b} {\vartheta[^\a_\b]\over \h^3}
Z^a_{int,k}\left[^{\a}_{\b}\right]|_{\cal M}~. \label{BGNDmobius}
\eea
In the presence of the background magnetic field $B_a$, the above
amplitudes become:
\bea {\cal A}^{ab}_k(B) &=& {i \over 4 \p^3 t} {\rm Tr} \left[(B_a
\l_a \g^k_a \otimes \g^k_b + \g^k_a \otimes B_b \l_b \g^k_b)
\sum_{\a\b} \h^{\a\b} {\vartheta[^\a_\b ]({i \e t \over 2})\over
\vartheta[^1_1]({i \e t \over 2})} \right]
Z^{ab}_{int,k}\left[^{\a}_{\b}\right]|_{\cal A}~, \nonumber \\
{\cal M}^{a}_k(B) &=& -{i \over 2 \p^3 t} {\rm Tr} \left[ B_a \l_a
\g_a^{2k} \sum_{\a\b} \h^{\a\b} {\vartheta[^\a_\b]({i \e t \over
2})\over \vartheta[^1_1]({i \e t \over 2})} \right]
Z^a_{int,k}\left[^{\a}_{\b}\right]|_{\cal M}~. \label{BGNDon} \eea
Notice that the only differences from (\ref{BGNDannulus},
\ref{BGNDmobius}) are in the contribution of the non-compact part
of the partition functions. This is expected since the presence of
the magnetic fields affect only the $x^2, x^3$ coordinates.
Therefore, the expressions (\ref{BGNDon}) are valid for all kinds
of orientifold models.

Since we are interested in the quadratic $B^2$ terms of the above
amplitudes, we expand the above formulae to quadratic order in the
background field
, using the following Taylor expansions:
\be \e  \simeq \left\{ \ba{ccccc} B_a \l_a \otimes 1 + 1 \otimes
B_b \l_b & \quad & \textrm{in} & \quad & {\cal A}^{ab} ,\\ 2 B_a
\l_a \ \ \ \ \ \ \ \ \ \ \ \ \ \ \ \ \ \ & \quad & \textrm{in} &
\quad & {\cal M}^{a} . \ea \right.\label{Shift-2}\ee
The zero-order $B$ terms are the amplitudes in the absence of the
magnetic field (\ref{BGNDannulus}, \ref{BGNDmobius}). These
expressions give the tadpole cancellation conditions in virtue of
the UV divergences.
%
%
The linear to $B$ terms appear from the $a=b=1$ sector in
(\ref{BGNDon}). This is a pour tadpole and vanishes when we have
tadpole cancellation. Therefore, it does not affect higher order
in $B$ amplitudes.
%
%
The second order-terms on $B$ are:
\bea {\cal A}_{2,k}^{ab}
&=& \p^2 i~
\Big[{\rm Tr} [\l_a^2 \g_a^k]{\rm Tr} [\g_b^k]+ {\rm Tr}
[\g_a^k]{\rm Tr} [\l_b^2 \g_b^k] 
+2{\rm Tr} [\l_a\g_a^k]{\rm Tr} [\l_b\g_b^k]\Big]
F^{ab}_k|_{\cal A} \label{BGNDannulusLAST} ~,\\
{\cal M}_{2,k}^a
&=& -4\p^2 i~{\rm Tr} [\l_a^2 \g_a^{2k}]~ F^{aa}_k |_{\cal M}
~,\label{BGNDmobiusLAST} \eea
defining $F^{ab}_k$ as a term which contains all the
spin-structure and the orbifold information:
\be F^{ab}_k|_\s={1\over 4\p^4}\sum_{\a\b}\h_{\a\b}~ \p i
\partial_\t
\left[ \log{\vartheta[^\a_\b](0|\t)\over \h(\t)}\right]
{\vartheta[^\a_\b](0|\t)\over \h^3(\t)} Z^{ab}_{int,k}[^a_b]|_\s~,
\label{Fab-k-SS}\ee
for both surfaces (the choice of $\t$ define the surface $\s$).
Note that the $a=b=1$ sector is not contained in the
(\ref{Fab-k-SS}).
This term can be formally written as the supertrace over states
from the open $ab$ $k$-orbifold sector:
\be F^{ab}_k|_{\s}= {|G|\over (2\pi)^2}Str^{ab}_{k, {\rm
open}}\left[{1\over 12}-s^2\right]e^{-tM^2/2}\Big|_\s~,
\label{FkSTRACEI}\ee
where the $s$ is the $4D$ helicity. Thus, for:
\begin{itemize}
\item $large$ $\t_2$ we have:
\be \lim_{\t_2\to\infty}F_k^{ab}=C_{k,IR}^{ab}+{\cal
O}[e^{-2\p\t_2}] ~,\label{FonIR}\ee
with $C_{k,IR}^{ab}={|G|\over (2\pi)^2}Str_k \left[{1\over
12}-s^2\right]_{open}$.
\item $small$ $\t_2$ we have
\be \lim_{\t_2\to 0}F_k^{ab}={1\over \t_2}\left[C_{k,UV}^{ab}
+{\cal O}[e^{-{\p\over 2\t_2}}]\right]~, \label{FonUV}\ee
where $C_{k,IR}^{ab}={|G|\over (2\pi)^2}Str_k \left[{1\over
12}-s^2\right]_{closed}$.
The helicity supertrace is now in the closed-string $k$-sector
mapped from the open $k$-sector dy a modular transformation.
\end{itemize}
Notice that in the annulus amplitude (\ref{BGNDannulusLAST}), the
two first terms are proportional to the square of the $B$ field.
This cases are proportional to annulus amplitudes ${\cal A}_2$,
where two vertex-operators (VOs) are on the same boundary. In the
last component of (\ref{BGNDannulusLAST}), the $B$ fields are
coming from the opposite D-branes and is proportional to ${\cal
A}_{11}$, with the VOs on different boundaries. The
(\ref{BGNDmobiusLAST}) is proportional to a M\"obius strip
amplitude with the insertion of two VOs.

The IR limit $t \to \infty$ can be found easily using the
(\ref{FonIR}). We regularize the integral by $\m \to 1/t^2$ and we
find the $\b$-function:
\bea b &=&-{2\over N}\sum_{k=1}^{N-1}\lim_{t \to
\infty}\left({\cal A}_{2,k}^{ab}(t)+ {\cal
M}_{2,k}^{a}(t)\right)\nonumber\\
&=& -{2\p^2i\over N} \sum_{k=1}^{N-1} \bigg[\Big({\rm Tr} [\l_a^2
\g_a^k]{\rm Tr} [\g_b^k]+ {\rm Tr} [\g_a^k]{\rm Tr} [\l_b^2
\g_b^k] \nonumber\\
&& +2{\rm Tr} [\l_a\g_a^k]{\rm Tr} [\l_b\g_b^k] \Big)
C^{ab}_{k,IR}|_{\cal A} - 4 {\rm Tr} [\l_a^2 \g_a^{2k}]
C^{a}_{k,IR} |_{\cal M} \bigg] ~.\eea
For the UV limit $t\to 0$, we use the (\ref{FonUV}) and we
regularize the integral by $\m \leq t$. The $A_2$ and $M$ together
are giving terms proportional to the tadpole cancellation
conditions
%
%
%
%
%
%
Therefore, when we have vanishing of RR and NSNS tadpoles, the
masses of the anomalous gauge bosons are given by ${\cal A}_{11}$:
\bea {1\over 2} M^2_{aa} &=&  {\p^2i \over N} \sum_{k=1}^{N-1}
{\rm Tr} [\l_a\g_a^k]^2
C^{ab}_{k,UV}|_{\cal A} ~,\label{Mass-U1-aa}\\
{1\over 2} M^2_{59} &=&  {\p^2i \over 2 N} \sum_{k=1}^{N-1} {\rm
Tr} [\l_5\g_5^k]{\rm Tr} [\l_9\g_9^k] C^{59}_{k,UV}|_{\cal A}~,
\label{Mass-U1-59}\eea
where $\a=5,9$. When we have non-vanishing NSNS tadpoles there is
an extra contribution to the mass formulas, proportional to the
non-vanishing tadpole.

The formulae (\ref{Mass-U1-aa}, \ref{Mass-U1-59}) still hold even
if we add Wilson lines.
Generically, adding a Wilson line we shift the windings or the
momenta in a coordinate with Neumann or Dirichlet boundary
conditions respectively. This breaks the gauge group.
In the transverse (closed) channel the shifts appears as phases
$e^{2\p i n \q}$ where $\q$ the shift and $n$ the momenta or
windings respectively to the above.
Since only the massless states contribute in the UV limit, the
effect of the Wilson line will appear only in the traces of the
$\g$ matrices.

The threshold correction \cite{Kiritsis:1997hj} is the finite part
of (\ref{BGNDannulusLAST}) and (\ref{BGNDmobiusLAST}). Generically
we have:
\be {16 \p^2 \over g^2}={16 \p^2 \over g_0^2} -{1\over
2N}\sum_{k=1}^{N-1} \int^{1/\m^2}_\m {dt\over t}\left({\cal
A}_{2}^{ab}+ {\cal M}_{2}^{a}\right) - b \log{\m^2\over
M^2}-{1\over 2}M^2_{ab}{1\over \m}~,\ee
where we separate the divergencies from the quadratic terms to
$B$. The above formulae for the $\b$-function, the corrections to
the gauge couplings and the masses of the anomalous $U(1)$s are
the same to the supersymmetric ones found in
\cite{Antoniadis:2002cs, Bachas:bh}.
Next, we will apply the above formulae to a non-supersymmetric
model that has been constructed by Scherk-Schwarz deformation
\cite{Anastasopoulos:2003ha}.

\subsection{A four-dimensional non-supersymmetric orientifold example}

In this section we will evaluate the masses of the anomalous
$U(1)$s in a $Z_2$ orientifold model where supersymmetry is broken
by a Scherk-Schwarz deformation. The spectrum is provided in
Table.\ref{SpectrumZN+SS1} \cite{Anastasopoulos:2003ha}.
We remind that the tadpole cancellation provides two different
solutions that depend on the inequivalent choices of $\g_h^2=\pm
1$ where $\g_h$ the action of $h$ on the Chan-Paton matrixes. The
16-dimensional 'shift' vector of the $Z_2$ orientifold is
(\ref{V-Z2}) 
%
%
The 'shift' vector of the SS deformation is
%
%
%
as it was defined in (\ref{V-h}). In both cases $a+b=16$, however
we implement for simplicity $a=b=8$. The massless spectrums are
provided in Table 1. The gauge group in both cases is the same.
The only difference appears in the exchange of the antisymmetric
reps with the bi-fundamental $(8,8)+(\overline{8},\overline{8})$
in the (99)/(55) matter sector. The spectrum is anomaly-free in
$4D$ since it is non-chiral.

The internal annulus partition functions for 99, 55 and 59 strings
are:
\bea Z^{99,55}_{int,k}[^\a_\b] &=& - \sum_{s,r=0}^1 (-1)^{\a s +\b
r} \bigg[ (-1)^{s\cdot m_4}P_{m_4} P_{m_5} \bigg]
{\vartheta[^\a_\b](0|\t) \over \h(\t)} (2 \sin {\p k\over 2})^2
\prod_{j=1}^2 {\vartheta[^{~~\a}_{\b+2v_j k}](0|\t) \over
\vartheta[^{~~1}_{1+2v_j k}](0|\t)} ~,\nn\\
Z^{59}_{int,k}[^\a_\b] &=& 2 \sum_{s,r=0}^1 (-1)^{\a s +\b r}
\bigg[(-1)^{s\cdot m_4}P_{m_4} P_{m_5}\bigg]
{\vartheta[^\a_\b](0|\t) \over \h(\t)} \prod_{j=1}^2
{\vartheta[^{~~\a+1}_{\b+2v_j k}](0|\t) \over
\vartheta[^{~~~0}_{1+2v_j k}](0|\t)}~. \label{Z-Internal}\eea
For $s=r=0$, we have the internal partition function of a
$T^2\times K^3 /Z_2$ orientifold. $s$ denotes the direct action of
the SS deformation and $r$ the twisted sector. The $(-1)^{s\cdot
m_4}P_{m_4} P_{m_5}$ is the lattice sum over momenta along the
first torus $T^2$:
\be (-1)^{s\cdot m_i} {P}_{m_i}({i\tau_2}/2) = {1
\over\eta({i\tau_2}/2)} \sum_{m_i} (-1)^{s\cdot m_i} q^{{\a'\over
4}\left({m_i\over R_i}\right)^2} ~,
\ee
For $s=1$ we have the SS deformation that shifts the $m_4$
momenta. As we mention before, $r=0,1$ denotes the $h$ untwisted
and twisted sectors respectively. However we will neglect the
twisted sector since it requires the insertion of anti-D-branes
\cite{Anastasopoulos:2003ha}.

To evaluate the masses of the anomalous bosons, we insert
(\ref{Z-Internal}) and (\ref{Fab-k-SS}) in the mass
formulae
. After some 'thetacology' we find $F^{\a\b}_{k=1}$ for
$\a,\b=5,9$.
%
%
%
%
%
%
%
%
In the UV region, only the first terms in both formulae contribute
to the mass of the anomalous $U(1)$s. Terms (that contains the SS
action $h$) after the Poisson re-summation become proportional to
$W_{\n_4+1/2}$ and do not contribute to the $C^{99,55,59}_{UV}$.
Since SS deformation does not contribute to the mass terms of the
anomalous $U(1)$s, we can directly evaluate their masses for both
two inequivalent solutions ($\g^2_h=\pm 1$):
\bea {1\over 2} M^2_{\a\a,ij} &=&  -{4 \p^2 \over 4} {\rm Tr}
[\l_i^a\g_g]{\rm Tr} [\l_j^a\g_g] {{\cal V}_1\over \p^2\a'}
\nn\\
&=&- {{\cal V}_1\over \a'} \left(-{i\over \sqrt{8}} \sin[2\p
V^a_i] \right) \left(-{i\over \sqrt{8}} \sin[2\p V^a_j] \right) =
{{\cal V}_1\over 8\a'}~,\label{Mass-U1-aa-Z2}\\
{1\over 2} M^2_{59,ij} &=&  {4 \p^2 \over 2\times 4} {\rm Tr}
[\l_i^5\g_g]{\rm Tr} [\l_j^9\g_g] {{\cal V}_1\over 2\p^2\a'}
= -{{\cal V}_1\over 32\a'}~,\label{Mass-U1-59-Z2}\eea
where $\a=5,9$. The mass-matrix has two massless gauge bosons
$-\tilde{A}_1+\tilde{A}_2$, $-A_1+A_2$ and two massive
$A_1+A_2+\tilde{A}_1+\tilde{A}_2$,
$-A_1-A_2+\tilde{A}_1+\tilde{A}_2$ with masses $3{\cal V}_1/
32\a'$, $5{\cal V}_1/ 32\a'$ respectively.

There are no anomalous $U(1)$s in these models since the spectrum
is non-chiral. However, the existence of the two massive gauge
bosons are the consequence of $6D$ anomalies \cite{Ibanez:2001nd,
Scrucca:2002is, Antoniadis:2002cs, Anastasopoulos:2003aj}. The
decompactification limit of the first torus (where the SS
deformation acts) leads to the {\cal N}=1 $6D$ $Z_2$ orientifolds
that contains two anomalous $U(1)$s that become massive via the
Green-Schwarz mechanism. Therefore, axions that participate in the
anomaly cancellation in the $6D$ model, contribute to the $4D$
masses of the anomalous $U(1)$s by volume dependant terms. The
ratio of the masses found in \cite{Anastasopoulos:2003aj} for the
$Z_2$ supersymmetric orientifold are the same to the above.

\subsection{Chapter Conclusions}

In this section we evaluated the general mass formula for the
anomalous $U(1)$s in non-supersymmetric orientifolds. We have
shown that the supersymmetric formulae of \cite{Antoniadis:2002cs}
are also valid in non-supersymmetric orientifolds provided that
the tadpoles cancel.

\newpage

\section{Anomalous $U(1)$s and spontaneous symmetry breaking}

In D-brane realizations of the Standard Model we must have at
least two Higgs in order to be able to give masses to all quarks
and leptons \cite{Antoniadis:2002cs}. Generically, each Higgs is
charged under the anomalous $U(1)$s.

We will analyze here the case of a single anomalous $U(1)$ coupled
to a Higgs field in order to discuss the relevant effects.
Consider a toy model with an anomalous $U(1)$ gauge field
$A'_{\m}$, chiral charged fermions and a complex Higgs. We also
have an axion $a$ to cancel the anomalies. The relevant part of
the low-energy effective Lagrangian can be written as:
\bea {\cal L} = 
& - &{1\over 4g^2_{A'}}F_{A'}^2 + M_s^2(\partial a+ A')^2
+ D_{\m} H D^{\m} H^* + V(|H|^2)\nn\\
& + & Q_{L} \bar{\psi}_{L} \sla{A'} \psi_{L} +   Q_{R}
\bar{\psi}_{R} \sla{A'} \psi_{R} + h \bar{\psi}_{L} \psi_{R}
H+c.c. \label{LUone} \eea
This Lagrangian (\ref{LUone}) is invariant under the ``anomalous"
$U(1)$ transformations.
\bea &&A'_{\m} \rightarrow A'_{\m} + \partial_\m \e ~~,~~~~~~
a \rightarrow a- \e \nn\\
&&\psi_L \rightarrow e^{iQ_{L}\e} \psi_L ~~~~,~~~~~~ \psi_R
\rightarrow e^{iQ_{R}\e} \psi_R \nn\\
&&H \rightarrow e^{i(Q_{R}-Q_{L})\e}H \label{u1trans}\eea
There are two sources of gauge symmetry breaking. One is the
stringy mass term and the other is the non-zero expectation value
of the Higgs. Writing $H=r e^{i \f}$, the Higgs potential fixes
the vacuum expectation value $\langle r\rangle=v$. The kinetic
term of the Higgs field gives an extra contribution to the $A'$
mass term:
\be v^2(\partial \f +\D Q A')^2 ~.\ee
To proceed with the one-loop calculation, it is necessary to add a
gauge fixing term
\bea {\cal L}_{gauge fixing} = \l \Big(\partial A' + {c M_s^2
\a\over {\l}} - {\Delta Q v^2 \f \over \l}\Big)^2~,
\label{gaugefixing} \eea
which keeps $A'_\m$ orthogonal to $a$ and $\f$. Redefining
$\tilde{a}=M a$ and $\tilde{\f}=v\f$ we can diagonalize the axions
by an $SO(2)$ rotation
\bea \left(\ba {c}
 a'\\
\f'\ea\right) = \left(\ba {cc}
 \cos \q' & \sin \q' \\
-\sin \q' & \cos \q' \ea\right) \left(\ba {c}
\tilde{a}\\
\tilde{\f}\ea \right) ~,\label{rotSU2} \eea
where $\cos\q'={cM_s \over \sqrt{c^2 M_s^2 + v^2 \Delta Q^2} }$
and $\sin\q'={\Delta Q v \over \sqrt{c^2 M_s^2 + v^2 \Delta
Q^2}}$. Now, the effective Lagrangian has the form:
\bea {\cal L} = 
& - &{1\over 4g^2_{A'}}F_{A'}^2 + m_{A'}^2 A'^2 
+(\partial b')^2 + m_{b'}^2 b'^2
+  (\partial \f')^2 \nn\\
& + & Q_L \bar{\psi}_L \sla{A'} \psi_L+ Q_R \bar{\psi}_R \sla{A'}
\psi_R + hv \bar{\psi}_L \psi_R e^{i (\sin\q' b'+\cos\q' \f')/v} +
c.c. \label{LUfinal} \eea
The masses are:
\bea m_{\psi} = hv ~&,&~~~~~~~~
m_{A'} = \sqrt{c^2 M_s^2 + v^2 {\Delta Q}^2} ~, \nn\\
m_{\f'} = 0      ~&,&~~~~~~~~ m_{a'} = \sqrt{c^2 M_s^2 + v^2
{\Delta Q}^2}/ \sqrt{\l}. \label{masses} \eea
We define $m_B=\m$ for simplicity.
The propagators are:
\bea &&D_{A'}^{\m \n}(k) = {-ig^{\m \n}\over {k^2-\m^2} }+
(1-\l^{-1}) {ik^\m k^\n \over (k^2-\m^2)(k^2-\m^2/{\l})}~~~,
\nn\\
&&G_{\f'}(k)= {i\over k^2} ~~~,~~~~~~~~~~~ G_{a'}(k) = {i\over
k^2-\m^2/{\l}} \label{propagators}\eea
We will gauge fix $\f'=0$ (physical gauge) and the Yukawa
couplings between the physical axion and the fermions is:
\be h_{eff}=h {cM_s\over \sqrt{c^2M_s^2+v^2\D Q^2}}~. \ee
In order to suppress this interaction we must have $cM_s\ll v \D
Q$.

\newpage

\section{A D-brane realization of the Standard Model}

Bottom to top model building shows that the SM can be embedded in
a product of unitary groups appearing on D-brane stacks as a
subgroup of $U(3)\times U(2)\times U(1)\times U(1)'$
\cite{Antoniadis:2000en}. However, for the rest of our study we
will omit the last single brane that provides the $U(1)'$ gauge
boson since it does not participate to the hypercharge. We will
concentrate onto $U(3)\times U(2)\times U(1)$\footnote{In fact the
minimal embedding is in $U(3)\times U(2)$, however such an
embedding has phenomenological problems: proton stability cannot
be protected and some SM fields cannot get masses.}. Each $U(n)$
factor arises from $n$ coincident D-branes. As $U(3)=SU(3)\times
U(1)$, a string with one end on this group of branes is a triplet
under $SU(3)$ with $Q_3=\pm 1$ abelian charge. Thus, $Q_3$ is
identified with the gauged baryon number. Similarly, the second
factor arises from two coincident D-branes (``weak" branes) and
the gauged overall abelian charge $Q_2$ is identified with the
weak-doublet number. Both collections have their own gauge
couplings $g_3$, $g_2$ that are functions of the string coupling
$g_s$ and possible compactification volumes. The necessity for at
least an extra $U(1)$ factor is due to the fact that we cannot
express the hypercharge as a linear combination of baryon and
weak-doublet numbers\footnote{It turns out that a complete
collection of SM D-branes (one that can accommodate all the
endpoints of SM strings) includes a fourth U(1)$_b$ component that
does not participate in the hypercharge. Such a D-brane wraps the
large dimensions, and consequently its coupling is ultra weak. It
is also anomalous and thus massive \cite{Antoniadis:2000en} . Due
to its weak coupling its contributions to magnetic moments are
negligible compared to the ones we consider. We will thus ignore
it in this paper.}. The U(1) brane can be in principle independent
of the other branes and has in general a different gauge coupling
$g_1$. In \cite{Antoniadis:2000en}, the $U(1)$ brane has been put
on top of either the color or the weak D-branes. Thus, $g_1$ is
equal to either $g_3$ or $g_2$.

Let us denote by $Q_3$, $Q_2$ and $Q_1$ the three $U(1)$ charges
of $U(3)\times U(2)\times U(1)$. These charges can be fixed so
that they lead to the right hypercharge. In order that we can
match the measured gauge couplings with the ones appropriate for
the brane-configuration and also avoid hierarchy problems we find
that we have to put the $U(1)$ brane on top of the color branes.
Consequently we set $g_1=g_3$. This fixes the string scale to be
between 6 to 8 TeV \cite{Antoniadis:2000en}. There are two
possibilities for charge assignments. Under $SU(3)\times
SU(2)\times U(1)_3 \times U(1)_2 \times U(1)_1$ the members of a
given quark and lepton family have the following quantum numbers:
\bea
&&Q ({\bf 3},{\bf 2};1,1+2z,0)_{1/6}~~~~~~~~~
L ({\bf 1},{\bf 2};0,1,z)_{-1/2}\nn\\
&&u^c ({\bf\bar 3},{\bf 1};-1,0,0)_{-2/3}~~~~~~~~~~~
l^c({\bf 1},{\bf 1};0,0,1)_1\nn\\
&&d^c ({\bf\bar 3},{\bf 1};-1,0,1)_{1/3}\label{charges}\eea
where $z=0,-1$. From (\ref{charges}) and the requirement that the
Higgs doublet has hypercharge 1/2, one finds two possible
assignments for it:
\bea H\ \ ({\bf 1},{\bf 2};0,1+2z,1)_{1/2}\quad &H'\ \ ({\bf
1},{\bf 2};0,-(1+2z),0)_{1/2} \label{Higgs} \eea
The trilinear Yukawa terms are
\bea \label{HY} z=0  \ :\qquad H'Qu^c\ &,& \quad H^\dagger Ll^c
     \ ,\quad  H^\dagger Qd^c\\
z=-1 \ :\qquad H' Qu^c\ &,& \quad H'^\dagger Ll^c
     \ ,\quad  H^\dagger Qd^c
\label{HtildeY} \eea
In each case, two Higgs doublets are necessary to give masses to
all quarks and leptons.
The $U(3)$ and $U(1)$ branes are D3 branes. The U(2) branes are D7
branes whose four extra longitudinal directions are wrapped on a
four-torus of volume 2.5 in string units \cite{Antoniadis:2000en}.
The spectator $U(1)_b$ brane is stretching in the bulk but the
fermions that end on it do not have KK excitations. Thus, the only
SM field that has KK excitations is a linear combination of the
hypercharge gauge boson and the two anomalous $U(1)$ gauge bosons.
The masses of KK states, are shifted from the basic state by
multiples of $0.8 M_s$.

We will now describe the structure of the gauge sector for the
D-brane configuration above. We denote by $A^i_\m$ the $U(1)_i$
gauge fields and $F^i_{\m\n}$ their corresponding field strengths.
Also we denote $G^\b_{\m\n}$ the field strengths of the
non-abelian gauge group where $\b$ runs over the two simple
factors. There is also a set of two axion fields $b^\a$ with
normalized kinetic terms. Starting from the kinetic terms of the
gauge fields and requesting for the cancellation of the $Q T^\a
T^\a$ mixed anomalies, we can write down the most general low
energy action
\bea {\cal L} = &-&{1\over 4} \sum_i F^i_{\mu\nu}F^{i,\mu\nu}
+\sum_i \bar{\y} Q_i \sla{A^i} \y
-{1\over 4} \sum_a Tr G^a_{\mu\nu}G^{a,\mu\nu}          \nonumber\\
&+& \sum_{\a, \b} \L_{\a,\b}{b^\a \over
M_s}\epsilon^{\mu\nu\rho\sigma} Tr[G^\b_{\mu\nu}
G^{\b}_{\rho\sigma}] + \sum_\a (\partial_{\mu} b^\a-M_s\l^{\a i}
A^i_{\mu})
(\partial^{\mu} b^\a-M_s\l^{\a j} A^{j,\mu})  \nn\\
&+& \sum_{\a,i,j} {C_{\a ij}\over M_s}
\epsilon^{\mu\nu\rho\sigma}\partial_{\mu} b^\a A^i_{\nu}
F^j_{\rho\sigma} + \sum_{i,j,k} {D_{ijk}\over
M_s}\epsilon^{\mu\nu\rho\sigma}
A^i_{\mu}A^j_{\nu} F^k_{\rho\sigma}\label{lor}\\
&+& \sum_\a Z_\a{b^\a\over M_s}\epsilon^{\mu\nu\rho\sigma}
Tr[R_{\mu\nu} R_{\rho\sigma}]\nn\eea
where charge operators $Q_i$ contain all coupling dependence. The
last term involves the curvature two-form $R_{\mu\nu}$ and is
responsible for the cancellation of the gravitational anomalies.
Under $U(1)$ gauge transformations (modified by the anomaly)
\bea A^i_\m \rightarrow A^i_\m+\partial_\m \e^i \quad, \quad b^\a
\rightarrow b^\a+\sum_i \l^{\a i} A^i_\m \label{trns}\eea
we have
\bea &&D_{ijk}=-D_{jik} \quad , \quad \sum_a \L_{\a,\b}
\l^{\a,\i}=Tr[Q^iT_\b T_\b] \nn\\
&&D_{ijk}=-\sum_a C_{\a ij} \l^{\a k}=Tr[Q^i Q^j Q^k] \quad, \quad
\sum_a Z_\a \l^{\a, i}=Tr[Q^i] \eea
The only free parameters which are not fixed by the anomalies are
$\l^{\a i}$. These define the mass matrix of gauge bosons
$M^2_{ij}=M^2_s\l^{\a i}\l^{bj}$. This matrix is symmetric and has
a zero eigenvalue corresponding to the non-anomalous hypercharge.
The $\l^{\a i}$ can be computed by a string calculation. The
parameters remaining in the mass matrix is the $2\times 2$
submatrix of the anomalous gauge bosons.

Now, we will describe the couplings of the gauge fields in more
details. The two first terms of (\ref{lor}) are written as
\be {\cal L} = -{1\over 4} \sum_i F^iF^i +\sum_i {g_i \over
\sqrt{i}} \bar{\y} Q_i \sla{A^i} \y \ee
where $g_i$ are the $SU(i)$ coupling constants and the charges
have the standard integral normalization (\ref{charges}). We will
set $x={g_3/\sqrt{3}\over g_2/\sqrt{2}}=\sqrt{5/3}$ as
$g_2/g_3\sim \sqrt{0.4}$ \cite{Antoniadis:2000en}. Doing a $O(3)$
rotation, we can go to a basis where the kinetic terms of the
$U(1)$ gauge fields are still diagonal, while one of them
corresponds to the hypercharge: $A_i =U_{ij} \widetilde{A}_j $
with $A_Y = \widetilde{A}_1 $.
This rotation is different in each theory.\\

For the $z=0$ case we use
\be U= \left( \ba {ccc} {2\sqrt{3}\over \sqrt{28+9x^2}} &
-{\sqrt{16+9x^2}\sin{\q}\over \sqrt{28+9x^2}} &
{\sqrt{16+9x^2}\sin{\q}\sqrt{3}\over \sqrt{28+9x^2}} \\
-{3x\over \sqrt{28+9x^2}} & -{2(-2\sqrt{28+9x^2}\cos{\q}
+3\sqrt{3}x\sin{\q})\over \sqrt{28+9x^2}\sqrt{16+9x^2}} &
{2(2\sqrt{28+9x^2}\sin{\q} +3\sqrt{3}x\cos{\q})\over
\sqrt{28+9x^2} \sqrt{16+9x^2}}\\
{4\over \sqrt{28+9x^2}} & {3x\sqrt{28+9x^2}\cos{\q}
+8\sqrt{3}\sin{\q}\over \sqrt{28+9x^2}\sqrt{16+9x^2}} &
{3x\sqrt{28+9x^2}\sin{\q} -8\sqrt{3}\cos{\q}\over
\sqrt{28+9x^2}\sqrt{16+9x^2}} \ea \right) \label{Uarrey} \ee
and the $U(1)$ charges:
\[Q_{Y} \sim Q_1-{Q_2\over 2} + {2Q_3\over 3}\]
\bea Q_{\a}\sim -\sqrt{3}x(16+9x^2)\sin{\q}Q_1
+2(2\sqrt{28+9x^2}\cos{\q}-3\sqrt{3}x\sin{\q})Q_2\nn\\
+(3x^2\sqrt{28+9x^2}\cos{\q}+8\sqrt{3}x\sin{\q})Q_3 \nn\eea
\bea Q_{b}\sim \sqrt{3}x(16+9x^2)\cos{\q}Q_1
+2(2\sqrt{28+9x^2}\sin{\q}+3\sqrt{3}x\cos{\q})Q_2\nn\\
+(3x^2\sqrt{28+9x^2}\sin{\q}-8\sqrt{3}x\cos{\q})Q_3
\label{rotchargone}\eea
We can obtain the $z=-1$ case from the one above by
$x\rightarrow-x$. The matrix $U$ is now
\be U= \left( \ba {ccc} {2\sqrt{3}\over \sqrt{28+9x^2}} &
-{\sqrt{16+9x^2}\sin{\q}\over \sqrt{28+9x^2}} &
{\sqrt{16+9x^2}\sin{\q}\sqrt{3}\over \sqrt{28+9x^2}} \\
{3x\over \sqrt{28+9x^2}} & {2(2\sqrt{28+9x^2}\cos{\q}
+3\sqrt{3}x\sin{\q})\over \sqrt{28+9x^2}\sqrt{16+9x^2}} &
-{2(-2\sqrt{28+9x^2}\sin{\q} +3\sqrt{3}x\cos{\q})\over
\sqrt{28+9x^2}\sqrt{16+9x^2}}
\\
{4\over \sqrt{28+9x^2}} & {-3x\sqrt{28+9x^2}\cos{\q}
+8\sqrt{3}\sin{\q}\over \sqrt{28+9x^2}\sqrt{16+9x^2}} &
-{3x\sqrt{28+9x^2}\sin{\q} +8\sqrt{3}\cos{\q}\over
\sqrt{28+9x^2}\sqrt{16+9x^2}} \ea \right) \label{Uarreysecond} \ee
and the charges:
\[ Q_{Y} \sim Q_1+{Q_2\over 2} + {2Q_3\over 3}\]
\bea Q_{\a}\sim  -\sqrt{3}x(16+9x^2)\sin{\q}Q_1
+2(2\sqrt{28+9x^2}\cos{\q}+3\sqrt{3}x\sin{\q})Q_2\nn\\
+(-3x^2\sqrt{28+9x^2}\cos{\q}+8\sqrt{3}x\sin{\q})Q_3\nn\eea
\bea Q_{b} \sim   \sqrt{3}x(16+9x^2)\cos{\q}Q_1
+2(2\sqrt{28+9x^2}\sin{\q}-3\sqrt{3}x\cos{\q})Q_2\nn\\
-(3x^2\sqrt{28+9x^2}\sin{\q}-8\sqrt{3}x\cos{\q})Q_3
\label{rotchargtwo}\eea
The parameter $\q$ can be used to diagonalize the mass matrix of
the two anomalous $U(1)$s $A_\a$ and $A_b$. The two eigenvalues
$\m_\a^2$, $\m_b^2$ and $\q$ parametrize effectively the $2\times
2$ mass matrix. The masses of the anomalous $U(1)$ gauge fields
have also contributions from the Higgs effect since the Higgses
are also charged under the anomalous $U(1)$s (appendix). However,
such corrections are of order of $m_Z/M_s$ and are thus
sub-leading for our purposes. String theory calculations indicate
that $\m_{\a,b}$ are a factor of 5-10 below the string scale
\cite{Antoniadis:2000en}. Thus they are expected to be in the TeV
range.

\subsection{Phenomenological aspects - Calculation of lepton
anomalous magnetic moment in the
         presence of an anomalous $U(1)$}

The recent precise measurement of the anomalous magnetic moment
(AMM) of muon $\a_{muon}=(g-2)/2$ from the Brookhaven AGS
experiment \cite{BNL} gave
\be \a_{muon}^{exp}=116 592 023 (151) \times 10^{-11}
\label{AMMexp} \ee
The difference between the experimental value (\ref{AMMexp}) and
the theoretical expectation, (for a review see \cite{theory}), due
to standard model (SM) is
%
%
\be \d\a_{muon} = \a_{muon}^{exp}-\a_{muon}^{SM} = (43\pm
16)\times 10^{-10}. \label{difference} \ee
The experimental precision is unprecented and it is going to reach
$\pm 4\times 10^{-10}$ soon. It becomes thus important to examine
the signals of physics  beyond the SM. Various explanations for a
discrepancy have been proposed building on earlier computations
\cite{early}. Many of those assume SUSY broken at a mass scale not
far above the weak scale \cite{kane,feng, martinwells,Byrne,
others}. Other approaches include large or warped extra
dimensional models, extended gauge structure and other
alternatives \cite{Nath,Graesser,Park,Song}.

Here, we compute such $(g-2)_{\rm anom}$ contributions from the
anomalous $U(1)$s and show that they are in the range implied by
the experimental result. We use (\ref{AMMexp}) to provide precise
constrains for the masses of the anomalous U(1)'s in the TeV
range.

To derive the AMM of a lepton, we consider the three-point
function of two leptons and a photon where a gauge boson or the
two scalars can be exchanged on the internal line:
\vspace{.8cm}
\begin{center}
\unitlength=1.2mm
\begin{fmffile}{fmfdf4}
\begin{fmfgraph*}(40,25)
\fmfpen{thick} \fmfleft{i1,i2} \fmflabel{$p'$}{i2} \fmflabel{$p
$}{i1} \fmfright{o2} \fmflabel{$A^\m$}{o2} \fmf{fermion}{i1,v1}
\fmf{fermion,label=$p-k$,l.side=right}{v1,v2}
\fmf{fermion,label=$p'-k$,l.side=right}{v2,v3}
\fmf{fermion}{v3,i2} \fmf{photon}{v2,o2} \fmffreeze
\fmf{photon,label=$k$,l.side=left}{v1,v3}
\end{fmfgraph*}
\end{fmffile}
\end{center}
We sandwich the above diagram between two on-shell spinors, so we
can use the Gordon decomposition and the mass-shell conditions.
Our goal is to write the expression in the form:
\be \bar{u}(p') \Big\{ \g_\m F_1(q^2)+{i \s_{\m\n} q^\n \over
2m}F_2(q^2) \Big\} u(p) \label{goal} \ee
where $q_\m=p'_\m-p_\m$. The $F_2(q^2=0)$ will give us a
correction of the AMM of the lepton which propagates. In the
present calculation, we have to include diagrams which are coming
from the non trivial couplings between the anomalous $U(1)$s and
leptons. The external vector gauge abelian field is the photon,
the internal propagating fields with momentum $k$ can be the
anomalous $U(1)$ gauge boson or the scalars (axions). We will
outline here these calculations. More details can be found in
appendix B.

As the anomalous $U(1)$ couples differently to left and right
leptons, it is neccesary to consider  diagrams where chirality is
conserved (L-L, R-R diagrams) and others where chirality is
different (L-R, R-L). The corresponding diagrams are
\vspace{.8cm}
\begin{center}
\unitlength=1.2mm
\begin{fmffile}{fmfdf5}
\begin{fmfgraph*}(40,25)
\fmfpen{thick} \fmfleft{i1,i2} \fmflabel{$\y_s$}{i2}
\fmflabel{$\y_l$}{i1} \fmfright{o2} \fmflabel{$A^\m$}{o2}
\fmf{fermion,label=$p$,l.side=right}{i1,v1}
\fmf{fermion,label=$p-k$,l.side=right}{v1,v2}
\fmf{fermion,label=$p'-k$,l.side=right}{v2,v3}
\fmf{fermion,label=$p'$,l.side=right}{v3,i2} \fmf{photon}{v2,o2}
\fmffreeze
\fmf{photon,label=$B_{anomalous}(k)$,l.side=left}{v1,v3}
\end{fmfgraph*}
\end{fmffile}
\end{center}
and in algebraic form:
\be \bar{u}(p')[\int {d^4k\over (2 \pi)^4} (iQ_s \g_\n P_s)
{i\over \sla{p'}-\sla{k}-m} \g_\m
                {i\over \sla{p}-\sla{k}-m} (iQ_l \g_\rho P_l) D^{\n\rho}(k)] u(p)
\label{oneloop} \ee
where $s, l = L, R$ label the chirality.

The propagator of $U(1)$ contains the arbitrary gauge fixing
parameter $\l$. In a non-chiral theory $\l$ disappears because of
the mass-sell conditions of the two spinors which sandwich the
diagrams (\ref{oneloop}). In a chiral theory, we need the
contribution of $b'$ with mass (\ref{masses}) in order to obtain a
gauge invariant result. We also have to add the one-loop diagrams
of $\phi'$. These diagrams are:
\vspace{.8cm}
\begin{center}
\unitlength=1.2mm
\begin{fmffile}{fmfdf9}
\begin{fmfgraph*}(40,25)
\fmfpen{thick} \fmfleft{i1,i2} \fmflabel{$\y$}{i2}
\fmflabel{$\y$}{i1} \fmfright{o2} \fmflabel{$A^\m$}{o2}
\fmf{fermion,label=$p$,l.side=right}{i1,v1}
\fmf{fermion,label=$p-k$,l.side=right}{v1,v2}
\fmf{fermion,label=$p'-k$,l.side=right}{v2,v3}
\fmf{fermion,label=$p'$,l.side=right}{v3,i2} \fmf{photon}{v2,o2}
\fmffreeze \fmf{scalar,label=$axion(k)$,l.side=left}{v1,v3}
\end{fmfgraph*}
\end{fmffile}
\end{center}
where ``axion" stands for $b'$ or $\f'$. In algebraic form they
are given by:
\be {m^2 \Delta Q^2 \over \m^2} \bar{u}(p') \int {d^4k\over (2
\pi)^4} \g_5 {i\over \sla{p'}
             -\sla{k}-m} \g_\m {i\over \sla{p}-\sla{k}-m} \g_5 G_{b'}(k) u(p)
\label{AxionBloop} \ee
for the $b'$ axion and
\be {(h c)^2 M_s^2 \over \m^2} \bar{u}(p') \int {d^4k\over (2
\pi)^4} \g_5 {i\over \sla{p'}
             -\sla{k}-m} \g_\m {i\over \sla{p}-\sla{k}-m} \g_5 G_{\phi '}(k) u(p)
\label{AxionFloop} \ee
for $\f'$. We expect the sum of the three diagrams to be
$\l$-independent. In the appendix we show it explicitly. In view
of this, we can use any gauge for the evaluation. For simplicity,
we choose the Feynman - t'Hooft gauge $\l=1$

The steps of this calculation are as follow:
\begin{itemize}
\item[a.] Express the denominator as a perfect square using the
Feynman parameter trick and shifting the loop momentum.
\item[b.] Move all the $\sla{p'}$ to the left, all the $\sla{p}$
to the right and make use of the on-shell spinor conditions.
\item[c.] Perform the momentum integral of the loop after  a Wick
rotation to Euclidean space.
\item[d.] Distinguish terms proportional to $p_\m$ and $p'_\m$.
\item[e.] Integrate the remaining variables that resulted from
Feynman parameter trick.
\end{itemize}
Following the steps above, we find for the anomalous $U(1)$
exchanged diagram (details can be found in Appendix B):
For L-L and R-R diagrams:
\be -{Q_L^2+Q_R^2 \over 16m\pi^2}(p_\m+p'_\m)\int_0^1 dx
{x(x^2-3x+2)\over x^2+(1-x){\m^2\over m^2}} \label{ammchiral} \ee
For mixed diagrams (L-R and R-L):
\be -{Q_L Q_R\over 16m\pi^2} (p_\m+p'_\m)\int_0^1 dx {2x(1-x)\over
x^2+(1-x){\m^2\over m^2}} \label{ammmixed} \ee
The axion $b'$ exchange diagram gives
\be {\Delta Q^2\over 16m\pi^2} {m^2\over \m^2}
(p_\m+p'_\m)\int_0^1 dx {x^3\over x^2+(1-x){\m^2\over m^2}}
\label{ammAxionB} \ee
The diagram for the axion $\f'$ has the same integral with
(\ref{ammAxionB}) in the limit $\m\rightarrow 0$. Since however
the axion is expected to get a small mass from non-perturbative
effects we will consider it with $m_{\f'}$ small. In this case we
obtain
\be {(h c)^2\over 16m\pi^2} {M_s^2\over \m^2} (p_\m+p'_\m)
\int_0^1 dx {x^3\over x^2+(1-x){m_{\f'}^2\over m^2}}
\label{ammAxionFmassive} \ee
As $M_s/\m\sim 1$, the limit of (\ref{ammAxionFmassive}) for
$m_{\f'}\rightarrow 0$ is:
\be {h^2\over 16m\pi^2} (p_\m+p'_\m){1 \over 2} \label{ammAxionF}
\ee

\subsection{Anomalous magnetic moment of muon in the D-brane
         realization of the standard model}

Using the results above we can now embark in the calculation of
the AMM of the muon in the D-brane realization of the SM. To do
this we have to include the contribution of (\ref{ammchiral}) and
(\ref{ammmixed}) for both anomalous $U(1)$s as well as the
(\ref{ammAxionB}) and (\ref{ammAxionF}) for the axion diagrams to
the SM result\footnote{We use for simplicity $m=m_{muon}$.}.
\bea \d\a &=& {h^2\over 16\p^2} +{1\over 8\p^2}\times \label{FullAMM}\\
&&\sum_{i=\a,b}\Big({m\over \m_i}\Big)^2 \int_0^1 dx {x(m^2\Delta
Q_i^2 x^2+\m_i^2(4 Q_{iL} Q_{iR}-(2-x)(Q^2_{iL}+ Q^2_{iR}))(1-x)
\over m^2 x^2+\m_i^2(1-x^2)}                 \nn \eea
In our case $\m_i \gg m$, therefore we expand the contributions
and keep the terms up to second order in $(\m_i / m)$. The final
result is
\be \a^{U(3)\times U(2) \times U(1)}_{muon}=\a^{SM}_{muon}
+\sum_{i=\a,b} {Q_{i L}^2-3Q_{i L}Q_{i R}+Q_{i R}^2\over 12\p^2}
\Big({m\over \m_i}\Big)^2 +{h^2\over 16\p^2} \label{AMM} \ee
where $Q_{\a L}, Q_{\a R}, Q_{b L}, Q_{b L}$ are the rotated by
(\ref{rotchargone}) or (\ref{rotchargtwo}), charges of
(\ref{charges}). We use as $Q_{iL}$ and $Q_{iR}$ the charges of
the $L$ and $l^c$ in (\ref{charges}).

\begin{figure}
\begin{center}
\epsfig{file=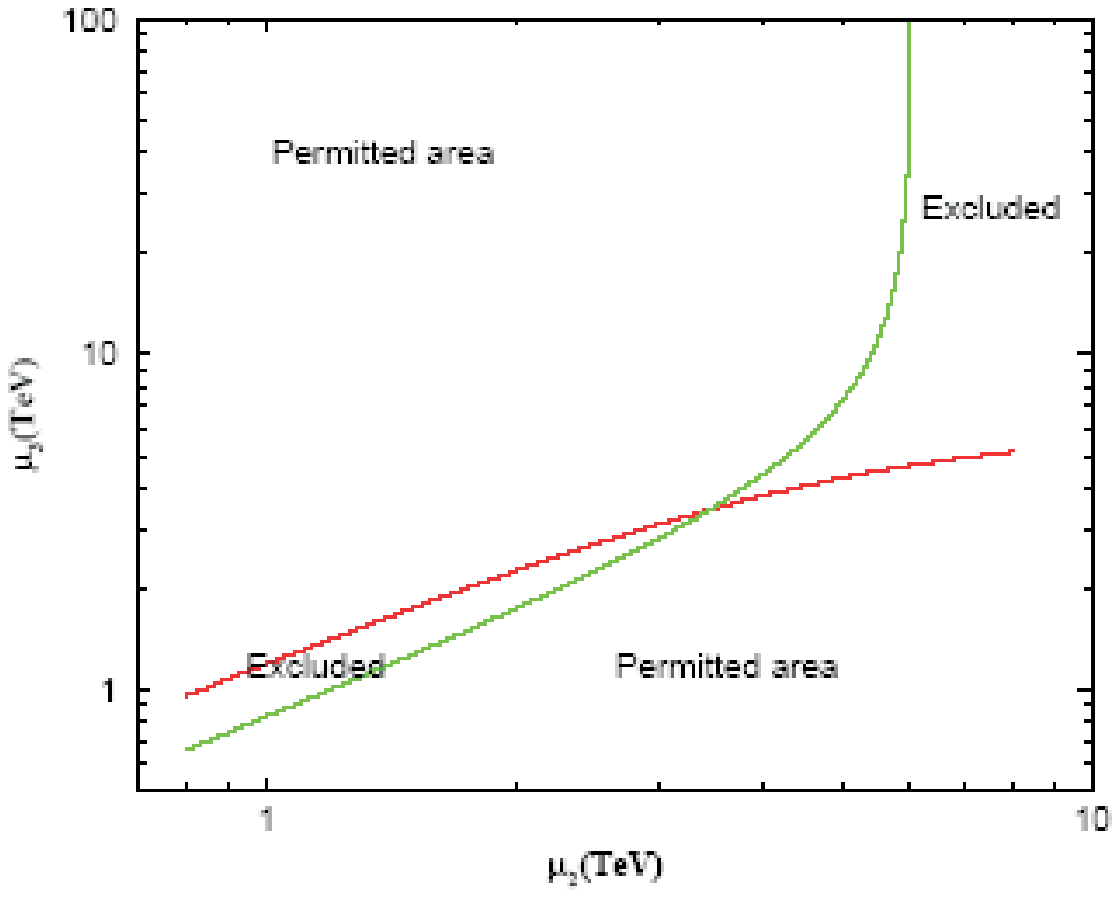,width=100mm}
\end{center}
\caption{The $z=0$ model. Between the two plots is the excluded
area, where the determinant of the second order equation is
negative. \label{restrictarea1}}
\end{figure}

Using the measured difference (\ref{difference}) we can express
one of the unknown variables as a function of the two others.
Thus, for $z=0$ we can find the $\m_\a$ and $\m_\b$ dependence of
$\tan \q$. We have to solve a second order equation:
\bea (12\p^2 \m_\a^2 \m_b^2 (\d\a-\a_{\f'})+m^2(817\m_\a^2
-1220\m_b^2))\tan^2\q
+26\sqrt{215}m^2(\m_\a^2 -\m_b^2)\tan\q &           \nonumber\\
+12\p^2 \m_\a^2 \m_b^2 (\d\a-\a_{\f'}) -1220 m^2 \m_\a^2 +817
m^2\m_b^2&=0
\nonumber\\
& \label{tanthetaFir} \eea
where we denote as $\a_{\f'}$ the contribution from the axion
$\f'$. As $\tan \q$ is real, the discriminant must be positive. We
can easily find the excluded area in the $\m_2$,$\m_3$ plane where
this discriminant is negative. In Fig.\ref{restrictarea1} we plot
this area for the z=0 model.

\begin{figure}
\begin{center}
\epsfig{file=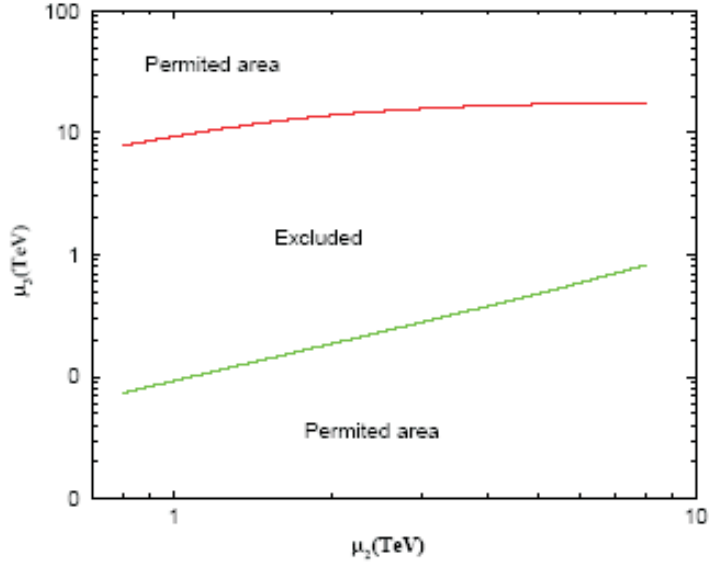,width=100mm}
\end{center}
\caption{The $z=-1$ model. Between the two plots is the excluded
area where the determinant of the second order equation is
negative. \label{restrictarea2}}
\end{figure}

For the $z=-1$ model we obtain
\bea (12\p^2 \m_\a^2 \m_b^2 (\d\a-\a_{\f'}) -m^2(10363\m_\a^2
+580\m_b^2))\tan^2\q
-362\sqrt{215}m^2(\m_\a^2 -\m_b^2)\tan\q &           \nonumber\\
+12\p^2 \m_\a^2 \m_b^2 (\d\a-\a_{\f'}) -m^2(580\m_\a^2
+10363m^2\m_b^2)&=0
\nonumber\\
& \label{tanthetaSec} \eea
and the allowed area is plotted in Fig.\ref{restrictarea2}. As
mentioned before the anomalous $U(1)$ masses are  expected to be
in the TeV range. Thus, there is little allowed space in this case
in order to reproduce the experimental result.

Until now we have evaluated diagrams of the lowest lying  string
states. The massive oscillator string states at level n have
masses equal to $\sqrt{n}M_s$. The ratio of the contribution of
such a state to that of a low lying state is expected to scale as
the square of the ratio of the masses. Thus corrections due to the
first massive level are in the 1-5\% range and higher levels are
further suppressed. There are also KK states that can contribute.
However their masses as mentioned earlier are as large as the
string scale and thus give suppressed contributions.

\subsection{Chapter Conclusions}

In this chapter we have analyzed contributions to the anomalous
magnetic moment of leptons in the minimal D-brane realization of
the Standard Model. We have shown that the two anomalous massive
gauge bosons present with masses in the TeV range, provide
contributions that have the correct order of magnitude to
accommodate the recent experimental data \cite{BNL}. Further
contributions from string oscillators and KK states are expected
to be sufficiently suppressed.

\newpage

\section{Conclusions}

In this thesis we have studied specific kind of open string
theories generated by orientifold models. We have provided the
general consistency conditions and we have derived the general
formulae for the corresponding massless spectrum of various open
string theories. This classification is very important in model
building procedures which target to embed the Standard Model in
string theory.

Applying the same ideas and technics, we could presumably provide
general formulae also in other string theory models with
intersecting branes, fluxes or asymmetric orientifolds.

As we have mentioned, all open string models that approach the
Standard Model contain anomalous $U(1)$ gauge fields. The anomaly
is cancelled via the Green-Schwarz mechanism that generates a mass
for the corresponding anomalous gauge boson. We have evaluated the
bare masses of the anomalous $U(1)$s in four-dimensional
supersymmetric orientifolds. However, we have found that there are
cases where even non-anomalous $U(1)$s acquire a mass and we have
showed that this is due to six-dimensional anomalies that upon
decompactifications affect the four-dimensional theory.

We have also evaluated the general formulae for the bare mass of
anomalous $U(1)$s in non-supersymmetric orientifolds. This is
important since the proper D-brane realization of the Standard
Model will have broken supersymmetry.

These results and formulae have direct implications for model
building both in string theory and field theory orbifolds. They
provide a necessary and sufficient condition for a non-anomalous
$U(1)$ to remain massless (the hypercharge for example). One has
just to check the associated higher dimensional anomalies in the
various decompactification limits.

We have studied other mass sources for the anomalous $U(1)$s. The
D-brane realizations of the Standard Model require that the
Higgses (usually there are more than one to give masses to all
quarks and leptons) are charged under the anomalous $U(1)$s. This
generates an extra mass source for the anomalous bosons.

Usually, there are mixings between the axions that cancel the
anomalies and the Higgses of the theory, therefore some of the
axions acquire masses. The study of these massive axions can
provide very interesting results that could eventually also be
tested at LHC, if the string scale is of order of a few TeV.

Finally, we have evaluated the contribution of the extra $U(1)$
fields to the anomalous moments of the leptons and it has been
shown that this imposes constraints on the magnitude of the string
scale.

\newpage


\centerline{\bf\Large Acknowledgments}

\vskip 1cm

First of all I would like to thank Prof. Elias Kiritsis who
offered me the possibility to do my Ph.D. under his supervision.
I am also grateful to Prof. Theodore Tomaras for being a great
teacher in my Diploma and in my Ph.D. time.
I would like to thank Dr. Amine Bouziane Hammou for helping me
very much and for very fruitful collaboration. Also I would like
to thank Dr. Nikos Irges for help and collaboration.

I would like to thank Prof. Massimo Bianchi, Prof. Mariano Quiros
and Dr. Marco Serone for useful discussions.

I would also like to thank Dr. Dominic Clancy, Dr. Amine Bouziane
Hammou, Gregoris Panotopoulos and Anne-Lise Chagneau for
proof-reading of my thesis.

I am also grateful to University of Crete and to Laboratoire de
Physique Th{\'e}orique Ecole Polytechnique for support and
hospitality.

I would like to thank the Laboratoire de Physique Th{\'e}orique de
l' Ecole Normal Sur\'eriere for hospitality.
I would also like to thank the Universitat Aut\`onoma de
Barcelona, the Universit\'a di Roma ``Tor Vergata", the Scuola
Internationale Superiore di Studi Advanzati (SISSA) for
hospitality during the last stage of this thesis.

This thesis was financially supported from the ``Herakleitos"
program of the Greek Ministry of Education.

Finally, I would like to thank my parents for supporting me all
these years of studies and all my friends.


\newpage

\bigskip\appendix

\centerline{\bf\Large Appendices}


\section{Definitions and identities}

The Dedekind function is defined by the usual product formula
(with $q=e^{2\pi i\tau}$)
\be \eta(\tau) = q^{1\over 24} \prod_{n=1}^\infty (1-q^n)\ . \ee
The Jacobi $\vartheta$-functions with general characteristic and
arguments  are
\bea \vartheta [^\a_\b] (z\vert\tau) &=& \sum_{n\in Z}
e^{i\pi\tau(n-\a/2)^2} e^{2\pi i(z- \b/2)(n-\a/2)}\nn\\
\vartheta [^\a_\b] (z\vert\tau) &=& \h ~e^{i\p \a(z+\b/2)}
q^{{\a\over 4}-{1\over 24}}\times \nn\\
&&~~~~~~~\prod_{n=1}^\infty \left(1+e^{2\p
i(z+\b/2)}q^{n+{\a-1\over 2}}\right) \left(1+e^{-2\p
i(z+\b/2)}q^{n+{\a-1\over 2}}\right) \eea
We define:
$\vartheta_1(z|\t) = \vartheta \left[^1_1 \right] (z|\t)$,
$\vartheta_2(z|\t) = \vartheta \left[^1_0\right] (z|\t) $,
$\vartheta_3(z|\t) = \vartheta \left[^0_0\right] (z|\t) $,
$\vartheta_4(z|\t) = \vartheta \left[^0_1\right] (z|\t) $.
The modular properties of these functions are:
\bea &&\h(\t+1) = e^{i\p /12}\h(\t)\ , ~ \ \ \vartheta [^\a_\b]
(z|\t+1)= e^{-{i\pi\over4} \a(\a-2)}\vartheta [^{\ \ \ \a}_{\a
+\b-1}](z |\t) \nn\\
&&\h(-1/\t) = \sqrt{-i\t} \h(\t)\ , \ \ \vartheta [^\a_\b]
\left({z \over \t} \Bigl| {-1 \over \t}\right)= \sqrt{-i \t} \
e^{i \p \left({\a \b\over 2} + {z^2 \over \t}\right)} \ \vartheta
[^{\ \b}_{-\a}] (z | \t ) \label{f8} \eea
A very useful identity that is valid for $\sum h_i=\sum g_i=0$ is
\be \sum_{\alpha,\beta=0,1}\h_{\a\b}~
\vartheta\left[^\a_\b\right](v)\prod_{i=1}^{3}
\vartheta\left[^{\a+h_i}_{\beta+g_i}\right]
(0)=\vartheta_1(-v/2)\prod_{i=1}^{3} \vartheta
\left[^{1-h_i}_{1-g_i}\right](v/2)~. \label{SuperID}\ee
%
%
%
%
%
%
%

\section{Partition functions and Lattices}

Let us define some of the objects that we used in this paper. The
oscillator dependant parts are:
\bea
T[^0_v]&=&\frac{1}{2} \sum_{a,b} (-1)^{a+b+ab}
\frac{\vartheta[^a_b]}{\eta} \prod_i -2\sin \p v_i \frac{\vartheta
[^{~~a}_{b+2v_i}]}{\vartheta [^{~~1}_{1+2v_i}]}. \label{Tov}\\
T[^g_v]&=&\frac{1}{2} \sum_{a,b} (-1)^{a+b+ab}
\frac{\vartheta[^a_b]}{\eta} \prod_i 
\frac{\vartheta
[^{a+2g_i}_{b+2v_i}]}{\vartheta [^{1+2g_i}_{1+2v_i}]}. \label{Tgv}
\eea
The lattice parts are:
\bea \L_{m+a,n+b} = {1 \over\eta(q) \eta(\bar{q})} \sum_{m,n}
q^{{\a'\over 4}\left({m+a\over R}+{n+b\over\a'}R\right)^2}
\bar{q}^{{\a'\over 4}\left({m+a\over R}+{n+b\over\a'}R\right)^2}
\label{LI}\eea
and the momentum and winding parts:
\bea P_{m}({i\tau_2}/2) &=& {1 \over\eta({i\tau_2}/2)} \sum_{m}
q^{{\a'\over 4}\left({m\over R}\right)^2} \label{PI}\\
W_{n}({i\tau_2}/2) &=& {1 \over\eta({i\tau_2}/2)} \sum_{n}
q^{{\a'\over 4}\left({n R \over \a'}\right)^2} \label{WI}
\eea

\section{Twisted Tadpoles}

Taking the UV limit of the transpose Klein Bottle and Annulus
(between similar branes) amplitudes, we have the tadpoles square
of an $O$-plane and a $D$-brane respectively.
%
%
%
%
Therefore, we can factorize and compute the contributions of these
hyperplanes. This can be a useful tool to evaluate the tadpole
conditions for a specific model. The contributions have
found\footnote{We remind that $v_a=(v_a^1,v_a^2,v_a^3)$.}:
\vskip 0.5cm
\begin{itemize}
\item Supersymmetric Models
\begin{itemize}
\item $v_a^3=0$
\begin{itemize}
\item $O$-Plane contributions (they all appear with opposite sign
in the NS and R sectors.):
\bea
\Omega \a~
\raisebox{-1.2ex}[0cm][0cm]{\epsfig{file=TadRPP_2.eps,width=12mm}}
~\a^2
&\sim & \sqrt{{\cal V}_3\over \prod_{l=1}^2 2 \sin 2\p v_a^l}
\prod_{l=1}^2 2 \cos \p v_a^l \nn\\
\Omega R_3\a~
\raisebox{-1.2ex}[0cm][0cm]{\epsfig{file=TadRPP_2.eps,width=12mm}}
~\a^2
&\sim & \sqrt{{\cal V}_3\over \prod_{l=1}^2 2 \sin 2\p v_a^l}
\prod_{l=1}^2 2 \sin \p v_a^l \nn\\
\Omega R_i\a~
\raisebox{-1.2ex}[0cm][0cm]{\epsfig{file=TadRPP_2.eps,width=12mm}}
~\a^2
&\sim & \e_{ij}\sqrt{1\over {\cal V}_3\prod_{l=1}^2 2 \sin 2\p
v_a^l} 2 \cos \p v_a^i 2 \sin \p v_a^j
\nn\\&&\textrm{where $i\neq j$} \nn
\eea
\item $D$-brane contribution (they also appear with opposite sign
in the NS and R sectors):
\bea
\textrm{D9}~
\raisebox{-1.2ex}[0cm][0cm]{\epsfig{file=TadBoundary.eps,width=12mm}}
~\a
&\sim & \sqrt{{\cal V}_3\over \prod_{l=1}^2 2 \sin \p v_a^l}
Tr[\g_{\a,9}] \nn\\
\textrm{D5}_3~
\raisebox{-1.2ex}[0cm][0cm]{\epsfig{file=TadBoundary.eps,width=12mm}}
~\a
&\sim & \sqrt{{\cal V}_3\over \prod_{l=1}^2 2 \sin \p v_a^l}
\prod_{l=1}^2 2 \sin \p v_a^l Tr[\g_{\a,5_3}] \nn\\
\textrm{D5}_i~
\raisebox{-1.2ex}[0cm][0cm]{\epsfig{file=TadBoundary.eps,width=12mm}}
~\a
&\sim & \sqrt{1\over {\cal V}_3\prod_{l=1}^2 2 \sin \p v_a^l} 2
\sin \p v_a^j Tr[\g_{\a,5_i}]
\nn\\&&\textrm{where $i\neq j$} \nn
\eea
\end{itemize}
\item $v_a^3\neq 0$:
\begin{itemize}
\item $O$-Plane contributions:
\bea
\Omega \a~
\raisebox{-1.2ex}[0cm][0cm]{\epsfig{file=TadRPP_2.eps,width=12mm}}
~\a^2
&\sim & \sqrt{1\over \prod_{l=1}^3 2 \sin 2\p v_a^l}
\prod_{l=1}^3 2 \cos \p v_a^l \nn\\
\Omega R_i\a~
\raisebox{-1.2ex}[0cm][0cm]{\epsfig{file=TadRPP_2.eps,width=12mm}}
~\a^2
&\sim & \sqrt{1\over \prod_{l=1}^3 2 \sin 2\p v_a^l} 2 \cos \p
v_a^i \prod_{l\neq i} 2 \sin \p v_a^l
\nn\\&&\textrm{where $i=1,2,3$} \nn
\eea
\item D-brane contributions:
\bea
\textrm{D9}~
\raisebox{-1.2ex}[0cm][0cm]{\epsfig{file=TadBoundary.eps,width=12mm}}
~\a
&\sim & \sqrt{1\over \prod_{l=1}^3 2 \sin \p v_a^l}
Tr[\g_{\a,9}] \nn\\
\textrm{D5}_i~
\raisebox{-1.2ex}[0cm][0cm]{\epsfig{file=TadBoundary.eps,width=12mm}}
~\a
&\sim & \sqrt{1\over \prod_{l=1}^3 2 \sin \p v_a^l} \prod_{l\neq
i} 2 \sin \p v_a^l Tr[\g_{\a,5_i}]
\nn\\&&\textrm{where $i=1,2,3$} \nn
\eea
\end{itemize}
\end{itemize}
\item Non-Supersymmetric Models. Breaking SUSY by SS deformation,
acting on the third torus where also $v_a^3=0$ some more tadpoles
are added on the above:
\bea
\Omega R_i h\a~ 
\raisebox{-1.2ex}[0cm][0cm]{\epsfig{file=TadRPP_2.eps,width=12mm}}
~\a^2
&\sim & \e_{ij}\sqrt{1\over {\cal V}_3\prod_{l=1}^2 2 \sin 2\p
v_a^l} 2 \cos \p v_a^i 2 \sin \p v_a^j
\nn\\&&\textrm{where $i\neq j$}\nn\\ 
%
%
\overline{\textrm{D5}}_i~
\raisebox{-1.2ex}[0cm][0cm]{\epsfig{file=TadBoundary.eps,width=12mm}}
~\a
&\sim & \e_{ij}\sqrt{1\over {\cal V}_3\prod_{l=1}^2 2 \sin \p
v_a^l} 2 \sin \p v_a^j Tr[\g_{\a,5_i}] \nn\\&& \textrm{where
$i\neq j$}
\nn\eea
These tadpoles have the same sign in both NS and R sectors.

\end{itemize}




\section{Correlation functions on the annulus}

We present here the derivation of the propagators that we will use
for the calculation of the annulus $\cal A$. This surface can be
defined as quotient of the torus ${\cal T}$ under the involution:
%
\be {\cal I}_{\cal A}(z)=1-\bar{z}. \label{involution} \ee

Thus, the correlators can be expressed in terms of the propagators
on the torus. For the bosonic case we have
\be \langle X(z)X(w)\rangle_{\cal T}= -{1 \over 4} \log
\left|\frac{\vartheta_1(z-w|\t)}{\vartheta'_1(0|\t)} \right|^2+ {
\p (z_2-w_2)^2 \over 2 \t_2} \equiv P_B(z,w) \label{bosonicTapp}
\ee
and symmetrizing under the involution:
\bea \langle X(z)X(w)\rangle_{\cal A} &=& {1 \over 2} [ P_B (z,w)+
P_B ({\cal I}_{\cal A} (z),w)+ P_B (z,{\cal I}_{\cal A} (w))+
P_B ({\cal I}_{\cal A} (z),{\cal I}_{\cal A} (w)] \nonumber\\
&=& P_B (z,w)+ P_B (z,1-\bar{w})~. \label{bosonicACapp}\eea
In the amplitude, the partial derivative of the above correlator
(\ref{bosonicACapp}) appears. Thus, we give the expression that we
use for $w=1/2$:
\be \langle \partial_z X(z)X(1/2)\rangle_{\cal A} = -{1 \over 2}
\left[\partial_z\log \vartheta_1(z-1/2|\t) + {2\p i z_2 \over
\t_2}\right]\label{PartialBosonicAC}\ee
for $z=z_1+iz_2$. We remind also that
$\partial_z=(\partial_{z_1}-i\partial_{z_2})/2$.
For the fermionic correlators on the torus we have the identity:
\be \langle\y(z)\y(w)\rangle^2\left[^{\a}_{\b}\right]=-{1\over
4}{\cal P}(z-w)-\p i\partial_{\t} \log{\vartheta
\left[^{\a}_{\b}\right] (0|\t)\over \h(\t)}
\label{FIdentityTapp}\ee
where ${\cal P}(z-w)$ is the Weierstrass function. Symmetrizing
the torus propagator under the involution we find that
(\ref{FIdentityTapp}) holds also for the annulus.

\section{Computations in Type I orientifolds}\label{AnnulusUVchapter}

In the appendix, we give some more details about the $6D$
computations of the mass term.

\subsection{Open strings attached on the same kind of branes}

The internal partition function of strings attached on the same
kind of branes is:
\be Z_{int,k}^{aa}[^\a_\b]= \prod^2_{j=1} (-2\sin \p k v_j)
{\vartheta[^{~~~\a}_{\b+2kv_j}](0|\t) \over
\vartheta[^{~~~1}_{1+2kv_j}](0|\t)} ~~~~~~~~\textrm{for a=5,9.}
\label{Zint-A} \ee
After the use of (\ref{SuperID}) and the fact that
$\vartheta[^1_1](0|\t)=0$, we find for the annulus amplitude:
\bea {\cal A}^{aa}_k &=& -{1\over 2N}\int [d\t] \t_2^{1+\d/2}[2\p
\h^3(\t)]^\d \left[ {1 \over 2\p \t^3} 4 \sin^2 {\p k\over
N}\right]\nonumber\\
&=&-{(2\p)^\d \over \p N} \sin^2 {\p k\over N} \int_0^{i \infty}
d\t_2 \t_2^{-2+\d/2} \h^{3\d}(\t_2). \label{Annulus11ab-2-A} \eea
We are interested in the UV limit of the above integral. The
annulus moduli is $\t_2=it/2$:
\bea {\cal A}^{aa,UV}_k &=&-{(2\p)^\d \over \p N} \sin^2 {\p
k\over N} 2^{1-\d/2} \int_0^1 dt ~ t^{-2+\d/2} \h^{3\d}(it/2)
\nonumber\\
&=& -{(2\p)^\d \over \p N} \sin^2 {\p k\over N} 2^{1-\d/2}
\int_0^1 dt ~ t^{-2+\d/2} \left[\left({2\over t}\right)^{1/2}
\h\left({2\over t}\right)\right]^{3\d} \nonumber\\
&=& -{4 \over \p^2 \d N} \left( {8 \over \d}\right)^\d \sin^2 {\p
k\over N} ~ \G(1+\d,\p\d/2). \label{Annulus11ab-UV-A} \eea
where $\G(a,x)$ is the incomplete $\G$-function and $\G(1,0)=1$.

\subsection{Open strings attached on different kind of branes}

Strings attached on different kind of branes have coordinates
$X^a$ with mixed Dirichlet-Neumann boundary conditions. Those
coordinates are half-integer moded and there are no windings or
momenta. The fermionic sectors interchange modes between R and NS
(since the R states should have same modes than the coordinates)
keeping the total fermionic pact unchanged. Thus, the internal
partition function for such strings is:
\be Z_{int,k}^{59}[^\a_\b]= \prod^2_{j=1} {\vartheta
[^{~~\a+1}_{\b+2kv_j}] (0|\t) \over
\vartheta[^{~~~0}_{1+2kv_j}](0|\t)}. \label{Zint-59-A} \ee
Following the same procedure, like in the case of the strings with
the same boundary conditions, we substitute (\ref{Zint-59-A}) in
(\ref{Fab-k-SS}) and after a bit of "thetacology" we find:
\be {\cal A}^{59}_k = -{1\over 2N}\int [d\t] \t_2^{1+\d/2}[2\p
\h^3(\t)]^\d \left[ {1 \over 2\p \t^3}\right]
\label{Annulus11ab-2-A_appendix} \ee
The integral is the same as in the case of the strings having the
same boundary conditions. Using the above result we find:
\bea {\cal A}^{59,UV}_k &=& -{1 \over \p^2 \d N} \left( {8 \over
\d}\right)^\d ~ \G(1+\d,\p\d/2). \label{Annulus11ab-59-UV-A} \eea

\section{D-terms and supersymmetry}\label{LagrangianBySUSY}

Consider a generic Lagrangian that depends on chiral fields
$\F_i$, $\bar{\F}_i$ and an abelian vector field:
\bea {\cal L}_{susy} &=& \int d^2\q d^2\bar{\q}~
K(\F_i,\bar{\F}_i,V) +{1\over 4}\int d^2\q~ f(\F_i)W^\a W_\a
+{1\over 4}\int
d^2\bar{\q} ~f(\bar{\F}_i) \bar{W}_{\dot{\a}} \bar{W}^{\dot{\a}}\nn\\
&&+\int d^2\q~ W(\F_i)+\int d^2\bar{\q} ~W(\bar{\F}_i) \nn\eea
where as usual $K$ the K\"ahler potential (arbitrary real
function), $f$ the gauge kinetic function and $W$ the
superpotential (holomorphic functions). The superfields have the
component expansions:
\bea \F_i&=&\f_i +\sqrt{2}\q \y_i +i\q\s^\m\bar{\q}\partial_\m\f_i
+\q\q F_i +{i\over \sqrt{2}} \q\q \bar{\q} \bar{\s}^\m\partial_\m
\y_i
+{1\over 4}\q\q\bar{\q}\bar{\q}\Box \f_i \nn\\
\bar{\F}_j&=&\bar{\f}_j +\sqrt{2}\bar{\q} \bar{\y}_j
-i\q\s^\m\bar{\q}\partial_\m\bar{\f}_j +\bar{\q}\bar{\q} \bar{F}_j
-{i\over \sqrt{2}} \bar{\q}\bar{\q} \partial_\m
\bar{\y}_j \bar{\s}^\m \q +{1\over 4}\q\q\bar{\q}\bar{\q}\Box \bar{\f}_j \nn\\
V&=&-\q\s^\m\bar{\q}A_\m +i\q\q\bar{\q}\bar{\l}
-i\bar{\q}\bar{\q}\q\l+ {1\over 2}\q\q\bar{\q}\bar{\q}D
\nn\eea
in the WZ gauge \cite{Wess:cp, Weinberg:mt}. Expanding the
Lagrangian in component fields we have:
\bea {\cal L}_{susy} &=&
-{1\over 4}A^\m A_\m \partial^2_V K|_0
+\left(-\partial_\m\f_i\partial^\m \bar{\f}_j
-{i\over2}\y_i\s^\m\partial_\m \bar{\y}_j
-{i\over2}\bar{\y}\bar{\s}^\m\partial_\m\y_j\right)
K_{i\bar{j}}\nn\\
&&+\left({i\over \sqrt{2}}\l\y_i+ {i\over 2}A^\m
\partial_\m \f_i\right)\partial_VK_i
-\left({i\over \sqrt{2}}\bar{\l}\bar{\y}_j+ {i\over 2}A^\m
\partial_\m \bar{\f}_j\right) \partial_VK_{\bar{j}}
+{1\over 2}\bar{\y}_j\bar{\s}^\m A_\m \y_i
\partial_VK_{i\bar{j}}
\nn\\
&&-{1\over 4} \Re(f_0)F^{\m\n}F_{\m\n} +i \Re(f_0) \bar{\l}
\bar{\s}^\m\partial_\m\l
-{1\over 4} \Im(f_0)F^{\m\n}\tilde{F}_{\m\n}\nn\\
&&-{1\over2}\y_i\y_l\bar{F}_j K_{il{\bar{j}}}
-{i\over2} \bar{\y}_j \bar{\s}_\m \y_i\partial^\m \f_l
K_{il{\bar{j}}}
+{i\over2} \partial^\m \bar{\f}_j \bar{\y}_m \bar{\s}_\m \y_l
K_{i{\bar{j}\bar{m}}}
+{1\over4}\bar{\y}_j \bar{\y}_m \y_i \y_l K_{il{\bar{j}\bar{m}}}
+{\cal V}[\f_i,\bar{\f}_j]
\nn\eea
The indexes $i, \bar{j}$ denote derivatives for $\f_i, \bar{\f}_j$
respectively. The $\Re,\Im$ denote Real and Imaginary parts. It is
taken $V=0=\q= \bar{\q}$ and $f_0\equiv f(\f_0)$. The $K$ are
functions of only the lowest component of the chiral fields (the
scalar fields). Notice that already, we can solve for the
auxiliary fields:
Where we define the real potential for the scalar fields:
\bea {\cal V}[\f_i,\bar{\f}_j] =W_i F_i+{1\over
4\Re{f_0}}(\partial_V K|_0)^2 \nn\eea

\subsection*{Green-Schwarz anomaly cancellation}

Consider now a K\"ahler potential and a gauge function suitable
for a model with anomalous $U(1)$s $V_\a$. The form of these
functions will be:
\bea K=K(\F_\a+\bar{\F}_\a+c^\a V_\a ; \bar{\F}_i e^{2q^\a_i
V_\a}\F_i)\quad ,\quad f=f(\F_\a)~~.\nn\eea
where $i$ for various chiral fields and $\a$ for the axions. We do
not include non anomalous $U(1)s$ for simplicity. We consider
diagonalized axions, one to one with the anomalous $U(1)$s. Notice
that the first combination gives:
\bea \F_\a+\bar{\F}_\a+c^\a V_\a&=&2s_\a +\sqrt{2}\q
\y_\a+\sqrt{2}\bar{\q} \bar{\y}_\a
-2\q\s^\m\bar{\q}\left(\partial^\m b_\a+{c^\a\over
2}A_\a^\m\right)
+\q\q F_\a +\bar{\q}\bar{\q} \bar{F}_\a\nn\\
&&+{i\over \sqrt{2}} \q\q  \bar{\q} (\bar{\s}^\m \partial_\m \y_\a
+\sqrt{2}c^\a \bar{\l}_\a)
-{i\over \sqrt{2}} \bar{\q}\bar{\q} (\partial_\m \bar{\y}_\a
\bar{\s}^\m + \sqrt{2}c^\a\l_\a)\q\nn\\
&&+{1\over 2}\q\q\bar{\q}\bar{\q}(\Box s_\a+c^\a D_\a)
\nn\eea
We have separated the lowest component of the axionic superfield
$\F_\a|_{\q=\bar{\q}=0}=\f_\a=s_\a+ib_\a$. The axion appears
always in the combination: $\partial^\m b_\a+{c^\a\over 2}A^\m_\a$
that is gauge invariant for $A^\m_\a\to A^\m_\a+\partial^\m \e_\a$
and $b_\a \to b_\a-{c^\a\over 2}\e_\a$.

In heterotic string theory, there is at most one anomalous gauge
boson and one axion is needed. In this case $s_0$ is the dilaton
and $b_0$ the dual to the model independent antisymmetric tensor.
In type I, there can be many anomalous $U(1)$s and the role of the
axions $b_\a=\d^k_\a b_k$ are played by the RR twisted fields ($k$
denotes the sector). The $s_\a=\d^k_\a m_k$ are the NSNS twisted
moduli corresponding to the blowup modes associated with the
singularities of the orbifold.

We will focus in an effective field theory that is coming from an
orientifold of Type IIB string theory. In this case the K\"ahler
potential is quadratic to chiral fields
\bea K={1\over 2}\sum_\a(\F_\a +\bar{\F}_\a +c^\a
V_\a)^2+\sum_i\bar{\F}_i e^{2q^\a_iV_\a}\F_i \eea
The lagrangian is simplified a lot:
\bea
{\cal L}_{susy}&=&
-{1\over 4} \Re(f_\a)F_\a^{\m\n}F_{\a\m\n} +i
\Re(f_\a)\bar{\l}_\a\bar{\s}^\m\partial_\m\l_\a
-{1\over 4} \Im(f_\a)F_\a^{\m\n}\tilde{F}_{\a\m\n}\nn\\
%
%
&& -\left(\partial_\m s_\a \partial^\m s_\a +\left(\partial_\m
b_\a+{c^\a\over 2}A_{\a\m}\right)^2
+{i\over2}\y_\a \s^\m\partial_\m \bar{\y}_\a
+{i\over2}\bar{\y}_\a\bar{\s}^\m\partial_\m \y_\a
\right)\nn\\
&& +{ic^\a\over \sqrt{2}}(\l_\a\y_\a-\bar{\l}_\a\bar{\y}_\a)\nn\\
%
%
&&-(\partial_\m+iq^\a_iA_{\a\m})\f_i
(\partial^\m-iq^\a_iA_\a^{\m}) \bar{\f}_i\nn\\
&&
-{i\over2} \bar{\y}_i \bar{\s}^\m
\bigg(\partial_\m+iq^\a_iA_{\a\m} \bigg)\y_i
-{i\over2}\y_i \s^\m \bigg(\partial_\m-iq^\a_iA_{\a\m}
\bigg)\bar{\y}_i
\nn\\
&&+\sqrt{2}iq^\a_i \bigg(\l_\a\y_i\bar{\f}_i
-\bar{\l}_\a\bar{\y}_i\f_i \bigg) +{\cal
V}[\f_\a,\bar{\f}_\a,\f_i,\bar{\f}_j]
\label{Lagrangian2Chiral1U1}
\eea
We should transform everything from the Weyl to the usual Dirac
basis. We have two Weyl spinors $\l_\a^D=\left(\ba {c} i\l_\a \\
-i\bar{\l}_\a \ea\right)$, $\y^D_\a=\left(\ba{c}\y_\a \\
\bar{\y}_\a\ea\right)$ (obviously the indices are not spinor)
and we construct Dirac spinors by $\y^D_i=\left(\ba{c}\y_{2i-1} \\
\bar{\y}_{2i}\ea\right)$ for $i=1 \ldots N$. In case $q_{2i-1}
\neq q_{2i}$ the model is chiral. If in addition $\sum_iq_i\neq 0$
the gauge boson $A_\a^\m$ is anomalous.
Putting all together:
\bea
{\cal L}_{susy}&=&
-{1\over 4} \Re(f_\a)F_\a^{\m\n}F_{\a\m\n} +i
\Re(f_\a)\bar{\l}_\a\bar{\s}^\m\partial_\m\l_\a
-{1\over 4} \Im(f_\a)F_\a^{\m\n}\tilde{F}_{\a\m\n}\nn\\
%
%
&& -\left(\partial_\m s_\a \partial^\m s_\a
+\left(\partial_\m b_\a+{c^\a\over 2}A_\a^\m\right)^2
+{i\over 2} \bar{\y}_\a^D \sla{\partial} \y_\a^D \right)\nn\\
&&+ {c^\a\over 2\sqrt{2}}\bigg(\bar{\l}_\a^D\y^D_\a+\bar{\y}_\a^D
\l^D_\a\bigg) \nn\\
%
%
%
&&-(\partial_\m+iq^\a_iA_{\a\m})\f_i (\partial^\m-iq^\a_iA_\a^\m)
\bar{\f}_i
-i\bar{\y}^D_i
\g^\m\bigg(\partial_\m+i(q^\a_{2i-1}P_L+q^\a_{2i}P_R)A_{\a\m}
\bigg) \y_i^D
\nn\\
&& +\sqrt{2}\Bigg[\bar{\l}_\a^D\bigg(
q^\a_{2i-1}\bar{\f}_{2i-1}P_L+q^\a_{2i}\f_{2i}P_R\bigg)\y_{i}^D
+\bar{\y_{i}}^D\bigg(q^\a_{2i-1}\f_{2i-1}P_R+q^\a_{2i}\bar{\f}_{2i}P_L
\bigg)\l_\a^D\Bigg]\nn\\
&& +{\cal V}[\f_\a,\bar{\f}_\a,\f_i,\bar{\f}_j] \nn
\eea
Notice the $1/2$ in front the Weyl spinor $\l_\a^D$,  $\y_\a^D$.
The mass-matrix of the fermions is not diagonal.

The potential ${\cal V}[\f_\a,\bar{\f}_\a,\f_i,\bar{\f}_j]$
provides the D-term:
\bea {1\over 4\Re{f_0}}(\partial_V K|_0)^2 = {1\over
2\Re{f_0}}\bigg(c^\a s_\a+\sum_i q^\a_i \bar{\f}_i
{\f}_i\bigg)^2\nn\eea
%
%


\section{The extended Standard Model fields}

In this appendix we provide some more details about the masses of
the fields and the gauge couplings. Based on (\ref{Higgs}) the
Higgs expectation values  have the form:
\bea h={v\over \sqrt{2}}
  \left(\ba {c} 1\\ 0 \ea\right) & , &
\bar{h}={v\over \sqrt{2}}
        \left(\ba {c} 1\\ 0 \ea\right) .
\label{vev} \eea
Thus, the covariant derivative of the Higgs (in the $z=0$ model)
is
\bea &&D^{\m} H={v\over \sqrt{2}}(\partial^\m-i{g_3 \bf{1} \over
\sqrt{2}}A_1^\m
          -i{g_2 \bf{1} \over 2}A_2^\m -i{g_2 \over 2} \t_\a W^\m_\a )
          \left(\ba {c} 1\\ 0 \ea\right) e^{i \f}\\
%
%
&&D^{\m} H'={v\over \sqrt{2}}(\partial^\m +i{g_2 \bf{1}\over
2}A_2^\m
           -i{g_2 \over 2} \t_\a W_\a^2 )
           \left(\ba {c} 1\\ 0 \ea\right) e^{i \f'}
\eea
where $W_\a$, $\a=1,2,3$ the $SU(2)$ gauge bosons. We normalize
all $U(N)$ generators according to $Tr T^\a T^b=\d^{\a b}/2$ and
measure the corresponding $U(1)_N$ charges with respect to the
coupling $g_N/\sqrt{2N}$, with $g_N$ the $SU(N)$ coupling constant
as in \cite{Antoniadis:2000en}. We have also $g_1=g_3$.

The mass matrix for the gauge bosons is
\be M=V^T m V \ee
where $V^T=(A_1,A_2,A_3,W_3,W_1,W_2)$ and
\be m={v^2\over 4} \left(\ba {cccccc}
      g_3^2 & {g_2 g_3 \over \sqrt{2}} & 0 &  {g_2 g_3 \over \sqrt{2}} & 0     & 0 \\
    {g_2 g_3 \over \sqrt{2}} & g_2^2          & 0 & 0                  & 0     & 0 \\
                           0         &    0   & 0 & 0                  & 0     & 0 \\
            {g_2 g_3 \over \sqrt{2}} &    0   & 0 & g_2^2              & 0     & 0 \\
                    0                &    0   & 0 & 0                  & g_2^2 & 0 \\
                    0                &    0   & 0 & 0                  & 0     & g_2^2

\ea\right). \label{bosonmassmatrix} \ee
Doing a rotation with the matrix (\ref{Uarrey}), we can go to a
basis where $\tilde{A_1}$ is the hypercharge. The other two $U(1)$
bosons $\tilde{A_2}, \tilde{A_3}$ are anomalous and we expect two
axions $\a_2, \a_3$ to cancel the anomalies. Inserting
\be {\cal L}_{axionic terms} = {1\over 2}(\partial \a_2 -M_2
\tilde{A_2})^2 +{1\over 2}(\partial \a_3 -M_3 \tilde{A_3})^2, \ee
two elements of the rotated mass matrix will be shifted. Since
$v\ll M_2,M_3\sim M_s$, we can perturbatively diagonalize this
matrix and find the new masses of these new fields. Finally, there
is a massless state (photon), a ``light" $Z$ boson with mass
\be m^2_Z={v^2 g_2^2 r^2 \over 2 t^2} -v^4{g_2^2 g_3^2 r^2 s^2
(M_2^2+M_3^2+(M_2^2-M_3^2)\cos 2\q) \over 64 t^4 M_2^2 M_3^2} +O
\Big[ {M_Z^6 \over M_s^4}\Big] \ee
and two heavy ones with masses:
\bea \m^2_2     =& M_2^2+v^2{8g_2^4 t^2 \cos^2\q
                   +g_3 \sin\q (-4g_2^2 t^3 \cos\q
                   +g_3 (130g_2^4+66 g_2^2 g_3^2 +9 g_3^4)\sin\q)
                        \over 2 s^2 t^2} +O \Big[ {M_Z^4 \over M_s^2}\Big] \nonumber\\
\m^2_3     =& M_3^2+v^2{g_3^2(130g_2^4+66 g_2^2 g_3^2 +9
g_3^4)\cos^2\q
                   +4 g_2^2 g_3 t^3 \cos\q \sin\q +8 g_2^4 t^2 \sin^2\q
                        \over 2 s^2 t^2} +O \Big[ {M_Z^4 \over M_s^2}\Big]
\eea where $t=\sqrt{14 g_2^2+3 g_3^2}$, $s=\sqrt{16 g_2^2+6
g_3^2}$, $f=\sqrt{11 g_2^2+3 g_3^2}$, $r=\sqrt{7g_2^2 +3g_3^2}$
and $\m_i=m_{A'_i}$, the masses of the new anomalous U(1)s.
The old fields as functions of the new rotated fields are:
\bea A_1 \approx&  {2 \sqrt{3} t g_2 A'_1
              - \sqrt{2} r s \sin\q A'_2
              + \sqrt{2} r s \cos\q A'_3
              - 6 g_2 g_3  W'_3  \over 2 t r}                    \nn\\
A_2 \approx&  {-\sqrt{6}g_3 s t A'_1
              + 4 g_2 r (2 t \cos\q -3 g_3 \sin\q) A'_2
              + 4 g_2 r (3 g_3 \cos\q +2 t \sin\q) A'_3
              + 3\sqrt{2} g_3^2 s W'_3 \over 2 r s t}            \nn\\
A_3 \approx&  {2 g_2 s t A'_1
              + \sqrt{6} r (g_3 t \cos\q +4 g_2^2\sin\q) A'_2
              + \sqrt{6} r (-4g_2^2 \cos\q +g_3 t \sin\q) A'_3
              - 2\sqrt{3} g_2 g_3 s W'_3  \over r s t}           \label{relations}\\
W_3 \approx& -{\sqrt{3} g_3 A'_1 + t W'_3 \over \sqrt{2} r}
\nonumber \eea
where $A'_1$ and $W'_3$ are the photon and the $Z^0$.

It is necessary to add a $R_{\x}$ gauge fixing term. This will
cancel some mixing terms which are coming from the kinetic terms
of the Higgses and it will maintain the manifest unitarity of the
theory with spontaneously broken gauge symmetry.
\bea
{\cal L}_{gauge fixing} &=& \l (\partial A'_1)^2                       \nonumber\\
                        & & +\m \Big(\partial A'_2-
                            v^2{ 2(\f-\f')g_2^2 t \cos\q -g_3(f^2 \f -3g^2_2 \f')\sin\q
                            \over 2 \m t s}
                            -{M_2\over 2\m}\a_2 \Big)^2                \nonumber\\
                        & & +\r \Big(\partial A'_3-
                            v^2{g_3(f^2 \f -3\f' g_2^2)\cos\q +2(\f-\f')g_2^2 t \sin\q
                            \over 2 \r t s}
                            -{M_3\over 2\r}\a_3 \Big)^2                \nonumber\\
                        & & +\s \Big(\partial W'_3+
                            v^2{(\f+\f')g_2 r \over 2 \sqrt{2} \s t}\Big)^2
\eea
The gauge fixing terms give masses to the axions and to the Higgs.
We can diagonalize perturbatively the mass-matrix of these fields.
Considering $\m=\l=\r=\s$ we find one massless and three massive
fields:
\bea &&m^2_{\tilde{a}_2} = {M_2^2 \over 4 \m}+O[M_Z^2] ~~,~~~~~
m^2_{\tilde{a}_3} = {M_3^2 \over 4 \m}+O[M_s^2]  \label{NHAmasses}\\
&& m^2_{\tilde{\f}}  = {1 \over 4 \m}{g_2^2 r^2 v^4 \over t^2}
                   +O \Big[ {M_z^2 \over M_s^2}\Big] ~~,~~~~~
m^2_{\tilde{\F}}  = 0                            \nn \eea
The old fields as a functions of the new ones are:
\bea \a_2 \approx & \tilde{a}_2
               -{v^4 (4 g_2^2 g_3 t^3 \cos2\q
               +(112 g_2^6-106 g_2^4 g_3^2-66 g_2^2 g_3^4-9 g_3^6)\sin2\q)
               \over 2 t^2 s^2 M_2 M_3} \tilde{a}_3
               +{v^2 g_3 s^2 \sin\q \over \sqrt{2} t M_2} \tilde{\f}
               +{v^2 (4 g_2^2 \cos\q - g_3 t \sin\q \over \sqrt{2} s M_2} \tilde{\F}
               \nonumber\\
\a_3 \approx & \tilde{a}_3
               -{v^2 g_3 s^2 \cos\q \over \sqrt{2} 2 t M_3} \tilde{\f}
               +{v^2 (g_3 t \cos\q + 4 g_2^2 \sin\q \over \sqrt{2} s^2 M_3} \tilde{\F}
               \nonumber\\
\f   \approx & {v^2(2 g_2^2 t \cos\q -g_3 f^2 \sin\q)\over t s
M_2}\tilde{a}_2
               +{v^2(g_3 f^2 \cos\q +2 g_2^2 t \sin\q)\over t s M_3}\tilde{a}_3
               +{1 \over \sqrt{2}}\tilde{\f}
               -{1 \over \sqrt{2}}\tilde{\F}
               \nonumber\\
\f'  \approx & {v^2 g_2^2 (-2 t \cos\q +3 g_3 \sin\q)\over t s
M_2}\tilde{a}_2
               -{v^2 g_2^2 (3 g_3 \cos\q +2 t \sin\q)\over t s M_3}\tilde{a}_3
               +{1 \over \sqrt{2}}\tilde{\f}
               +{1 \over \sqrt{2}}\tilde{\F}
\eea
From the trilinear Yukawa couplings we can find how leptons couple
to the new Higgses and axions. 
%
%
%
%
%
%
%
%
%

\section{The evaluation of lepton vertex functions}

Here we will give some details about the calculation of the lepton
AMM. Our goal is to separate from the vertex functions, terms
proportional to $\s^{\m\n} q_m$. As the vertex functions are
sandwiched by two on-shell spinors we can use the Gordon
decomposition and try to distinguish terms proportional to $p^\m$
and $p'^\m$. We will begin with (\ref{oneloop}) for the anomalous
$U(1)$ diagram. We rewrite it here:
\be \bar{u}(p')[\int {d^4k\over (2 \pi)^4} (iQ_s \g_\n P_s)
{i\over \sla{p'}-\sla{k}-m} \g_\m
                {i\over \sla{p}-\sla{k}-m} (iQ_l \g_\rho P_l) D^{\n\rho}(k)] u(p)
\label{oneloopApp} \ee
where $s, l = L, R$ denote the chiralities. The propagator of the
$U(1)$ $D^{\m\n}$ contains the gauge fixing parameter $\l$. This
parameter is expected to disappear from physical gauge invariant
couplings. We will verify explicitly here that $\l$ disappears
from the sum of all the vertex functions. The $D^{\m\n}$ consist
of two terms, one independent and one dependent on $\l$. First, we
will calculate the correction from the $\l$-independent part. In
this case we have a fraction with three factors in the
denominator. Using the Feynman parameter trick we write the
denominator as follows:
\be {1\over ((p'-k)^2-m^2)((p-k)^2-m^2)((k^2-\m^2)} =
          2! \int_0^1 dx \int_0^{1-x} dy {1\over D^3}
\label{FeynmanTrickApp} \ee
where
\be D = k^2-2k(px+p'y)+p^2x+p'^2y-m^2(x+y)-\m^2(1-x-y) \ee
In order to express the denominator as a function of the norm of
the momentum, we shift $k$ to $k+px+p'y$. We find $D = k^2-\Delta$
where
\be \Delta = m^2(x+y)+\m^2(1-x-y) \label{denominatorApp} \ee

Next, we will express the numerator of (\ref{oneloopApp}) in terms
of $k^\m$ in order to integrate on the internal momenta. Because
of the symmetry, two identities are useful here:
\be \int {d^4k\over (2 \pi)^4} {k^\m \over D^3} = 0
\label{idoddApp} \ee
\be \int {d^4k\over (2 \pi)^4} {k^\m k^\n \over D^3} = \int
{d^4k\over (2 \pi)^4} {{1\over 4}k^2 g^{\m\n} \over D^3}
\label{idevenApp} \ee
We keep only terms proportional to even powers of $k^\m$. We will
separate chiral and mixed diagrams:

\noindent (1) $L-L$, $R-R$ diagrams. The numerator of
(\ref{oneloopApp}) with $s=l$ has the form
\be \g_\n {1\pm\g_5 \over 2} (\sla{A}+m)\g_\m(\sla{C}+m) \g^\n
{1\pm\g_5 \over 2} \label{numeratorStyleApp} \ee
which, after some algebra becomes
\be {1\over 2}\g_\n \sla{A}\g_\m \sla{C} \g^\n + {1\over 2} m^2
\g_\n \g_\m \g^\n. \label{numIApp} \ee
Terms which contain one $\g_5$ are orthogonal to $\g_{\m\n}$ and
we can ignore them. Also the second term in (\ref{numIApp}) does
not contribute since it is proportional to $\g_\m$. Thus, only the
first term remains. Shifting $k$ to $k+px+p'y$ we obtain
\be \g^\n ((1-y)\sla{p'}-x \sla{p}-\sla{k}) \g_\m ((1-x)\sla{p}-y
\sla{p'}-\sla{k}) \g_\n \label{numIIApp} \ee
Moving all $\sla{p'}$ to the left, all $\sla{p}$ to the right,
using (\ref{idoddApp}), (\ref{idevenApp}) and on-shell conditions,
we find
\be 4m[(1-2x-y+xy+x^2)p_{\m} + (1-x-2y+xy+y^2)p_{\m}]
\label{numeratorResalt1App} \ee
Here there is a symmetry under the reflection $x\leftrightarrow
y$. Thus, we can make the coefficients of $p_{\m}$ and $p'_{\m}$
equal adding the ``reflected" terms and divide the result by 2.
Now, only the integrals on $x$ and $y$ remain. Integrating on $x$
and making a change of variables, we find:
\be -{Q_s^2\over 16m\pi^2}(p_\m+p'_\m)\int_0^1 dx
{x(x^2-3x+2)\over x^2+(1-x){\m^2\over m^2}} \label{ammchiralApp}
\ee
Our main interest is for $\m\gg m$. Expanding, we find:
\be {Q_s^2\over 16m\pi^2}(p_\m+p'_\m)
     \Big(-{2\over 3}\Big({m\over \m}\Big)^2
         +\Big(-{19\over 12} -2\log\Big({m\over \m}\Big)\Big)\Big({m\over \m}\Big)^4
         +O\Big({m\over \m}\Big)^5\Big)
\label{ammchiralexpandApp} \ee

\noindent (2) $L-R$ and $R-L$ diagrams. The only difference from
the above lies in the numerator. Working similarly, for $s\neq l$
in (\ref{oneloopApp}) we find
\be 4m[(1-2x)p_{\m} + (1-2y)p'_{\m}] \label{numeratorResalt2App}
\ee
and finally
\be -{Q_L Q_R\over 16m\pi^2} (p_\m+p'_\m)\int_0^1 dx {2x(1-x)\over
x^2+(1-x){\m^2\over m^2}} \label{ammmixedApp} \ee
The expansion for $\m\gg m$ gives:
\be {Q_L Q_R\over 16m\pi^2}(p_\m+p'_\m)
     \Big(2\Big({m\over \m}\Big)^2
         -2\Big(-{11\over 3} -4\log\Big({m\over \m}\Big)\Big)\Big({m\over \m}\Big)^4
         +O\Big({m\over \m}\Big)^5\Big)
\label{ammmixedexpandApp}\ee

We will now calculate the contribution of the second
($\l$-dependent) term of the massive gauge field's propagator
(\ref{propagators}) in (\ref{oneloop}). The denominator contains
four factors. We will use again the Feynman parameter trick.

Due to the projection operators, there are terms with two, one and
no $\g_5$. Terms with one $\g_5$ do not contribute to (\ref{goal})
being orthogonal  to both $\g_\m$, $\s_{\m\n}$. Terms without
$\g_5$ vanish using mass-shell conditions of the fermions that
sandwich the diagram. Only terms with two $\g_5$s remain. After a
lot of Diracology we obtain
\bea -(1&-&\l^{-1}){\Delta Q^2 (p_\m+p'_\m) \over 16 \p^2}
\int_0^1 dx \int_0^x dy \int_0^y dz  \times \nonumber\\
& \Bigg( & -{m (-1 + 3 z) \over m^2 y^2 + \m^2 \Big( x-y+ {1-x
\over \l} \Big)} +{m^3 z y^2 \over \Big( m^2 y^2 + \m^2 \Big( x-y+
{1-x \over \l} \Big) \Big)^2} \Bigg) \label{secldeptermApp} \eea

Now, we will calculate the axion diagrams (\ref{ammAxionB}) and
(\ref{ammAxionF}). The $\b'$ axion diagram is equal to
\be {m^2 \Delta Q^2 \over \m^2} \bar{u}(p') \int {d^4k\over (2
\pi)^2} \g_5 {i\over \sla{p'}
             -\sla{k}-m} \g_\m {i\over \sla{p}-\sla{k}-m} \g_5 G_{b'}(k) u(p)
\label{AxionBloopApp} \ee
The only difference with the $U(1)$ diagram (\ref{oneloopApp}) is
in the numerator. So, we focus on it and the result is
\be 2[(x^2+yx)p_{\m} + (y^2+xy)p'_{\m}]. \label{numAxionApp}\ee
Thus, the (\ref{AxionBloopApp}) contribution is
\be {\Delta Q^2\over 16m\pi^2} {\m^2\over m^2} (p_\m+p'_\m)
\int_0^1 dx {x^3\over x^2+(1-x){\m^2\over \l m^2}}
\label{ammAxionBApp} \ee
In the entire contribution only (\ref{secldeptermApp}) and
(\ref{ammAxionBApp}) are $\l$ dependent. Adding these two terms
and calculating the $\l$ derivative  using Mathematica we find
zero. Thus, $\l$ disappears as it should and we can use the
Feynman - t'Hooft gauge for simplicity. As we are interested in
$\m\gg m$, we expand (\ref{ammAxionBApp}):
\be {\Delta Q^2\over 16m\pi^2} {\m^2\over m^2} (p_\m+p'_\m)
     \Big(\Big(-{11\over 6} -2\log\Big({m\over \m}\Big)\Big)\Big({m\over \m}\Big)^4
     +O\Big({m\over \m}\Big)^5\Big).
\label{ammAxionexpandBApp} \ee

Let  us now turn to the $\f'$ diagram. The corresponding integral
is the $\m\rightarrow 0$ limit of the the integral in
(\ref{ammAxionBApp}). However we will consider a more general case
where $\m$ is small. Keeping the same coupling constant as the
above we have
\be {h^2\over 16m\pi^2}(p_\m+p'_\m) \int_0^1 dx {x^3\over
x^2+(1-x){m_{\f'}^2\over m^2}} \label{ammAxionFmassiveApp} \ee
Considering $m_{\f'}$ very small we can expand
(\ref{ammAxionFmassiveApp}) and we find
\be {h^2\over 16m\pi^2}(p_\m+p'_\m) \Big({1\over 2} + \Big(1+
log\Big({m_{\f'}\over m} \Big) \Big) \Big({m_{\f'}\over m}\Big)^2
\Big) +O \Big( {m_{\f'}\over m} \Big)^3 \label{ammAxionexpandFApp}
\ee
In the last formula there is $h$ which is computable from SM. From
(\ref{masses}) is obvious that we need to estimate the expectation
value of the Higgs $v$. Using the mass of $Z^0$
$M_{Z^0}=91.19GeV$, the electron charge $e$ and the value of
$\sin^2\q_W=0.23$ from SM we find $v=2 M_{Z^0} \sin\q_W
\sqrt{1-\sin^2\q_W}/ e$ so
\be h={e m_{muon} \over 2 M_{Z^0} \sin\q_W \sqrt{1-\sin^2\q_W}}
\label{hcoupling} \ee

\section{Massless spectrums of some orientifolds}\label{MasslessSpectrums}


%
\begin{table}[h]
\begin{center}
\footnotesize
\renewcommand{\arraystretch}{1.25}
\begin{tabular}{|c|c|c|}
\hline
Twist Group   & & \\
\cline{1-1} Gauge Group & \raisebox{2.5ex}[0cm][0cm]{ (99)/(55)
matter} &
\raisebox{2.5ex}[0cm][0cm]{ (95) matter}  \\
\hline\hline $Z'_6 $ & $(\overline{4},1,8)+
(1,4,\overline{8})+(6,1,1) $ &
$(\overline{4},1,1;\overline{4},1,1)+(1,4,1;1,4,1)$  \\
\cline{1-1} $U(4)_9^2\times U(8)_9\times$ & $ +(1,\overline{6},1)+
(4,1,8) + (1,\overline{4},\overline{8}) $ &
$ +(1,\overline{4},1;1,1,8)+(1,1,8;1,\overline{4},1) $\\
$U(4)^2_5\times U(8)_5$ & $ +(\overline{4},4,1) + (4,4,1) +
(\overline{4},\overline{4},1) $ &
$+(4,1,1;1,1,\overline{8})+(1,1,\overline{8};4,1,1)$ \\
& $ + (1,1,28) + (1,1,\overline{28}) $& \\
%
%
%
%
\hline\hline $Z_6 $ & $2 (15,1,1)+ 2(1,\overline{15},1) $ &
$(6,1,1;6,1,1)+(1,\overline{6},1;1,\overline{6},1)$  \\
\cline{1-1} $U(6)_9^2\times U(4)_9\times$ & $ +2
(\overline{6},1,4)+ 2(1,6,\overline{4}) $ &
$ +(1,6,1;1,1,\overline{4})+(1,1,\overline{4};1,6,1) $\\
$U(6)^2_5\times U(4)_5$ & $ +(\overline{6},1,\overline{4}) +
(1,6,4) + (6,\overline{6},1) $ &
$+(\overline{6},1,1;1,1,4)+(1,1,4;\overline{6},1,1)$ \\
\hline
\end{tabular}\end{center}
\caption{The transformations of the massless fermionic states in
two D=4 orientifolds.}\label{SpectrumD=4}
\end{table}

\begin{table}[h]
\begin{center}
\footnotesize
\renewcommand{\arraystretch}{1.25}
\begin{tabular}{|c|c|c|}
\hline
Twist Group   & & \\
\cline{1-1} Gauge Group & \raisebox{2.5ex}[0cm][0cm]{ (99)/(55)
matter} &
\raisebox{2.5ex}[0cm][0cm]{ (95) matter}  \\
\hline\hline $Z_2 $ & $2 \times 120+ 2 \times \overline{120} $  &
$(16; \overline{16})+(\overline{16}; 16)$  \\
\cline{1-1}
$U(16)_9\times U(16)_5$ &  & \\
\hline\hline $Z_3$ &$(8,16_v)+
(\overline{8},16_v)$ &  -  \\
\cline{1-1}
$U(8)\times SO(16)$ & $ +(28,1)+ (\overline{28},1) $   & \\
\hline\hline $Z_4 $ & $(28,1)+ (\overline{28},1) $  &
$(8,1;\overline{8},1)+(\overline{8},1;8,1)$  \\
\cline{1-1} $U(8)_9\times U(8)_9\times$ & $ +(1,28)+
(1,\overline{28}) $ &
$ +(1,8;1,\overline{8})+(1,\overline{8};1,8) $\\
$U(8)_5\times U(8)_5$ & $ +(8,\overline{8}) + (\overline{8},8) $ & \\
\hline\hline $Z_6$ &$(\underline{6, 1}, 1)
+(\underline{\overline{6},1},1)$
 &  $(4,1,1;\overline{4},1,1)+(\overline{4},1,1;4,1,1)$ \\
\cline{1-1} $( U(4)^2\times U(8))_9\times$ & $+
(\underline{4,1},\overline{8})+(\underline{\overline{4},1},8)$
 & $+(1,4,1;1,\overline{4},1)+(1,\overline{4},1;1,4,1)$  \\
$+( U(4)^2\times U(8))_5$    &
& $+(1,1,8;1,1,\overline{8})+(1,1,\overline{8};1,1,8)$ \\
\hline
\end{tabular}\end{center}
\caption{The transformations of the massless fermionic states in
all the D=6 orientifolds. The underlined numbers denote all the
possible permutations.\label{SpectrumD=6}}
\end{table}
\newpage
\begin{table}[h]
\footnotesize \renewcommand{\arraystretch}{1.25}
\begin{tabular}{|c c c|}
%
%
\hline \hline
& & \\
& \raisebox{2.5ex}[0cm][0cm]{\textbf{Z}$_2$} &  \\
\hline \hline \hline
$\g_h^2=-1$ & & \\
$\big\{U(a)\times U(b)\big\}_{9,5}$ &
\raisebox{2.5ex}[0cm][0cm]{~~~~~~~~~~~~~~ (99)/(55) matter
~~~~~~~~~~~~~~} & \raisebox{2.5ex}[0cm][0cm]{~~~~ (59) matter
~~~~}
\\
\hline \hline
Scalars & adjoint + $(a,b) + c.c.$ & $(a,1;\bar{a},1)+
(1,b;1,\bar{b}) + c.c.$\\
\hline
Fermions & $(\Yasymm,1)+(1,\Yasymm)+2(a,\bar{b}) + c.c.$ &
$(a,1;1,\bar{b})+
(1,b;\bar{a},1) + c.c.$\\
\hline \hline
$\g_h^2=+1$ & & \\
$\big\{U(a)\times U(b)\big\}_{9,5}$ &
\raisebox{2.5ex}[0cm][0cm]{(99)/(55) matter} &
\raisebox{2.5ex}[0cm][0cm]{(59) matter}  \\
\hline \hline
Scalars &  adjoint + $(\Yasymm,1)+(1,\Yasymm)+c.c.$ &
$(a,1;\bar{a},1)+
(1,b;1,\bar{b}) + c.c.$\\
\hline
Fermions & $(a,b)+2(a,\bar{b})+ c.c.$ & $(a,1;1,\bar{b})+
(1,b;\bar{a},1) + c.c.$\\
\hline
%
%
%
%
%
%
%
%

\hline \hline
& & \\
& \raisebox{.8ex}[0cm][0cm]{\textbf{Z}$_3$} &  \\
\hline \hline \hline
$\g_h^2=-1$ & & \\
$U(a)\times U(b)\times U(8)$ &
\raisebox{.8ex}[0cm][0cm]{~~~~~~~~~~~~~~(99)/(55)~matter~~~~~~~~~~~~~~}
& ~~~~~~~~~~~~~~~~~~~~~~~~~~~~\\
\hline \hline
& & \\
\raisebox{.8ex}[0cm][0cm]{Scalars} & \raisebox{.8ex}[0cm][0cm]{
adjoint $+(a,b,1)+ (\bar{a},1,8)+ (1,b,\bar{8})+c.c.$} &
\\
\hline
& $2 \left( (a,\bar{b},1)+ (1,1,\YasymmS)\right)+(\YasymmS,1,1)+$
& \\
\raisebox{.8ex}[0cm][0cm]{Fermions} & $+(1,\YasymmS,1)+
(\bar{a},1,\bar{8})+ (1,\bar{b},8)+c.c.$ &
\\
\hline \hline
$\g_h^2=+1$ & & \\
$U(a)\times U(b)\times$ & (99) matter & \\
$SO(c)\times SO(d)$ &&\\
\hline \hline
& adjoint $+(\YasymmS,1,1,1)+ (\bar{a},1,c,1)$ & \\
\raisebox{.8ex}[0cm][0cm]{Scalars} & $+(1,\YasymmS,1,1)+
(1,\bar{b},1,d)+c.c.$ &
\\
\hline
& $2 \left((a,\bar{b},1,1)+ (1,1,c,d)\right)
+(\bar{a},\bar{b},1,1)$
& \\
\raisebox{.8ex}[0cm][0cm]{Fermions} & $+(a,1,1,d)+ (1,b,c,1)+c.c.$
&
\\
\hline
%
%
\end{tabular}\caption{The $h$ action on the Chan-Paton charges
breaks the gauge group of the six-dimensional supersymmetric
orientifolds compactified on $K3$. For $Z_2$ we have $a+b=16$ and
for $Z_3$: $a+b=8$.\label{SpectrumZN+SS1}}

\end{table}
\newpage
%
%
%
%
\begin{table}[h]\footnotesize \renewcommand{\arraystretch}{.5}
\begin{tabular}{|c c c|}
\hline \hline
& & \\
& & \\
& & \\
& \raisebox{.8ex}[0cm][0cm]{\textbf{Z}$_4$} &  \\
& & \\
\hline \hline \hline
$\g_h^2=-1$ & & \\
$\big\{U(a)\times U(b)\times$ &
~~~~~~~~~~~~~~(99)/(55)~matter~~~~~~~~~~~~~~
& ~~~~~~~~~~~~~~~(59) matter ~~~~~~~~~~~~~~~\\
$U(c)\times U(d)\big\}_{9,5}$ &&\\
\hline \hline
&  adjoint $+(\bar{a},\bar{b},1,1)+(a,1,\bar{c},1)$ &
$(a,1_3;\bar{a},1_3)+(1,b,1_2;1,\bar{b},1_2)+$ \\
\raisebox{.8ex}[0cm][0cm]{Scalars} & $+(1,b,1,\bar{d})+
(1,1,c,d)+c.c.$ &
$(1_2,c,1;1_2,\bar{c},1)+(1_3,d;1_3,\bar{d})+c.c.$
\\
\hline
& $2\times \left((a,\bar{b},1,1)+(1,1,c,\bar{d})\right)$ &
\\
Fermions &
$+(\YasymmS,1,1,1)+(\bar{a},1,1,\bar{d})+(1,\YasymmS,1,1)$ &
\raisebox{.8ex}[0cm][0cm]{
$(a,1_3;1,\bar{b},1_2)+(1,b,1_2;\bar{a},1_3)$+} \\
& $(1,\bar{b},c,1)+(1,1,\YasymmS,1)+(1,1,1,\YasymmS)+c.c.$ &
\raisebox{.8ex}[0cm][0cm]{$(1_2,c,1;1_3,\bar{d})+
(1_3,d;1_2,\bar{c},1)+c.c.$} \\
\hline \hline
$\g_h^2=+1$ & & \\
$\big\{U(a)\times U(b)\times$ &
(99)/(55) matter & (59) matter \\
$U(c)\times U(d)\big\}_{9,5}$ &&\\
\hline \hline
&  adjoint $+(\YasymmS,1_3)+(\bar{a},1,c,1)+(1,\YasymmS,1_2)$ &
$(a,1_3;\bar{a},1_3)+(1,b,1_2;1,\bar{b},1_2)+$ \\
\raisebox{.8ex}[0cm][0cm]{Scalars} & $+(1,\bar{b},1,d)+
(1_2,\bYasymmS,1)+ (1_3,\bYasymmS)+c.c.$ &
$(1_2,c,1;1_2,\bar{c},1)+ (1_3,d;1_3,\bar{d})+c.c.$
\\
\hline
& $2 \left((a,\bar{b},1,1)+ (1,1,c,\bar{d})\right)+
(\bar{a},\bar{b},1,1)$ &
$(a,1_3;1,\bar{b},1_2)+ (1,b,1_2;\bar{a},1_3)+$\\
\raisebox{.8ex}[0cm][0cm]{Fermions} & $+(a,1,1,\bar{d})+
(1,b,\bar{c},1)+ (1,1,c,d)+c.c.$ &
$(1_2,c,1;1_3,\bar{d})+ (1_3,d;1_2,\bar{c},1)+c.c.$ \\
\hline
%
%
%
%
%
%
%
%
\hline \hline
& & \\
& & \\
& & \\
& \raisebox{.8ex}[0cm][0cm]{\textbf{Z}$_6$} &  \\
& & \\
\hline \hline \hline
$\g_h^2=-1$ & & \\
$\big\{U(a)\times U(b)\times$ & & \\
$ ~~U(c)\times U(d)\times $ & \raisebox{.8ex}[0cm][0cm]{(99)/(55)
matter} & \raisebox{.8ex}[0cm][0cm]{ (59) matter } \\
$~~~~~ U(e)\times U(f) \big\}_{9,5} $& & \\
\hline \hline
&  adjoint $+(\bar{a},\bar{b},1_4)+ (a,1,\bar{c},1_3)+$ &
$(a,1_5;\bar{a},1_5)+ (1,b,1_4;1,\bar{b},1_4)+$ \\
Scalars & $(1,b,1,\bar{d},1_2)+ (1_2,c,1,\bar{e},1)+$ &
$(1_2,c,1_3;1_2,\bar{c},1_3)+ (1_4,e,1;1_4,\bar{e},1)$
\\
& $(1_3,d,1,\bar{f})+ (1_4,e,f)+c.c.$ &
$(1_3,d,1_2;1_3,\bar{d},1_2)+ (1_5,f;1_5,\bar{f})+c.c.$ \\
\hline
& $2 \left( (a,\bar{b},1_4)+ (1_2,c,\bar{d},1_2)+
(1_4,e,\bar{f})\right)+$ & $(a,1_5;1,\bar{b},1_4)+
(1,b,1_4;\bar{a},1_5)+$\\
& $(\bar{a},1_2,\bar{d},1_2)+ (1,\bar{b},c,1_3)+
(1_2,\bar{c},1_2,f)$ &

$(1_2,c,1_3;1_3,\bar{d},1_2)+(1_4,e,1;1_5,\bar{f})$ \\
\raisebox{.8ex}[0cm][0cm]{Fermions}& $+(1,b,1_4;\bar{a},1_5)+
(1_3,\bar{d},e,1)+ (\YasymmS,1_5) $ &
$(1_3,d,1_2;1_2,\bar{c},1_3)+(1_5,f;1_4,\bar{e},1)$\\
& $(1,\YasymmS,1_4)+ (1_4,\bYasymmS,1)+ (1_5,\bYasymmS)+c.c.$ &
$+c.c.$
\\
\hline
%
%
%
%
%
\hline
$\g_h^2=+1$ & & \\
$\big\{U(a)\times U(b)\times$ & & \\
$ ~~U(c)\times U(d)\times $ & \raisebox{.8ex}[0cm][0cm]{(99)/(55)
matter} & \raisebox{.8ex}[0cm][0cm]{ (59) matter } \\
$~~~~~ U(e)\times U(f) \big\}_{9,5} $& & \\
\hline \hline
& adjoint $+(\bar{a},1,\bar{c},1_3)+ (1,\bar{b},1,d,1_2)$ &
$(a,1_5;\bar{a},1_5)+ (1,b,1_4;1,\bar{b},1_4)$ \\
Scalars & $(1_2,\bar{c},1,e,1)+ (1_3,\bar{d},1,f)+(\YasymmS,1_5)$
& $(1_2,c,1_3;1_2,\bar{c},1_3)+ (1_4,e,1;1_4,\bar{e},1) $
\\
& $+ (1,\YasymmS,1_4)+ (1_4,\bYasymmS,1)+ (1_5,\bYasymmS)$ &
$ (1_3,d,1_2;1_3,\bar{d},1_2) + (1_5,f;1_5,\bar{f})$ \\
\hline
& $2\times \left(
(a,\bar{b},1_4),~(1_2,c,\bar{d},1_2),~(1_4,e,\bar{f})\right)$ &
$(a,1_5;1,\bar{b},1_4)+ (1,b,1_4;\bar{a},1_5)$
\\
Fermions & $(\bar{a},\bar{b},1_4)+ (a,1_2,\bar{d},1_2)+
(1,b,\bar{c},1_3)$ &
$(1_2,c,1_3;1_3,\bar{d},1_2)+ (1_2,c,1;1_3,\bar{d})$ \\
& $(1_2,c,1_2,\bar{f})+ (1_3,d,\bar{e},1)+ (1_4,e,f)$ &
$(1_3,d,1_2;1_2,\bar{c},1_3)+ (1_3,d;1_2,\bar{c},1)$ \\
\hline \hline
\end{tabular}\caption{The $h$ action on the Chan-Paton charges
breaks the gauge group of the six-dimensional supersymmetric
orientifolds compactified on $K3$. For $Z_4$ we have $a+b=c+d=8$
and for $Z_6$: $2a+2b=c+d=2e+2f=8$.\label{SpectrumZN+SS2}}
\end{table}
\newpage
%
%
%
%
%
%

\end{document}